\newcommand{\Dslash}{\rlap{/}\kern-2.0pt D}
\newcommand{\psibar}{\overline{\psi}}
\newcommand{\Psibar}{\overline{\Psi}}
\newcommand{\tr}{{\rm Tr}}
\newcommand{\qbar}{\overline{q}}

\def\qbq{\langle \overline{q} q \rangle}
\def\mres{m_{\rm res}}
\def\dmi{\delta m_i}
\def\dmqbq{\delta m_{\langle \, \overline{q} \, q \, \rangle}}
\def\mpi2{m_\pi^2}
\def\chipp{\chi_{\pi \pi}}

\def\chiaa{\chi_{A A}}

\def\chimpp{\chi_{\pi_{\rm (mp)} \pi}}

\def\corrpp{\langle \pi^a(x) \pi^a(0) \rangle}
\def\corrss{\langle \sigma(x) \sigma(0) \rangle_{\rm c}}
\def\corraa{\langle A_0^a(x) A_0^a(0) \rangle}

\def\mppi{\pi_{\rm (mp)}^a}
\def\qw{Q^{\rm (w)}}
\def\qmp{Q^{\rm (mp)}}

\documentstyle[prd,aps,preprint,tighten,epsf]{revtex}
\begin{document}



\preprint{CU-TP-980, BNL-HET-00/20, RBRC-113}

\title{Quenched Lattice QCD with Domain Wall Fermions and the
  Chiral Limit} 

\author{
T.~Blum$^a$,
P.~Chen$^b$,
N.~Christ$^b$,
C.~Cristian$^b$,
C.~Dawson$^c$,
G.~Fleming$^b$,
A.~Kaehler$^b$,
X.~Liao$^b$,
G.~Liu$^b$,
C.~Malureanu$^b$,
R.~Mawhinney$^b$,
S.~Ohta$^{ad}$,
G.~Siegert$^b$,
A.~Soni$^c$,
C.~Sui$^b$,
P.~Vranas$^e$,
M.~Wingate$^a$,
L.~Wu$^b$,
Y.~Zhestkov$^b$}

\address{
\vspace{0.5in}
$^a$RIKEN-BNL Research Center,
Brookhaven National Laboratory,
Upton, NY 11973}

\address{$^b$Physics Department,
Columbia University,
New York, NY 10027}

\address{$^c$Physics Department,
Brookhaven National Laboratory,
Upton, NY 11973}

\address{$^d$Institute for Particle and Nuclear Studies,
KEK,
Tsukuba, Ibaraki, 305-0801, Japan }

\address{$^e$Physics Department,
University of Illinois,
Urbana, IL 61801}

\date{July 25, 2000}
\maketitle

\begin{abstract}

Quenched QCD simulations on three volumes, $8^3 \times$, 
$12^3 \times$ and $16^3 \times 32$ and three couplings, $\beta=5.7$, 
5.85 and 6.0 using domain wall fermions provide a consistent picture 
of quenched QCD.  We demonstrate that the small induced effects of 
chiral symmetry breaking inherent in this formulation can be 
described by a residual mass ($\mres$) whose size 
decreases as the separation between the domain walls ($L_s$) 
is increased.  However, at stronger couplings much larger 
values of $L_s$ are required to achieve a given physical value 
of $\mres$.  For $\beta=6.0$ and $L_s=16$, we find $\mres/m_s=0.033(3)$, 
while for $\beta=5.7$, and $L_s=48$, $\mres/m_s=0.074(5)$, where 
$m_s$ is the strange quark mass.  These values are significantly 
smaller than those obtained from a more naive determination in our 
earlier studies.  Important effects of topological near zero modes 
which should afflict an accurate quenched calculation are easily visible 
in both the chiral condensate and the pion propagator.  These effects can be 
controlled by working at an appropriately large volume.  A non-linear
behavior of $m_\pi^2$ in the limit of small quark mass suggests
the presence of additional infrared subtlety in the quenched 
approximation.  Good scaling is seen both in masses and in $f_\pi$
over our entire range, with inverse lattice spacing varying 
between 1 and 2 GeV.
\end{abstract}


\newpage



\section{Introduction}
\label{sec:intro}

Since spontaneous chiral symmetry breaking is a dominant property of the QCD vacuum
and is responsible for much of the low energy physics seen in Nature, having a
first principles formulation of lattice QCD which does not explicitly
break chiral symmetry has been an important goal.  Both Wilson and
staggered fermions recover chiral symmetry in the continuum limit
but with these techniques the chiral and continuum limits cannot be 
decoupled.  For the QCD phase transition, which is dominantly a chiral 
symmetry restoring transition, a formulation that is free of violations 
of chiral symmetry due to lattice artifacts, should give a phase 
transition more closely approximating that of the continuum limit.
For the measurement of matrix elements of operators in hadronic 
states, a formulation that respects chiral symmetry on the lattice 
substantially reduces operator mixing through 
renormalization.  Lastly, since much of our 
analytic understanding of low-energy QCD is formulated in 
terms of low-energy effective field theories based on chiral symmetry, 
a lattice formulation preserving chiral symmetry allows controlled 
comparison with analytic expectations.

Building on the work of Kaplan \cite{Kaplan:1992bt}, who showed how to
produce light chiral modes in a $d$ dimensional theory as surface
states in a $d+1$ dimensional theory, a number of attractive lattice 
formulations have been developed which achieve a decoupling of the 
continuum and chiral limits.  Here we will use Kaplan's approach as 
was further developed by Narayanan and Neuberger \cite{Narayanan:1993wx,Narayanan:1993ss,Narayanan:1994sk,Narayanan:1995gw} 
and by Shamir \cite{Shamir:1993zy}.  It is Shamir's approach, commonly 
known as the domain wall fermion formulation, which we adopt.  
(For reviews of this topic see 
Refs.~\cite{Narayanan:1993gq,Creutz:1994px,Shamir:1995zx,Blum:1998ud,Neuberger:1999ry} and for more extensive recent references see Ref.~\cite{Chen:2000zu}.)
For a physical
four-dimensional problem, the domain wall fermion Dirac operator, $D$,
is a five-dimensional operator with free boundary conditions for the
fermions in the new fifth dimension.  The desired light, chiral
fermions appear as states exponentially bound to the four-dimensional
surfaces at the ends of the fifth dimension.  The remaining modes for
$D$ are heavy and delocalized in the fifth dimension.

An additional important feature of the domain wall fermion Dirac
operator in the limit $L_s\rightarrow \infty$ is the existence of an
``index'', an integer that is invariant under small changes in the 
background gauge field.  Here $L_s$ is the extent of the lattice in the
fifth dimension.  This property, true for all but a set of gauge 
fields of measure zero, can be readily seen using the overlap formalism
\cite{Narayanan:1993wx,Narayanan:1993ss,Narayanan:1994sk,Narayanan:1995gw}.
In the smooth background field limit, this index is the normal topological 
charge but, even for rough fields, it signals the presence of massless 
fermion mode(s) when non-zero.  These zero modes can easily be recognized
in numerical studies with semiclassical gauge field backgrounds 
\cite{Narayanan:1997sa,Chen:1998ne,Edwards:1998dj,Edwards:1998gk,Edwards:1998sh,Gadiyak:2000kz}.

These powerful theoretical developments in fermion formulations require
additional study to demonstrate their merit for numerical work.  For
the case of domain wall fermions, a growing body of numerical results
are available.  Both quenched~\cite{Edwards:1998sh,Blum:1997mz,Blum:1997jf,Mawhinney:1998ut,Fleming:1998cc,Kaehler:1998xx,Blum:1999xi,Fleming:1999eq,Wu:1999cd,AliKhan:1999zn,Edwards:1999bm,Aoki:1999uv,Aoki:2000pc,Chen:1998xw} and
dynamical~\cite{Chen:1998xw,Vranas:1997tj,Vranas:1998vm,Vranas:1999nx,Vranas:1999pm,Chen:2000zu} 
domain wall fermion simulations have been conducted and the domain wall
approach is readily adapted to current algorithms for lattice QCD.
(Much work is also being done on the numerical implementation of the
overlap formulation and its variations 
\cite{Neuberger:1997fp,Edwards:1998wx,Hernandez:1998et,Borici:1999zw,Edwards:1999yi,Giusti:1999be,Neuberger:1999rz,Dong:2000mr,Edwards:2000qv,Hernandez:2000sb,Narayanan:2000qx}.) 
A fundamental question, which is a major part of this paper, involves 
quantifying the residual chiral symmetry breaking effects of finite 
extent in the fifth dimension.

Due to current limits on computer speed, some lattice QCD studies are
only practical when the fermionic determinant is left out of the
measure of the path integral.  The resulting quenched theory does not
suppress gauge field configurations with light fermionic modes, in
contrast with the original theory where, for small quark mass, the
determinant strongly damps such configurations.  The measurement of
observables involving fermion propagation through configurations with
unsuppressed light fermionic modes can in principle lead to markedly
different infrared behavior than that found in full QCD, in the limit of
small quark masses.  Domain wall fermions, which produce light chiral
modes at finite lattice spacing and preserve the global symmetries of
continuum QCD, should produce a well-defined chiral limit for full
QCD.  The central question addressed in this paper is whether a 
well-controlled chiral limit also exists within the quenched 
approximation.  A thorough theoretical and numerical understanding of 
the quenched chiral limit is essential if the good chiral properties 
of domain wall fermions are to be exploited in quenched lattice 
simulations.

Here we present results from extensive simulations of quenched QCD with
domain wall fermions, primarily at two lattice spacings, $a^{-1} \sim 1
$ and 2 GeV.  Many different values for the fifth dimensional extent,
$L_s$, and the bare quark mass, $m_f$ have been used.  Hadron masses,
$f_\pi$ and the chiral condensate, $\qbq$, are the primary hadronic
observables we have studied.  In calculating physical observables using
domain wall fermions, four-dimensional quark fields $q(x)$ are defined
from the five-dimensional fields $\Psi(x,s)$ by taking the left-handed
fields from the four-dimensional hypersurface with smallest coordinate
in the fifth dimension and the right-handed fields from the
hypersurface with the largest value of this coordinate.  We also 
present results from measuring the lowest eigenvectors and eigenvalues 
of the hermitian domain wall fermion operator.

Here we list the major topics in each section of this paper.  Section
\ref{sec:dwf} defines our conventions and gives details of the
hermitian domain wall fermion operator.  Section \ref{sec:sim_results}
discusses our simulation parameters and fitting procedures and includes
tables of run parameters and hadron masses for $m_f \ge 0.01$.  In
Section \ref{sec:zero_modes_qbq} a precise understanding of how finite
$L_s$ effects enter $\qbq$ is developed and measurements of $\qbq$
which show the role of fermionic zero modes are reported.  We study the
pion mass in the chiral limit in Section \ref{sec:chiral_m_pi}, which
requires understanding zero mode effects.  Section \ref{sec:m_res}
contains two determinations of the residual chiral symmetry breaking
for finite $L_s$; one from measuring appropriate pion correlators and
the other from the explicitly measured small eigenvalues and
eigenvectors of the hermitian domain wall fermion operator.  Our
determination of $f_\pi$, an important check of the chiral properties
of domain wall fermions, is discussed in Section \ref{sec:hadron},
along with the scaling of hadron masses.

Because of the length of this paper and the number of topics covered,
we now give a brief summary of our major results, organized to
correspond to the expanded discussion in Sections
\ref{sec:zero_modes_qbq}, \ref{sec:chiral_m_pi} \ref{sec:m_res} and
\ref{sec:hadron}.


\underline{Zero mode effects in $\qbq$}:
As already mentioned, the domain wall fermion operator $D$ has an
Atiyah-Singer index $\nu$ for $L_s \rightarrow \infty $. However, in 
quenched QCD, $\nu$ plays no role in the generation of gauge field
configurations.  For $\nu \ne 0$, both $D$ and the hermitian domain
wall fermion operator $D_H$ \cite{Furman:1995ky} have zero modes.
Since $\qbq$ is an appropriately restricted trace of $D_H^{-1}$ it
should diverge as $\langle |\nu|\rangle / m_f V$ for small $m_f$ if the
ensemble average of $|\nu|$ is non-zero.  Here $V$ is the four-dimensional,
space-time volume of the lattice being studied.  For large but finite 
$L_s$, the residual chiral symmetry breaking should cut off this 
divergence.

\begin{figure}
\epsfxsize=\hsize
\epsfbox{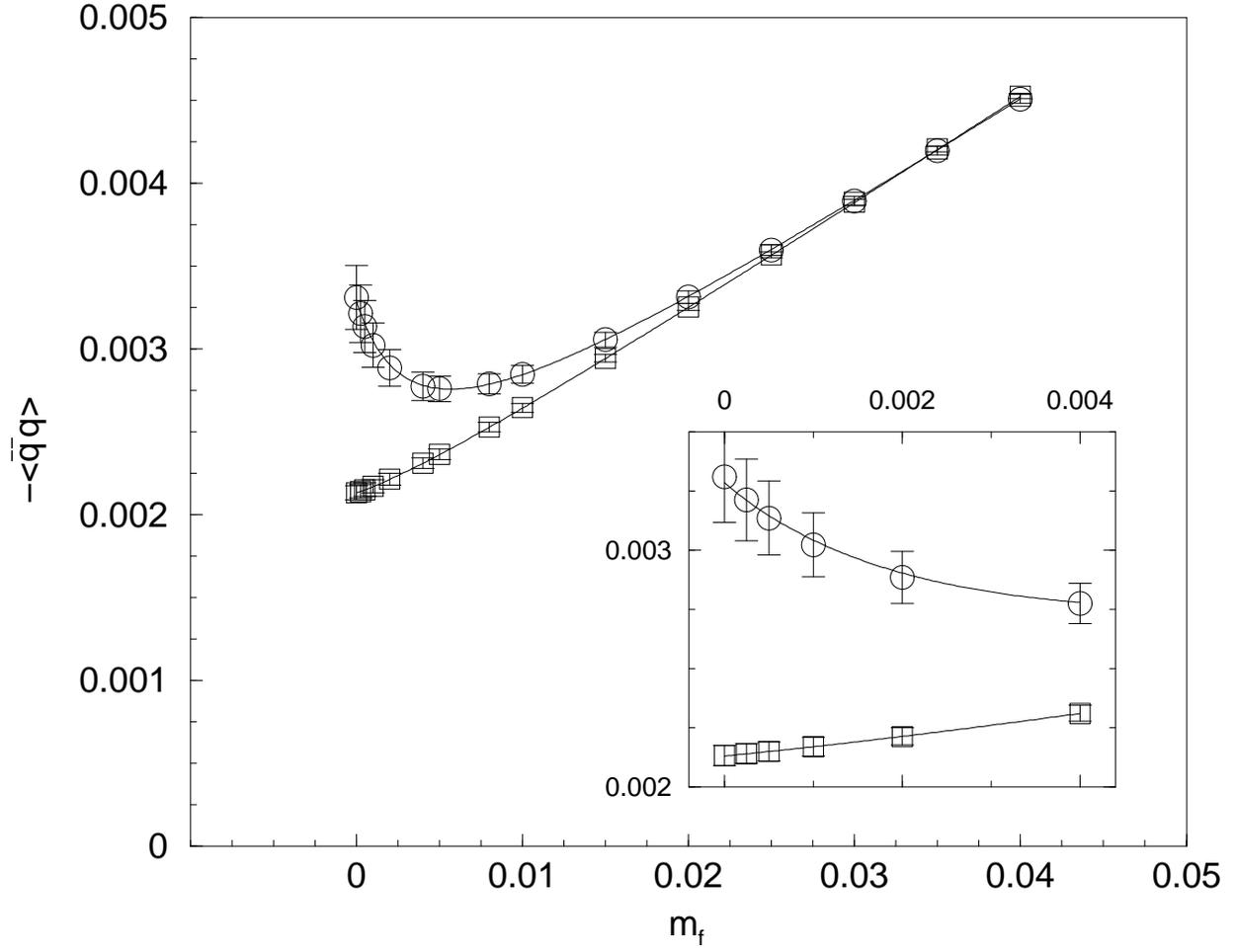}
\caption{$\qbq$ for quenched simulations done on $8^3 \times 32$
lattices ($\circ$) and $16^3 \times 32$ lattices ($\Box$) at $\beta
= 5.7$ with $L_s = 32$.  The smaller volume shows a pronounced
rise as $m_f \rightarrow 0$ as is expected if unsuppressed zero modes
are present.  For the larger volume, the effect of
topological near zero modes is reduced if not eliminated.  This is
expected since $\langle | \nu | \rangle/V$ should fall as $1/\sqrt{V}$.}
\label{fig:qbq_b5_7_ls32}
\end{figure}

Figure \ref{fig:qbq_b5_7_ls32} shows $\qbq$ versus the quark mass $m_f$
for $a^{-1} \sim 1$ GeV on two different volumes of linear dimensions of
about 1.6 and 3.2 Fermi.  A divergence for $m_f \rightarrow 0$ is clearly 
visible on the smaller volume, but not on the larger.  This is expected 
since $\langle | \nu |\rangle/V$ should go as $1/\sqrt{V}$ and is 
clear evidence for unsuppressed zero modes in quenched QCD, first 
reported in Ref.~\cite{Fleming:1998cc}.  Notice that there may be 
other problems with the chiral limit of $\qbq$ that are masked by 
this $1/m_f$ divergence.


\underline{The chiral limit of $m_\pi$}:
With this clear evidence for zero mode effects in $\qbq$, one might
expect to see zero mode contributions in any quark propagator 
$D^{-1}(x,y)$ if at both $x$ and $y$ a single zero eigenvector has 
reasonable magnitude.  For sufficiently large volume, needed to see
asymptotic behavior in the limit of large $|x-y|$, there should be 
no zero mode effects.  Our results for the zero mode effects on the 
pion mass are presented in Figure
\ref{fig:mpi2_vs_mf_b5_7_8nt32_ls48} which shows $\mpi2$ versus $m_f$
for $8^3 \times 32$ lattices with $L_s = 48$ and Figure
\ref{fig:mpi2_vs_mf_b5_7_16nt32_ls48_mres}, where all the parameters are
the same except that the volume was increased to $16^3 \times 32$.  The
pion mass is determined from three different correlators which are each
affected differently by zero modes.  For the smaller volume, the pion
masses measured disagree for small $m_f$, while they agree for the
larger volume.

Notice that on the larger volume shown in Figure
\ref{fig:mpi2_vs_mf_b5_7_16nt32_ls48_mres}, where zero-mode effects 
are not apparent, $\mpi2$ shows signs of curvature in $m_f$ with 
the three $m_f=0$ values lying below the extrapolation from larger 
masses.  In addition, this simple large-mass linear extrapolation 
vanishes at a value of $m_f$ that is more negative than the point 
$m_f + \mres=0$ (shown in the graph by the star) also suggesting 
downward concavity.  While the discrepancy between this $x$-intercept 
and the point $m_f + \mres=0$, may be caused by $O(a^2)$ effects, 
we find a considerably larger discrepancy when making a similar 
comparison at $\beta=6.0$.  Thus, we have evidence that $m_\pi^2$ 
does not depend linearly on $m_f$ in the chiral limit.


\underline{Determining the residual mass}:
In the limit of small lattice spacing, the dominant chiral symmetry 
breaking effect, due to the mixing between the domain walls, is the 
appearance of a residual mass, $\mres$ in the low energy effective 
Lagrangian.  The Ward-Takahashi identity for domain wall fermions
\cite{Furman:1995ky} has an additional contribution representing this
explicit chiral symmetry breaking due to finite $L_s$.  Matrix elements
of this additional term between low energy states determine the residual
quark mass.  Figure \ref{fig:mres_vs_ls_b6_0} shows our results 
for $\mres$ for
$16^3 \times 32$ lattices at $\beta = 6.0$ as a function of $L_s$.
$\mres$ is clearly falling with $L_s$ and reaches a value of $ \sim 2$
MeV for $L_s \ge 24$.  Our data does not resolve the precise behavior
of $\mres$ for large $L_s$, but the very small value makes this less
important for current simulations.  A similar study on lattices with
$a^{-1} \sim 1$ GeV but with larger $L_s=48$ finds a value of 
$\mres/m_s=0.074(5)$ or $\mres \approx 8$ MeV.

We have also used the Rayleigh-Ritz method, implemented using the 
technique of Kalk-Reuter and Simma \cite{Kalkreuter:1996mm}, to 
determine the low-lying eigenvalues and eigenvectors for the 
hermitian domain wall fermion operator.  The results exhibit the 
approximate behavior expected from low-energy excitations in the 
domain wall formulation.  We use the resulting eigenvalues to 
provide an independent estimate of the residual mass which is 
nicely consistent with the more precise value determined from 
pseudoscalar correlators.


\underline{Results for $f_\pi$, hadron masses and scaling}:
With this detailed understanding of the chiral limit of quenched
lattice QCD with domain wall fermions, we have calculated $f_\pi$ using
both pseudoscalar and axial-vector correlators.  The results for
lattices with $a^{-1} \sim 2 $ GeV are shown in Figure
\ref{fig:fpi_b6_0_2parm}, where good agreement between the two
methods is seen.  To do this comparison, the appropriate Z-factor for
the local axial current must be determined and a consistent value for
$\mres$ must be known.  The good agreement in the figure is a significant 
test of these measurements as well as the chiral properties of domain
wall fermions.  We find very good scaling in the ratio $f_\pi/m_\rho$ 
for $a^{-1} \sim 1$ to 2 GeV.  For $m_N/m_\rho$ scaling is 
within 6\%.  We also find that $-\qbq = (256(8){\rm MeV})^3$ from 
our $\beta = 6.0$ simulations.


\section{Domain wall fermions}
\label{sec:dwf}

In this section we first define our notation, including the domain wall
fermion Dirac operator, and then derive the precise form of the
Banks-Casher relation for domain wall fermions, to second order in the
quark mass.  In this paper, the variable $x$ specifies the coordinates
in the four-dimensional space-time volume, with extent $L$ along each of
the spatial directions and extent $N_t$ along the time direction, while
$s=0,1, \dots, L_s-1$ is the coordinate of the fifth direction, with
$L_s$ assumed to be even.  The space-time volume $V$ is given by
$V = L^3 N_t$.  The domain wall fermion operator acts on a
five-dimensional fermion field, $\Psi(x,s)$, which has four spinor
components.  A generic four-dimensional fermion field, with four spin
components, will be denoted by $\psi(x)$, while the specific
four-dimensional fermion field defined from $\Psi(x,s)$ will be denoted
by $q(x)$.  The space-time indices for vectors will be enclosed in
parenthesis while for matrices they will be given as subscripts.
Our general formalism follows that developed by Furman and 
Shamir~\cite{Furman:1995ky}.


\subsection{Conventions}

The domain wall fermion operator is given by
\begin{equation}
  D_{x,s; x^\prime, s^\prime} = \delta_{s,s^\prime}
    D^\parallel_{x,x^\prime} + \delta_{x,x^\prime} D^\bot_{s,s^\prime}
\label{eq:d}
\end{equation}
\begin{eqnarray}
D^\parallel_{x,x^\prime} & =&
  {1\over 2} \sum_{\mu=1}^4 \left[ (1-\gamma_\mu)
  U_{x,\mu} \delta_{x+\hat\mu,x^\prime} + (1+\gamma_\mu)
  U^\dagger_{x^\prime, \mu} \delta_{x-\hat\mu,x^\prime} \right]
  \nonumber \\
  & + & (M_5 - 4)\delta_{x,x^\prime}
\label{eq:D_parallel}
\end{eqnarray}
\begin{eqnarray}
D^\bot_{s,s^\prime} 
     &=& {1\over 2}\Big[(1-\gamma_5)\delta_{s+1,s^\prime} 
                 + (1+\gamma_5)\delta_{s-1,s^\prime} 
                 - 2\delta_{s,s^\prime}\Big] \nonumber\\
     &-& {m_f\over 2}\Big[(1-\gamma_5) \delta_{s, L_s-1}
       \delta_{0, s^\prime}
      +  (1+\gamma_5)\delta_{s,0}\delta_{L_s-1,s^\prime}\Big].
\end{eqnarray}
\label{D_perp}
Here, $U_{x,\mu}$ is the gauge field at site $x$ in direction $\mu$,
and $s$ and $s^\prime$ lie in the range $0 \le s,s^\prime \le L_s-1$.
The five-dimensional mass, representing the height of the domain wall
in Kaplan's original language, is given by $M_5$, while $m_f$ directly
couples the two domain walls at $s=0$ and $s = L_s - 1$.  Since the
light chiral modes should be exponentially bound to the domain walls,
$m_f$ mixes the two chiralities and is therefore the input bare quark
mass.  The value of $M_5$ must be chosen to produce these light surface
states and, in the free field case, $0 < M_5 < 2$ produces a single
fermion flavor with the left-hand chirality bound to $s = 0$ and the
right to $s = L_s-1$.  In order to use our pre-existing,
high-performance Wilson fermion operator computer program as part of
our domain wall fermion operator, we have used the operator $D$ above, 
which is the same as $D_F^\dagger$ of Ref.~\cite{Furman:1995ky}.

Following Ref.~\cite{Furman:1995ky}, we define the four-dimensional quark
fields $q(x)$ by
\begin{eqnarray} 
q(x)    &=& P_{\rm L} \Psi(x,0) + P_{\rm R} \Psi(x, L_s-1) \nonumber \\
\bar{q}(x) &=& \Psibar(x,L_s-1) P_{\rm L} + \Psibar(x, 0) P_{\rm R}
\label{eq:quark_projection}
\end{eqnarray} 
where we have used the projection operators $P_{R,L} = ( 1 \pm
\gamma_5) / 2 $.  Symmetry transformations of the five-dimensional
fields yield a four-dimensional axial current
\begin{equation} 
  {\cal A}^a_\mu(x) = \sum_{s = 0}^{L_s - 1 } {\rm sign}
    \left( s  - \frac{L_s - 1} {2} \right) j^a_\mu(x,s).
  \label{eq:axial_cc}
\end{equation}
Here
\begin{equation}
  j^a_\mu(x, s) = \frac{1}{2} \left[ \overline{\Psi}(x+\hat{\mu}, s )
    (1 + \gamma_\mu ) U^\dagger_{x+\hat{\mu}, \mu} t^a \Psi(x,s)
    - \overline{\Psi}(x,s)(1 - \gamma_\mu) U_{x,\mu} t^a
    \Psi(x+\hat{\mu}, s) \right]
\end{equation}
while the flavor matrices are normalized to obey 
${\rm Tr}\,(t^a t^b ) = \delta^{ab}$.  The divergence of this current 
satisfies
\begin{equation}
  \Delta_\mu {\cal A}^a_\mu(x) = 2m_f J^a_5(x) + 2 J^a_{5q}(x)
  \label{eq:axial_cc_diverg}
\end{equation}
where $\Delta_\mu f(x) = f(x) - f(x-\hat{\mu}) $ is a simple finite
difference operator and the pseudoscalar density $J^a_{5}(x)$ is
\begin{eqnarray}
  J^a_{5}(x) & = &
	       - \Psibar(x, L_s - 1)  P_{\rm L} t^a \Psi(x, 0)
	       + \Psibar(x, 0)  P_{\rm R} t^a \Psi(x, L_s-1)
                 \nonumber \\
	     & = & \overline{q}(x) t^a \gamma_5 q(x).
                 \label{eq:ward_tak_mass_term}
\end{eqnarray}
This equation differs from the corresponding continuum expression by
the presence of the $J^a_{5q}(x)$ term, which is built from point-split
operators at $L_s/2$ and $L_s/2 - 1 $ and is given by
\begin{equation}
  J^a_{5q}(x) = - \Psibar(x, L_s/2-1)  P_{\rm L} t^a \Psi(x, L_s/2)
		+ \Psibar(x, L_s/2)  P_{\rm R} t^a \Psi(x, L_s/2-1).
  \label{eq:midpt_term}
\end{equation}
We will refer to this term as the ``mid-point'' contribution to the
divergence of the axial current.

This mid-point term adds an additional term to the axial Ward-Takahashi
identities and modifies observables, like the pion mass, which are
controlled by these identities.  The Ward-Takahashi identity is
\begin{equation}
  \Delta_\mu \langle {\cal A}^a_\mu(x) O(y) \rangle = 
	2m_f \langle J^a_5(x) O(y) \rangle + 2 \langle J^a_{5q}(x) O(y)
	\rangle + i \langle \delta^a O(y) \rangle.
\label{eq:ward_tak_id}
\end{equation}
For operators, $O$ made from the fields $q(y)$ and $\overline{q}(y)$,
it has been shown \cite{Furman:1995ky} that the $J^a_{5q}$ term in 
Eq.\ \ref{eq:ward_tak_id} vanishes for flavor non-singlet currents when $L_s \rightarrow \infty$.  For the singlet current, this extra term
generates the axial anomaly.  The mid-point term represents the 
contribution of finite $L_s$ effects on the low-energy physics of 
domain wall fermions.


\subsection{Definition of the residual mass and the chiral limit}
\label{sec:mres_def}

For domain wall fermions, the axial transformation which leads to the
Ward-Takahashi identity of Eq.\ \ref{eq:ward_tak_id} rotates the
fermions in the two half-spaces along the fifth direction with opposite
charges.  For $m_f=0$, the action is not invariant under this
transformation due to the coupling of the left- and right-handed light
surface states at the midpoint of the fifth dimension.  This results in
the additional term in the divergence of the axial current, as given in
Eq.\ \ref{eq:midpt_term}.  In the $L_s \rightarrow \infty$
limit where the explicit mixing between the $s \sim 0$ and $s \sim L_s-1$ 
states vanishes, this extra ``mid-point'' contribution will be zero and 
a continuum-like Ward-Takahashi identity will be realized.

Since we must work at finite $L_s$ it is useful to characterize
the chiral symmetry breaking effects of mixing between the domain
walls as precisely as possible.  We do this by adopting the language
of the Symanzik improvement program \cite{Symanzik:1983dc,Symanzik:1983gh}.  
Here we use an effective continuum Lagrangian ${\cal L}_n$ to 
reproduce to $O(a^n)$ the amplitudes predicted by our lattice theory 
when evaluated at low momenta and finite lattice spacing.  Clearly 
${\cal L}_0$ is simply the continuum QCD Lagrangian, while 
${\cal L}_1$ will include the dimension-five, clover term: $\overline{\psi}\sigma^{\mu\nu}\psi F_{\mu\nu}$ \cite{Sheikholeslami:1985ij}.  
The chiral symmetry breaking effects of mixing between the domain walls 
will appear to lowest order in $a$ as an additional, dimension three operator 
$\propto a^{-1} e^{-\alpha L_s}\; \overline\psi\psi$.  This term represents the 
residual mass term that remains even after the explicit input chiral 
symmetry breaking parameter $m_f$ has been set to zero.  The next chiral 
symmetry breaking contribution from domain wall mixing will be $O(a^2)$ 
smaller, appearing as a coefficient of order $a^1$ for the clover term.

We define the chiral symmetry breaking parameter $\mres$ so the 
complete coefficient of the mass term in ${\cal L}_0$ is proportional 
to the simple sum $m_f+\mres$.  While this is a precise definition of 
$\mres$, valid for finite lattice spacing, a precise determination of 
$\mres$ in a lattice calculation will be impeded by the need to 
quantitatively account for the additional chiral symmetry breaking 
effects of terms of higher order in $a$.

Close to the continuum limit, for long distance amplitudes, the
Ward-Takahashi identity given in Eq.~\ref{eq:ward_tak_id} must agree
with the corresponding identity in the effective continuum theory.  
Thus, for the non-singlet case, the sum of the first two terms on the 
right-hand side of  Eq.~\ref{eq:ward_tak_id} must be equivalent to an 
effective quark mass, $m_{\rm eff} = m_f +\mres$, times the pseudo-scalar 
density $J_{5}^a$.  Thus, the residual mass, $\mres$ appears in the 
low energy identity:
\begin{equation}
  J^a_{5q} \approx \mres J^a_5
  \label{eq:mres}
\end{equation}
where this equality will hold up to $O(a^2)$ in low-momentum amplitudes.

Thus, close to the continuum limit, $\mres$ in Eq.~\ref{eq:mres} is 
a universal measure of the chiral symmetry breaking effects of
domain wall fermions for all low energy matrix elements, with corrections 
coming from terms of higher order in the lattice spacing.  However, 
away from the continuum limit the $O(a^2)$ terms may be appreciable.  
In addition, if there are high energy scales entering an observable, 
such a low energy description is not valid and the explicit chiral 
symmetry breaking effects of finite $L_s$ can be more complicated 
than a simple additive shift of the input quark mass by $\mres$.

Many aspects of the chiral behavior of the domain wall theory can be easily 
understood by reference to the more familiar Wilson fermion formulation.
For finite $L_s$ the domain wall formulation can be viewed as an 
``on- and off-shell improved'' version of Wilson fermions.  
The low energy effective Lagrangian for 
domain wall fermions is the same as that for the Wilson case except the 
coefficients of the chiral symmetry breaking terms are expected to decrease exponentially with $L_s$.   Viewed in this way, one might expect to achieve
a vanishing pion mass by fine-tuning $m_f$ to a critical value, $m_{fc}$ in
very much the same way as one fine-tunes $\kappa$ to $\kappa_c$ for Wilson 
fermions.  As the above discussions demonstrates, $m_{fc} = \mres + O(a^2)$.
Just as in the Wilson case, this limit can be interpreted as the approach 
to the critical surface of the Aoki phase 
\cite{Vranas:1999nx,Aoki:1999uv,Aoki:2000pc}.


\subsection{The hermitian domain wall fermion operator}

A hermitian operator $D_H$ can be constructed \cite{Furman:1995ky} from $D$ 
through
\begin{equation}
  D_H \equiv \gamma_5 R_5 D
\end{equation}
where $(R_5)_{s s^\prime} \equiv \delta_{s, L_s - 1 - s^\prime}$ is
the reflection in the fifth dimension around the five-dimensional 
midpoint, $s=(L_s-1)/2$.  Writing out $D_H$ gives
\begin{eqnarray}
D_H & = &  \gamma_5 \, D^\parallel_{x,x^\prime} \,
  \delta_{s + s^\prime, L_s-1} \nonumber \\
    & + & \gamma_5 \left[ P_L \, \delta_{s + s^\prime, L_s}
        +                 P_R \, \delta_{s + s^\prime, L_s - 2}
	-                     \delta_{s + s^\prime, L_s - 1} \right.
      \label{eq:dherm} \\
    & - & \left. m_f ( P_L \, \delta_{s,0} \delta_{s^\prime,0}
      +       P_R \, \delta_{s,L_s-1} \delta_{s^\prime,L_s-1} ) \right]
	  \delta_{x,x^\prime} \nonumber
\end{eqnarray}
while as an explicit matrix in the $s,s^\prime$ indices:
\begin{equation}
D_H =
\left( \begin{array}{cccccc}
 -m_f \gamma_5 P_L & & & & \gamma_5 P_R & \gamma_5(D^\parallel - 1 ) \\
  & & & \gamma_5 P_R & \gamma_5(D^\parallel - 1 ) & \gamma_5 P_L \\
  & & \cdots  & \cdots & \cdots  & \\
  & \cdots  & \cdots & \cdots  & & \\
  \gamma_5 P_R & \gamma_5(D^\parallel - 1 ) & \gamma_5 P_L  & & & \\
  \gamma_5(D^\parallel - 1 ) & \gamma_5 P_L  & & & & -m_f \gamma_5 P_R\\
  \end{array} \right).
\end{equation}
The eigenfunctions and eigenvectors of $D_H$ will be denoted by
\begin{equation}
  D_H \Psi_{\Lambda_H} = \Lambda_H \Psi_{\Lambda_H}
\end{equation}
with the five-dimensional propagator given by
\begin{equation}
  S^{(5)}_{x,s;x^\prime, s^\prime} = \sum_{\Lambda_H}
    \frac{\Psi_{\Lambda_H}(x,s) \Psi^\dagger_{\Lambda_H}
      (x^\prime, \tilde{s}) \, \gamma_5 \, (R_5)_{\tilde{s}, s^\prime}}
      {\Lambda_H}.
\end{equation}
(Grassmann variables in the Euclidean path integral will be
denoted by $\overline{\Psi}$ and $\Psi$, while the eigenfunctions
of $D_H$ will be denoted $\Psi^\dagger_{\Lambda_H}$ and
$\Psi_{\Lambda_H}$.)

We will find it convenient to define three additional matrices 
\begin{equation}
  (\Gamma_5)_{s, s^\prime} = \delta_{s, s^\prime} \, {\rm sign}
    \left(\frac{L_s - 1}{2} - s \right)
\end{equation}
\begin{equation}
  \qw_{s, s^\prime} = P_L \, \delta_{s, 0} \,
      \delta_{s^\prime, 0}
    + P_R \, \delta_{s, L_s-1} \,  \delta_{s^\prime, L_s -1}
\end{equation}
and
\begin{equation}
 \qmp_{s, s^\prime} =
      P_L\, \delta_{s, L_s/2} \, \delta_{s^\prime, L_s/2}
    + P_R \, \delta_{s, L_s/2 - 1} \, \delta_{s^\prime, L_s/2 -1}.
\end{equation}
The transformation which generates the current in Eq.\
\ref{eq:axial_cc} is
\begin{eqnarray}
  \Psi & \rightarrow & \exp( i \alpha^a t^a \Gamma_5 ) \Psi \nonumber \\
  \overline{\Psi} & \rightarrow & \overline{\Psi}
    \exp(- i \alpha^a t^a \Gamma_5 ).
  \label{eq:axial_sym}
\end{eqnarray}
The matrices $\qw$ and $\qmp$ are the two parts of $D_H$ which 
correspond to terms in $D=\gamma_5 R_5 D_H$ which are not invariant 
under the transformation in Eq.\ \ref{eq:axial_sym}.  
The matrix $\qw$ underlies the explicit mass term and, in the original 
operator $D$, explicitly mixes the $s=0$ and $s=L_s-1$ walls.
Likewise, the matrix $\qmp$ is a ``mid-point'' matrix with non-zero
elements only in the center of the fifth dimension.  It represents
the component of the operator $D$ which connects the left and right 
half regions.  
These two contributions provide the terms on the right hand side of
Eq.\ \ref{eq:axial_cc_diverg} and one easily finds
\begin{equation}
  \{ \Gamma_5, D_H \} =  2m_f \qw + 2 \qmp.
\label{eq:sym_constraint}
\end{equation}
Since it is expected that there are eigenvectors of $D_H$ which
are exponentially localized on the domain walls, we see that with
$m_f = 0$ and  the limit $L_s \rightarrow \infty$ taken, $D_H$
anticommutes with $\Gamma_5$ in the subspace of these eigenvectors.
This is the property expected for massless, four-dimensional
fermions in the continuum in Euclidean space.

Using the matrix, $\qw$, we can write a simple form for the
four-dimensional chiral condensate, $\qbq$ 
\begin{eqnarray}
  -\qbq & = & -\frac{1}{12 V} \sum_x \langle \overline{q}(x) q(x)
		\rangle \label{eq:qbd_def} \\
    & = & - \frac{1}{12 V}  \sum_x \langle \,
        \overline{\Psi}(x,s) (R_5 \qw)_{s,s'} \Psi(x,s^\prime) \rangle \\
    & = & \frac{1}{12 V}  \left\langle \sum_{x, \Lambda_H}
	\frac{\Psi^\dagger_{\Lambda_H}(x,s) \, \gamma_5 \,
	\qw_{s, s^\prime}
	\Psi_{\Lambda_H}(x,s^\prime)} {\Lambda_H} \right\rangle \\
    & = & \frac{1}{12 V}  \left\langle \sum_{\Lambda_H}
	\frac{ \langle \Lambda_H | \gamma_5 \qw | \Lambda_H
	\rangle } { \Lambda_H} \right\rangle \label{eq:dh_spectral_qbq}
\end{eqnarray}
where in the last line a bra/ket notation has been used.  The
large angle brackets indicate the average over an appropriate
ensemble of gauge fields.

We define the pion interpolating field as $\pi^a(x) \equiv
i \overline{q}(x) t^a \gamma_5 q(x)$ and then find that the
pion two-point function is given by (no sum on $a$)
\begin{equation}
  \corrpp  = \left\langle
    \sum_{\Lambda_H, \Lambda_H^\prime}
      \frac{ \Psi^\dagger_{\Lambda_H^\prime} (x,s)
	     \, \qw_{s,s^\prime} \,
	     \Psi_{\Lambda_H}(x,s^\prime) \;
             \Psi^\dagger_{\Lambda_H} (0,\tilde{s})
	     \, \qw_{\tilde{s},\tilde{s}^\prime} \,
	     \Psi_{\Lambda_H^\prime}(0,\tilde{s}^\prime) }
	     {\Lambda_H \Lambda_H^\prime}
    \right\rangle.
  \label{eq:dwf_corrpp_spec}
\end{equation}
Note that the generators, $t^a$, do not appear in the spectral sum,
since they merely serve to specify the contractions of the quark
propagators and that $\pi^a(x) = i J^a_5(x)$. To investigate the extra
term in the axial Ward-Takahashi identity, Eq.\ \ref{eq:ward_tak_id},
we will also have need to measure the correlation function between
interpolating pion fields defined on the domain walls and the mid-point
contribution to the divergence of the axial current, $J^a_{5q}$.  We
define a mid-point pion interpolating field by $\mppi(x) = i
J^a_{5q}(x)$ and the spectral decomposition for the correlator between
interpolating pion operators on the wall and the midpoint is
\begin{equation}
  \langle \mppi(x) \pi^a(0)  \rangle = \left\langle
    \sum_{\Lambda_H, \Lambda_H^\prime}
      \frac{ \Psi^\dagger_{\Lambda_H^\prime} (x,s) \, \qmp_{s,s^\prime}
	     \, \Psi_{\Lambda_H}(x,s^\prime) \;
             \Psi^\dagger_{\Lambda_H} (0,\tilde{s})
	     \, \qw_{\tilde{s}, \tilde{s}^\prime} \,
	     \Psi_{\Lambda_H^\prime}(0,\tilde{s}^\prime) }
	     {\Lambda_H \Lambda_H^\prime}
    \right\rangle.
\end{equation}

We define a local axial current as $A^a_\mu(x) \equiv \overline{q}(x)
t^a \gamma_\mu \gamma_5 q(x)$ and note that it is different from ${\cal
A}^a_\mu$ defined in Eq.\ \ref{eq:axial_cc}.  The two-point function of
the zeroth component of this current, $\corraa$, has a form similar to
Eq.\ \ref{eq:dwf_corrpp_spec} with a factor of $\gamma_0$ multiplying
each $\qw$ and an overall minus sign.  Finally, our scalar density is
$\sigma(x) \equiv \overline{q}(x) q(x)$ and the connected correlator
$\corrss$ also has the form of Eq. \ref{eq:dwf_corrpp_spec} with a
factor of $\gamma_5$ multiplying each $\qw$ and an overall minus sign.


\section{Hadron Masses for \lowercase{$m_f \ge 0.01$}}
\label{sec:sim_results}

In this section we present the results for $m_\pi$, $m_\rho$ and $m_N$ 
obtained for reasonably heavy input quark mass, $m_f \ge 0.01$ where the 
lower limit corresponds to $m_{\rm quark} \approx m_{\rm strange}/4$.
The more challenging study of $m_\pi$ for $m_f \rightarrow 0$ is 
described later, in Section~\ref{sec:chiral_m_pi}.  This section is
organized as follows.  We begin by describing the Monte Carlo
runs on which the results in this paper are based.  Next the methods
used to determine the hadron masses are discussed, both the propagator 
determinations and our fitting procedures.  Finally, we present the 
results of those calculations for the easier, large mass case, 
$m_f \ge 0.01$.


\subsection{Simulation summary}

The results reported in this paper were obtained from ensembles of
gauge field configuration generated from pure gauge simulations using
the standard Wilson action\cite{Wilson:1974sk} at three values of the
coupling parameter, $\beta=6/g^2$: 5.7, 5.85 and 6.00.  Thus, these
ensembles follow the distribution, $\exp{\{{6/g^2} \sum_{\cal P} {\rm
tr} U_{\cal P}\}}$ where the sum ranges over all elementary plaquettes
${\cal P}$ in the lattice and $U_{\cal P}$ is the ordered product of
the four link matrices associated with the edges of the plaquette
${\cal P}$. Some of the $\beta=5.7$ simulations and a portion of those
at $\beta=5.85$ were performed using the hybrid Monte Carlo `$\Phi$'
algorithm~\cite{Gottlieb:1987mq}.  These runs were performed on an
$8^3\times 32$ space-time volume with a domain wall height $M_5 =
1.65$.  Each hybrid Monte Carlo trajectory consisted of 50 steps with a
step size $\Delta t = 0.02$.  These runs are summarized in
Table~\ref{tab:run_parameters_hmc}.  In each case the first 2,000
hybrid Monte Carlo trajectories were discarded for thermalization
before any measurements were made.  After these thermalization
trajectories, successive measurements of hadron masses and the chiral
condensate, $\qbq$ were made after each group of 200 trajectories.

A second set of simulations were performed using the heatbath method of
Creutz~\cite{Creutz:1980zw}, adapted for $SU(3)$ using the two-subgroup
technique of Cabibbo and Marinari~\cite{Cabibbo:1982zn} and improved
for a multi-processor machine by the algorithm of Kennedy and
Pendleton~\cite{Kennedy:1985nu}.  The first 5,000 sweeps were discarded
for thermalization.  These runs are described in
Table~\ref{tab:run_parameters_hb} where the values of $M_5$ used are
also given.  Finally, the single $\beta=5.85$ run with $M_5 = 1.9$ was
performed using the MILC code \cite{MILC:1991aa}.  Here four 
over-relaxed heatbath sweeps \cite{Brown:1987rr,Creutz:1987xi} with 
$\omega=2$ were followed by one Kennedy-Pendleton sweep, with 50,000 
initial sweeps discarded for thermalization.

A portion of the $\beta=5.7$ masses described here appeared earlier in
Ref.~\cite{Mawhinney:1998ut} while the first of the $\beta=6.0$ results 
appear in Refs.~\cite{Wingate:1999yr} and \cite{Wu:1999cd}.


\subsection{Mass measurement techniques}

We follow the standard procedures for determining the hadron masses
from a lattice calculation, extracting these masses from the
exponential time decay of Euclidean-space, two-point correlation
functions.  In our calculation the source may take two forms.  The
first is a point source
\begin{equation}
O_\Gamma^a(x) = \bar q(x) t^a \Gamma q(x)
\label{eq:pt_src_M}
\end{equation} 
which is usually introduced at the origin.  The flavor index $a$ is 
introduced to make clear that we do not study the masses of flavor
singlet states.  For the nucleon state we use a combination of three 
quark fields:
\begin{equation}
O_{\rm P}(x) = \epsilon_{abc} u_a(x) [u_b(x) C \gamma^5 d_c(x)]
\label{eq:pt_src_N}
\end{equation} 
where for simplicity we have written the source for a proton in terms
of up and down quark fields, $q=u$ and $d$.  Here $C$ is the $4 \times
4$ Dirac charge-conjugation matrix, $\epsilon$ the anti-symmetric
tensor in three dimensions and the color sum over the indices $a$, $b$
and $c$ is shown explicitly.  Only these point sources are used in the
$\beta=5.7$ running.

The second variety of source used in this work is a wall source.  Such
a source is obtained by a simple generalization of
Eqs.~\ref{eq:pt_src_M} and \ref{eq:pt_src_N} in which we replace the
quark fields evaluated at the same space-time point
$x=(\vec r, t)$ with distributed fields, each of which is summed over
the entire spatial volume at a fixed time $t$.  Gauge covariance is
maintained by introducing a gauge field dependent color matrix
$V[U](\vec r,t_i)$ which transforms the spatial links in the time slice
$t=t_i$ into Coulomb gauge.  Thus, to construct our wall sources we
simply replace the quark field $q(\vec r,t_i)_c$ by the non-local field
\begin{equation}
q^w(t_i)_c = \sum_{\vec r} V(\vec r, t_i)_{c,c'} \, q(\vec r,t_i)_{c'}
\label{eq:wall_source}
\end{equation}
where $c$ and $c'$ are color indices.  We use these wall sources for the 
$\beta=5.85$ calculations and a combination of both wall and point 
sources in the $\beta=6.0$ studies.  The use of wall sources for these 
weaker coupling runs is appropriate since the physical hadron states 
are larger in lattice units and better overlap is achieved with the 
states of interest by using these extended sources.

In all cases we use a zero-momentum-projected point sink for the second
operator in the correlation function.  This is obtained by simply
summing the operators in Eqs.~\ref{eq:pt_src_M} and \ref{eq:pt_src_N}
over all spatial positions $\vec r$ in a fixed time plane $t=t_f$.
Thus, for example, we will extract the mass $m_\Gamma$ of the lightest
meson with quantum numbers of the Dirac matrix $\Gamma$ from the large
$t_f-t_i$ expression:
\begin{equation}
\Big\langle \sum_{\vec r} O_\Gamma^a(\vec r,t_f) \bar q^w(t_i) \Gamma
t^a q^w(t_i) \Big\rangle
\sim A ( e^{-m_\Gamma (t_f-t_i)} + e^{-m_\Gamma (N_t-t_f+t_i)}).
\label{eq:exp}
\end{equation}
A similar equation is used for the nucleon correlation function except
that the second exponent representing the state propagating through the
antiperiodic boundary condition connecting $t=0$ and $t=N_t-1$ is
reversed in sign and has exchanged upper and lower components for a
spinor basis in which $\gamma^0$ is diagonal.

For both the calculation of the quark propagators from which these
hadron correlators are constructed and the evaluation of the chiral
condensate, $\qbq$, we invert the five-dimensional domain wall fermion
Dirac operator of Eq.~\ref{eq:d}, using the conjugate gradient method
to solve an equation of the form $D y = h$.  This iterative method is
run until a stopping condition is satisfied, which requires that the
norm squared of the residual be a fixed, small fraction $\epsilon$ of
the norm squared of the source vector $h$.  At the $n^{th}$
iteration, we determine the residual $r_n$ as a cumulative 
approximation to the difference vector obtained by applying the 
Dirac operator to the present approximate solution $y_n$ and 
$h$:  $r_n = D y_n - h$.  We stop the process when 
$|r_n|^2/|h|^2 < \epsilon$.

For the calculation of $\qbq$ we use $\epsilon=10^{-6}$ for the runs of
Table~\ref{tab:run_parameters_hmc} and $\epsilon=10^{-8}$ for those in
Table~\ref{tab:run_parameters_hb}.  For the computation of hadron
masses in the runs of Table~\ref{tab:run_parameters_hmc} we use
$\epsilon=10^{-8}$ when $L_s$ has the values 10, 16, 24 and 48, the
condition $\epsilon=10^{-10}$ for the case $L_s=32$.  For the
hadron masses computed in the runs in Table~\ref{tab:run_parameters_hb}
we used $\epsilon=10^{-8}$ for $\beta=5.7$ and 6.0, and
$\epsilon=10^{-7}$ for $\beta=5.85$.  Tests showed that zero-momentum 
projected hadronic propagators eight time slices from the source, 
calculated with a stopping condition of $10^{-6}$, differed by 
less than 1\% from the same propagators calculated with a stopping 
condition of $10^{-12}$ for $m_f \ge 0.01$ \cite{Malureanu:1998thesis}.  
For a quark mass $m_f = 0.01$ and a $16^3 \times 32$ volume with 
$L_s = 16$, typically $\approx 1,500$ conjugate gradient iterations 
were required to meet the stopping condition.  For our very light 
quark masses ($m_f \le 0.001$) up to 10,000 iterations were required 
for convergence.

The final step in extracting the masses of the lowest-lying hadron
states from the exponential behavior of the correlation functions given
in Eq.~\ref{eq:exp} is to perform a fit to this exponential form over a
time range chosen so that this single-state description is accurate.
Choosing $t_i=0$, we use the appropriate $t_f \rightarrow N_t-t_f$
symmetry of Eq.~\ref{eq:exp} to fold the correlator data into one-half
of the original time range $0 \le t_f < N_t$.  We then perform a
single-state fit of the form in Eq.~\ref{eq:exp} for the time range
$t_{\rm min} \le t_f \le t_{\rm max} \le N_t/2$.  Typically $t_{\rm
max}$ is simply set to the largest possible value, $t_{\rm
max}=N_t/2$.\footnote{For the $\beta=5.7$ runs we used smaller values of
$t_{\rm max}$ for the $\pi$ and $\rho$ fitting, typically 12 or 14, in
order to avoid the effects of rounding errors.  These finite-precision
errors, caused by a poor choice of initial solution vector, were seen
at the largest time separations for the very rapidly falling
propagators found at this strong coupling.}

The lower limit, $t_{\rm min}$, is decreased to include as large a 
time range as possible so as to extract the most accurate 
results.  However, $t_{\rm min}$ must be sufficiently large 
that the asymptotic, single-state formula in Eq.~\ref{eq:exp} 
is a good description of the data in the time range
studied.  These issues are nicely represented by the effective mass,
$m_{\rm eff}(t)$, with the parameters $A$ and $m\equiv m_{\rm eff}(t)$ 
in Eq.~\ref{eq:exp} determined to exactly describe
the hadron correlator at the times $t$ and $t+1$.  To the extent that
$m_{\rm eff}(t)$ is independent of $t$, the data are in a time range which is
consistent with the desired single state signal.  As an illustration,
this effective mass is plotted in Figure~\ref{fig:m_eff}
for the $\pi$, $\rho$ and nucleon states in the $16^3
\times 32$, $\beta=6.0$, $L_s=16$, $m_f=0.01$ calculation.  Good
single-state fits are easy to identify from the plateau regions for the
case of $m_\pi$ and $m_\rho$.  For the nucleon the rapidly increasing
errors at larger time separations for this relatively light
quark mass make it more difficult to determine a plateau.  Better
nucleon plateaus are seen for larger values of $m_f$.

The actual fits are carried out by minimizing the correlated $\chi^2$ to
determine the particle mass and propagation amplitude.  We then choose
$t_{\rm min}$ as small as possible consistent with two criteria.
First, the fit must remain sufficiently good that the $\chi^2$ per
degree of freedom does not grow above 1-2.  Second, we require that the
mass values obtained agree with those determined from a larger value of
$t_{\rm min}$ within their errors.

In order to keep the fitting procedure as simple and straight forward
as possible, we choose values for $t_{\rm min}$ which can be used for
as large a range of quark masses, domain wall separations and particle
types as possible.  Given the large number of Monte Carlo runs and
variety of masses and $L_s$ values it is possible to employ an
essentially statistical technique to determine $t_{\rm min}$.  In
choosing the appropriate $t_{\rm min}$ we examine two distributions.
The first distribution is a simple histogram of values of $\chi^2/ {\rm
dof}$ obtained for all quark masses and a particular physical quantum
number.  We require that for our choice of $t_{\rm min}$, this
distribution is sensibly peaked around the value 1 or lower.  An
example is shown in Figure~\ref{fig:fit_hist} for the $\beta=6.00$
$\rho$ mass determined from a wall source for three values of $t_{\rm
min}$: 5, 7 and 9.

In the second distribution we first determine a fitted mass $m_i(t)$
and the corresponding error $\sigma_i(t)$ for the state $i$, where the
lower bound on the fitting range is given by $t$.  We then choose a
$t^\prime > t$ and examine a measure of the degree to which $m_i(t)$
and $m_i(t^\prime)$ agree.  The measure we choose is
\begin{equation}
\delta_i (t',t) = {m_i(t')-m_i(t) \over \sigma_i(t')}.
\end{equation}
In Figure~\ref{fig:fit_hist} we show the distribution of values of
$\delta_i(t',t_{\rm min})$ for the $\rho$ meson for all $t^\prime >
t_{\rm min}$ and three choices for $t_{\rm min}$: 5, 7 and 9.  The
distributions include $\rho$ mesons with all values of $m_f\ge 0.01$
and all values for $L_s$ used in the calculations.

In our sample Figure~\ref{fig:fit_hist}, we have a reasonable
distribution of $\chi^2/{\rm dof}$ values for all three choices of $t_{\rm
min}$ with only a slight improvement visible as $t_{\rm min}$ increases
from 5 to 9.  Likewise the distribution of mass values found at
$t^\prime > t_{\rm min}$ is in reasonable agreement for each value of
$t_{\rm min}$ with a slight bias toward larger values being visible at
the lowest value $t_{\rm min}=5$.  Examining this figure and
corresponding figures for the $\pi$, for our quoted masses, we chose
$t_{\rm min}=7$ for these states.  The fact that
Figure~\ref{fig:fit_hist} does not sharply discriminate between these
three possible choices of $t_{\rm min}$ implies that we will get
essentially equivalent results from each of these three values.

Our choices of $t_{\rm min}$ are as follows.  For $\beta=5.7$, 
where only point sources are used, $t_{\rm min}$ was chosen 
to be 7 for the $\pi$, $\rho$ and nucleon.  For $\beta=5.85$, 
hadron masses were determined only from the doubled $12^3 \times 64$ 
configurations using wall sources and the value $t_{\rm min}=6$ 
for the $\rho$ and 7 for the $\pi$ and nucleon.  Finally for
$\beta=6.0$ the most accurate mass values were determined using wall
sources and it is these mass results which we quote below.  Here
$t_{\rm min}$ was chosen to be 7 for the $\pi$ and $\rho$ and 8 for the
nucleon.  We were able to extract quite consistent results with larger
errors using point sources.  Here the needed value of $t_{\rm min}$ was
10 for the $\pi$ and $\rho$ and $t_{\rm min}=8$ for the nucleon.
Finally, the errors are determined for each mass by a jackknife
analysis performed on the resulting fitted mass.


\subsection{Hadron mass results}
\label{ssec:hdm_mass_rslts}
The hadron masses that result from the fitting procedures described 
above are given in 
Tables~\ref{tab:mhad_832_b57k165ns10}-\ref{tab:mhad_1632_b60k18ns24}.  
Omitted from this tabulation are the masses for the more difficult 
cases $m_f=0.0$ and 0.001 which are discussed later in 
Section~\ref{sec:chiral_m_pi}.  In each case the pion mass was determined
from the $\corraa$ correlator.  While the results presented in these 
tables will be used in later sections of this paper, there
are some important aspects of these results which will be discussed
in this section.  In particular, the dependence on volume and the
$m_f$ dependence of the $\rho$ and nucleon masses will be examined.

We begin by examining the dependence of the $\rho$ and nucleon
on the input quark mass, $m_f$.  In Figures~\ref{fig:m_vs_mf_5.7},
\ref{fig:m_vs_mf_5.85} and \ref{fig:m_vs_mf_6.0} we plot the $\rho$ and 
nucleon masses as a function of $m_f$,   As the figures show, each 
case is well described by a simple linear dependence on $m_f$.  The
data plotted in these figures appear in 
Tables~\ref{tab:mhad_832_b57k165ns32} \ref{tab:mhad_1232_b585k19ns20} 
and \ref{tab:mhad_1632_b60k18ns16}, respectively.     
Also plotted in Figure~\ref{fig:m_vs_mf_6.0} are our results for $m_\rho$
with non-degenerate quarks.  The coincidence of these two results 
implies the familiar conclusion that to a good approximation the 
meson mass depends on the simple average of the quark masses of 
which it is composed.  For simplicity in obtaining jackknife errors, 
we have included in these linear fits only that data associated with 
ensembles of configurations on which all relevant quark mass values 
were studied.  Added configurations where only particular quark 
masses had been evaluated were not included.

A simple linear fit provides a good approximation to all the masses
considered in this section, in particular for $m_f \ge 0.01$.
In Table~\ref{tab:extrap_832_b5_7} we assemble the fit parameters 
for the $\beta=5.7$, $8^3 \times 32$ masses, while 
Tables~\ref{tab:extrap_1632_b5_7}, \ref{tab:extrap_1232_b5_85} 
and \ref{tab:extrap_1632_b6_0} contain the fit parameters for 
the $\beta=5.7$, $16^3 \times 32$, $\beta=5.85$, $12^3\times 32$ 
and $\beta=6.0$, $16^3 \times 32$ calculations, respectively.
The parameters presented in these three tables were obtained by
minimizing a correlated $\chi^2$ which incorporated the 
effects of the correlation between hadron masses obtained with
different valence quark masses, $m_f$, but determined on the
same ensemble of quenched gauge configurations.  The errors
quoted follow from the jackknife method and the small values of
$\chi^2/{\rm dof}$ shown demonstrate how well these linear fits work.
Because of the visible curvature in the pion mass for our $\beta=5.7$ 
and 6.0 results, the linear fits for $m_\pi^2$ were made to the lowest 
three mass values.  For the $\rho$ and nucleon and all three masses 
at $\beta=5.85$ we fit to the masses obtained for the full range
of $m_f$ values.

Next we consider the effects of finite volume by comparing the $8^3
\times 32$ and $16^3 \times 32$, volumes used in the $\beta=5.7$,
$L_s=48$ calculation.  The value of $m_\pi = 0.383(4)$ found at the
lightest $m_f=0.02$ mass value for the $16^3 \times 32$ implies a
Compton wavelength of 2.6 in lattice units.  This lies between 1/4 and
1/3 of the linear dimension of the smaller lattice, suggesting that we
should not expect large finite volume effects.  This is borne out by
comparing the data in Tables~\ref{tab:mhad_832_b57k165ns48} and
\ref{tab:mhad_1632_b57k165ns48_b} where the two sets of masses agree
within errors.

This apparent volume independence within our errors can be nicely
summarized by comparing the coefficients of the linear fits of the
$\rho$ and nucleon.  Writing the two $a$ and $b$ coefficients from the
tables as a pair [a,b], we can compare the $16^3 \times 32$ values from
Table~\ref{tab:extrap_1632_b5_7} $[0.775(18), 2.20(7)]$ and $[1.03(4),
4.13(17)]$ for the $\rho$ and nucleon with the corresponding numbers
for the $8^3 \times 32$ numbers from Table~\ref{tab:extrap_832_b5_7}:
$[0.790(13), 2.18(5)]$ and $[1.13(4), 3.79(15)]$.  For $m_\pi$
the results on the two volumes agree to within the typical 1\%
statistical errors.  However, for the case of the $\rho$ and nucleon
masses, finite volume effects may be visible on the two standard
deviation or 1-2\% level for the more accurate masses obtained for $m_f
\ge 0.06$

Since in lattice units the $\rho$ mass decreases by about a factor
of two as we change $\beta$ from 5.7 to 6.0, the $16^3$ spatial
volume used at $\beta=6.0$ should be equivalent to the $8^3$ volume
just discussed at $\beta=5.7$.  Thus, we expect that the $\rho$ and
nucleon masses that we have found on this $16^3$ volume will differ 
from their large volume limits by an amount on the order of a few 
percent while the finite-volume pion masses may be accurate on the
0.5\% level.


\section{Zero modes and the chiral condensate}
\label{sec:zero_modes_qbq}


\subsection{Banks-Casher formula for domain wall fermions}
\label{sec:dwf_banks_casher}

In the previous section, our results for quark masses $m_f \ge 0.01$
were given, where the smallest values of $m_f$ gave $m_\pi/m_\rho \sim
0.4$.  Since the domain wall fermion operator with $m_f = 0 $ should
give exact fermionic zero modes as $L_s \rightarrow \infty$,
observables determined from quark propagators at finite $L_s$, when
small quark masses are used, should show the effects of
topological near-zero modes.  For quenched simulations, where zero
modes are not suppressed by the fermion determinant, these modes can be
expected to produce pronounced effects.  One important practical
question is the size of the quark mass where the effects are
measurable.   To begin to investigate this we now turn to the
simplest observable where they can occur, $\qbq$.

Before considering the domain wall fermion operator, we review the
spectral decomposition of the continuum four-dimensional, 
anti-hermitian Euclidean Dirac operator 
$\Dslash^{(4)}$.\footnote{The naive lattice fermion operator $\Dslash =
\gamma_u ( U_{x,\mu}\delta_{x+\hat{\mu},x^\prime }-U_{x^\prime,\mu}^\dagger
\delta_{x-\hat{\mu}, x^\prime}) $ and the lattice staggered
fermion operator have eigenvalues and eigenvectors which also obey Eq.\
\ref{eq:anti_herm_d4_evalues}.} The eigenfunctions and corresponding
eigenvalues of such an anti-hermitian operator satisfy
\begin{equation}
  (\Dslash^{(4)} + m)\psi_\lambda =
    (i\lambda + m)\psi_\lambda.
\end{equation}
with $\lambda$ real and
\begin{equation}
  \gamma_5\psi_\lambda = \left\{
  \begin{array}{cc}
    \pm\psi_\lambda & \lambda = 0 \\
    \mbox{\ }\psi_{-\lambda} &  \lambda \ne 0.
  \end{array} \right.
  \label{eq:anti_herm_d4_evalues}
\end{equation}
(We use $\lambda$ to label eigenfunctions and eigenvalues of
the anti-hermitian operator, saving $\lambda_H$ for the ``hermitian'' 
case defined below.)
In the continuum, the presence of zero modes is guaranteed by the
Atiyah-Singer index theorem for a gluonic field background with
non-zero winding number \cite{Schwarz:1977az,Brown:1977bj}.

The four-dimensional quark propagator, $S^{(4)}_{x,y}$, can be written
as
\begin{equation}
  S^{(4)}_{x,y} = 
           \sum_\lambda \frac{ \psi_\lambda(x)
	   \psi^\dagger_\lambda(y)}{i\lambda + m}
\end{equation}
leading directly to the Banks-Casher relation \cite{Banks:1980yr}
(with our normalization for the chiral condensate)
\begin{eqnarray}
  -\langle \bar{\psi} \psi \rangle & = &
    \frac{1}{12V} \frac { \langle | \nu | \rangle }{m} +
          \frac{m}{12V} \left \langle \sum_{\lambda \ne 0}
          \frac{1}{\lambda^2 + m^2} \right \rangle \\
	  & = & {2m \over 12} \int^\infty_0 \, d \lambda
	       \frac{ \rho(\lambda) }
	       { \lambda^2 + m^2 }
\label{eq:banks_casher}
\end{eqnarray}
where $\nu$ is the winding number and $ \rho(\lambda)$ is the average
density of eigenvalues.  For quenched QCD, $\rho(\lambda)$ has no
dependence on the quark mass.  For both quenched and full QCD,
one expects that $|\nu| \sim \sqrt{V}$, as is the case for a dilute 
instanton gas model.  Thus, zero modes lead to a divergent $1/m$ term 
in $\qbq$ whose coefficient decreases as $1/\sqrt{V}$.  (This contrasts 
with the behavior seen \cite{Chen:1998xw} above the deconfinement 
transition where it can be shown that the $1/m$ term remains non-zero 
for quenched QCD in the infinite volume limit \cite{Chandrasekharan:1998yx}.)  
Before discussing the results of our simulations, we first address 
how this simple expectation of a $1/m$ term in $\qbq$ due to zero 
modes should appear for the domain wall fermion operator.

We will find it useful to compare the spectrum and properties of the
hermitian domain wall fermion operator $D_H$ with the hermitian
four-dimensional operator, $D^{(4)}_{H}$, defined by
\begin{equation}
  D^{(4)}_{H} = \gamma_5 ( \Dslash^{(4)} + m ).
\end{equation}
The eigenvalues, $\lambda_H$, and eigenvectors, $\psi_{\lambda_H}$, for
this operator can be given in terms of $\lambda$ and $\psi_\lambda$
given above.  If $\lambda = 0$ we immediately get an eigenvalue
$\lambda_H = \pm m$ for the hermitian operator, and an eigenvector
with the definite chirality $+1$ or $-1$.
For $\lambda \ne 0 $, the eigenvectors of $D^{(4)}_{H}$ are linear
combinations of $\psi_\lambda$ and $\psi_{-\lambda}$ and the
corresponding eigenvalues are $\lambda_H = \pm \sqrt{ \lambda^2 + m^2 }
$.  Since $( \Dslash^{(4)} + m) ^{-1} = (D^{(4)}_{H})^{-1} \gamma_5$,
we have
\begin{eqnarray}
  - \langle \bar{\psi} \psi \rangle
    & = & \frac{1}{12V} \left \langle {\rm Tr} (\Dslash^4 + m)^{-1}
	  \right \rangle \\
    & = & \frac{1}{12V} \left \langle {\rm Tr} \, ( \gamma_5
	  (D^{(4)}_{H})^{-1} ) \right \rangle \\
    & = & \frac{1}{12V} \left \langle \sum_{x,\lambda_H}
	  \frac{\psi_{\lambda_H}(x)^\dagger \gamma_5
	  \psi_{\lambda_H}(x)}{\lambda_H} \right \rangle
	  \label{eq:dh4_spec_qbq}.
\end{eqnarray}
Since for $D^{(4)}_{H}$
\begin{equation}
  \sum_x \psi_{\lambda_H}^\dagger(x) \gamma_5 \psi_{\lambda_H}(x) =
   \frac{m}{\lambda_H}
\end{equation}
Eq.\ \ref{eq:dh4_spec_qbq} also reduces to the Banks-Casher relation,
Eq.~\ref{eq:banks_casher}.  For finite mass, the zero-mode hermitian
eigenfunctions are chiral, while other eigenfunctions have
a chirality proportional to the mass.  This will be important in
our comparisons with domain wall fermions.

For large $L_s$, it is expected that the spectrum of light eigenvalues
of the hermitian domain wall fermion operator, $D_H$, should reproduce
the features of the operator $D^{(4)}_{H}$.  Since $D_H$ depends
continuously on $m_f$, for small $m_f$ its $i$th eigenvalue must have
the form
\begin{equation}
  \Lambda_{H,i}^2 = a^\prime_i + b^\prime_i m_f + c^\prime_i m_f^2
		  + \cdots.
\end{equation}
To make a connection with the normal continuum form for the eigenvalues
we reparameterize $\Lambda_{H,i}^2$ as 
\begin{equation}
  \Lambda_{H,i}^2 = n_{5,i}^2( \lambda^2_i + ( m_f + \dmi)^2)
    + \cdots.
 \label{eq:dherm_lambda_vs_mf}
\end{equation}
Here $n_{5,i}$ is an overall normalization factor and we have defined
$\dmi$, which enters as a contribution to the total quark mass for the
$i$th eigenvalue.  For $\dmi = - m_f$, $\Lambda_{H,i}^2$ is at its
minimum.  Modes which become precise zero modes when $L_s
\rightarrow \infty$ will have non-zero values for $\lambda_i$ and
$\delta m_i$ for finite $L_s$.  We will refer to such modes as
topological near-zero modes.

From perturbation theory in $m_f$, one can easily see that
\begin{equation}
  \frac{d \Lambda_{H,i}}{dm_f} = \langle \Lambda_{H,i} | \gamma_5
    \qw | \Lambda_{H,i} \rangle
\label{eq:m_f_pt}
\end{equation}
while the chain rule applied to Eq.\ \ref{eq:dherm_lambda_vs_mf}
gives
\begin{equation}
  \frac{d \Lambda_{H,i}(m_f)}{dm_f} = \frac{n_{5,i}^2 (m_f +
    \dmi ) } {\Lambda_{H,i}}.
\label{eq:chain_rule}
\end{equation}
Combining this with Eq.\ \ref{eq:dh_spectral_qbq} gives
\begin{equation}
  -\qbq = {1\over 12 V}\left \langle \sum_i \frac{m_f + \dmi}
    {\lambda_i^2 + ( m_f + \dmi)^2} \right \rangle
  \label{eq:dwf_banks_casher}
\end{equation}
which agrees with the Banks-Casher form, Eq.\ \ref{eq:banks_casher}
with the addition of the $i$ dependent mass contribution $\delta m_i$.
Thus, the parameter $\lambda_i$ in Eq.\ \ref{eq:dherm_lambda_vs_mf}
should be identified with the eigenvalues of the continuum
anti-hermitian operator $\Dslash^{(4)}$.  As indicated by 
Eqs.~\ref{eq:m_f_pt} and \ref{eq:chain_rule}, $\dmi$ should 
represent a contribution to the eigenvalue from the chiral 
symmetry breaking effects of coupling of the domain walls, 
present for finite $L_s$.

These arguments show that the domain wall fermion chiral condensate
will grow as $1/m_f$ for gauge field configurations with topology,
provided $L_s$ is large enough to make $\dmi$ and $\lambda_i$ small.
The continuum expectation of a $1/m_f$ divergence is modified at small
$m_f$ by the non-zero values of $\dmi$ and $\lambda_i$ for topological
near-zero modes.  For a single configuration, the precise departure
from a $1/m_f$ divergence is dominated by the eigenvalues with the
smallest values for $\dmi$ and $\lambda_i$;  for an ensemble average,
the departure from $1/m_f$ behavior depends on the distribution of
values of $\delta m_i$.  With this understanding of $\qbq$ for domain
wall fermions, we turn to our simulation results.


\subsection{Quenched measurements of $\qbq$ }

In this section we discuss our results for $\qbq$ for quenched QCD
simulations with domain wall fermions.  Tables
\ref{tab:run_parameters_hmc}, \ref{tab:run_parameters_hb} and
\ref{tab:mass_ranges} give details about the runs where $\qbq$ was
measured.  The most important aspect of the run parameters is the small
values for $m_f$ used, including $m_f = 0.0$ where finite $L_s$ keeps
$\Lambda_H$ non-zero, allowing the conjugate gradient inverter to be
used.  Of course the number of conjugate gradient iterations becomes
quite large.

Equation\ \ref{eq:dwf_banks_casher} shows that we should expect large 
values for $-\qbq$ for small $m_f$ for configurations with topological
near-zero modes.  Figure \ref{fig:qbq_b5_7_ls32_48} shows $-\qbq$ for
$8^3 \times 32$ lattices at $\beta = 5.7$ with both $L_s = 32$ and
48.   The quark masses used cover the ranges
$0-0.04$ and $0.00025-0.008$, defined in Table~\ref{tab:mass_ranges}.  
Both values for $L_s$ show an increase in $-\qbq$ for very 
small quark mass, an effect expected from the presence of 
a non-zero value for $\langle | \nu | \rangle$.  (This effect 
was first reported for domain wall fermions based on quenched
simulations done on $8^3 \times 32$ lattices with $\beta = 5.85$, $M_f
= 1.65$ and $L_s = 32$ and listed in Table \ref{tab:run_parameters_hmc}
\cite{Fleming:1998cc}.)

Motivated by the form of Eq.\ \ref{eq:dwf_banks_casher} we have
fit $-\qbq$ to the following phenomenological form
\begin{equation}
  -\qbq = \frac{ a_{-1}}{ m_f + \dmqbq} + a_0 + a_1 m_f
  \label{eq:qbq_phenom}
\end{equation}
where $a_{-1}$, $a_0$, $a_1$ and $\dmqbq$ are parameters to be
determined.  $\dmqbq$ represents a weighted average of $\delta m_i$
over the eigenvalues which dominate $\qbq$ for small $m_f$.  The
measurements of $\qbq$ for different values of $m_f$ are strongly
correlated, being done on the same gauge field configurations with,
generally, the same random noise estimator used to determine $\qbq$ for 
all the masses.  The common noise source makes the signal for the $1/m_f$
divergence particularly clean, since the overlap of the topological
near-zero mode eigenvectors with the random source does not fluctuate
on a single configuration.  This strong correlation precludes doing a
correlated fit of $\qbq$ to $m_f$, since the correlation matrix is too
singular.  Thus, the fits in this section are uncorrelated fits of
$\qbq$ to $m_f$.

Table \ref{tab:qbq_phenom} gives the results for fits to the form of
Eq.\ \ref{eq:qbq_phenom} for our $\beta = 5.7$, 5.85 and 6.0
simulations.  All the fits have a value $\chi^2/{\rm dof}$ less than 0.1, 
a consequence of doing uncorrelated fits to such correlated data.  In
Figure \ref{fig:qbq_b5_7_ls32_48}, one sees that the fit represents 
the data quite well.  Continuing with $8^3 \times 32$ lattices at
$\beta = 5.7$, Table \ref{tab:qbq_phenom} shows the fit parameters are
very similar for $L_s = 32$ and 48, except for $\dmqbq$, which drops
from 0.0040(4) to 0.0017(2).  This indicates a decrease in $\dmi$ as
$L_s$ increases.

Figure \ref{fig:qbq_b6_0_ls16_24} is a similar plot of $-\qbq$ for $16^3
\times 32$ lattices with $\beta = 6.0$ for $L_s = 16$ and 24.  The rise
in $\qbq$ for small $m_f$ exhibits the same general structure as for
the $\beta = 5.7$ data in Figure \ref{fig:qbq_b5_7_ls32_48}, but the
effect is larger.  Here $\dmqbq$ falls from 0.00056(3) for $L_s = 16$
to 0.00011(1) for $L_s = 24$.

To further demonstrate that the divergence for small $m_f$ is due to
eigenfunctions of $D_H$ that represent zero modes of a definite
chirality, Figure \ref{fig:qbq_qbg5q_evol_b6_0} shows the evolution of
both $-\qbq$ (solid lines) and $-\langle \overline{q} \gamma_5 q \rangle$
(dotted lines).  These evolutions are for $16^3 \times 32$ lattices at
$\beta = 6.0$ with $L_s = 16$.  Eigenfunctions with a positive
chirality contribute equally to $\qbq$ and $\langle \overline{q}
\gamma_5 q \rangle$, while negative chirality eigenfunctions contribute
with an opposite sign to $\langle \overline{q} \gamma_5 q \rangle$.
The topological near-zero modes should be approximately chiral and, for
smaller values of $m_f$, one see large fluctuations in $\qbq$ and
$\langle \overline{q} \gamma_5 q \rangle$.  Some of the fluctuations
have the same sign and some are of opposite sign.  Thus, we have
configurations with eigenfunctions which are very good approximations
to the exact zero modes expected as $L_s \rightarrow \infty$.

As mentioned earlier, $\langle | \nu | \rangle/V$ should decrease with
volume, with the asymptotic dependence given by $1/\sqrt{V}$.  To
investigate this numerically, we have measured $\qbq$ on both $8^3
\times 32$ and $16^3 \times 32$ lattices at $\beta = 5.7$ and 5.85 with
$L_s = 32$ and show the $\beta = 5.7$ results in Figure
\ref{fig:qbq_b5_7_ls32}.  The graph clearly shows that the $1/m_f$
divergence is drastically suppressed by the larger volume.  The
coefficient of the $1/m_f$ term falls from $6.0(6) \times 10^{-6}$ to
$2.5(4) \times 10^{-6}$ as the volume is changed by a factor of 8.
This may be somewhat misleading, since $\dmqbq$ also changes by a
factor of about 2, likely due to the phenomenological nature of the fit
and the small effects of the $1/m_f$ pole for the larger volume.
Putting aside this systematic difficulty, the $1/m_f$ coefficient
decreases by a factor of $1/\sqrt{5.8}$, showing the general behavior
expected but not in precise agreement with the expected asymptotic
form.  For $\beta = 5.85$, where the physical size of the lattices is
smaller, the $1/m_f$ coefficient falls from $3.8(3) \times 10^{-6}$ to
$0.60(9) \times 10^{-6}$, a factor of 6.3.  We have not seen the
expected $1/\sqrt{V}$ dependence for the $1/m_f$ coefficient, but it
does decrease with volume in accordance with general ideas.  It is
possible that on the larger $16^3\times 32$ volume, the $1/m_f$ rise 
is not large enough to allow its coefficient to be determined without 
systematic errors.

Thus, we have clear evidence for topological near-zero modes in our
quenched simulations using domain wall fermions.  They are revealed
through a large $1/m_f$ rise in our values for $-\qbq$, the presence of
configurations where $\qbq$ and $\langle \overline{q} \gamma_5 q
\rangle$ are large and of opposite sign and the volume dependence of
the coefficient of the $1/m_f$ term.  We have extracted a quantity,
$\dmqbq$, from a phenomenological fit to $\qbq$, which represents the
effects of finite $L_s$ on the eigenmodes with small eigenvalues which
dominate $\qbq$ for $m_f \rightarrow 0$.  Physical values for $\qbq$ in
the chiral limit, without the contribution of the topological near-zero
modes, will be presented in Section \ref{sec:hadron}.  We now turn to a
discussion of how these zero modes, and the expected light modes
responsible for chiral symmetry breaking, are evident in measurements
of the pion mass.


\section{The pion mass in the chiral limit}
\label{sec:chiral_m_pi}

For domain wall fermions with $L_s = \infty$, the chiral limit is
achieved by taking $m_f = 0$.  For our quenched simulations at finite
$L_s$, we must investigate the chiral limit in detail to demonstrate
that the changes from the $L_s \rightarrow \infty$ limit are under
control and of a known size.  As is discussed in Sec.~\ref{sec:mres_def}, 
for low energy QCD physics the dominant effect of finite $L_s$ should 
be the appearance of an additional chiral symmetry breaking term 
in the effective Lagrangian describing QCD.  This term has the form 
$\mres \, \overline{q}(x) q(x)$ and in the continuum limit its presence 
will make $m_\pi$ vanish at $m_f = -\mres$ up to terms of order $a^2$.
Our investigation of the chiral limit is made more difficult since
there are other issues affecting this limit, beyond having $L_s$
finite.  For domain wall fermion quenched simulations, the chiral limit
may be distorted by:
\begin{enumerate}
\item Order $a^2$ effects.  Since we are working at finite lattice 
  spacing chiral symmetry will not be precisely restored even for 
  $m_f +\mres=0$.  In particular, additional chiral symmetry breaking
  will come from the effects of higher dimension operators suppressed by
  factors of $a^n$ for $n\ge 2$.  Thus, we cannot not expect $m_\pi$ to
  vanish precisely at the point $m_f = -\mres$, but perhaps at a nearby
  point, removed from $-\mres$ by a terms of $O(a^2)$.
\item  Finite $L_s$. The residual mass, $\mres$, should represent 
  the finite $L_s$ effects for physics describable by a 
  low-energy effective Lagrangian.  However, there will be additional
  effects of finite $L_s$ for observables sensitive to ultraviolet 
  phenomena.  Further, a quantity with sufficiently severe infrared 
  singularity may show unphysical sensitivity to those $L_s$-dependent 
  eigenfunctions $|\Lambda_{H,i}\rangle$ (and the parameters 
  $n_{5,i}$, $\lambda_i$ and $\delta m_i$ of the previous section) with 
  small eigenvalues $\Lambda_{H,i}$.
\item  Topological near-zero modes.  The previous section has shown
  these dominate $\qbq$ for small quark masses ( $m_f \le 0.01$ ) for
  the volumes we are using.  From the Ward-Takahashi identity, these
  effects must also by present in the pion correlator $\corrpp$.
\item Finite volume.  For staggered fermions, where the remnant chiral
  symmetry at finite lattice spacing requires $\mpi2 = 0$ when the
  input quark mass is zero, the finite volumes used in simulations
  have been seen to make $\mpi2$ non-zero when extrapolated to the
  chiral limit from above \cite{Aoki:1994gi,Mawhinney:1996qy}.  
  Such an effect may also be expected to occur for domain wall fermions.
\item Analytic results argue for the presence of ``quenched chiral
  logs'' with the dependence of $\mpi2$ on the quark mass in
  quenched QCD different from that of full QCD
\cite{Sharpe:1990me,Bernard:1992mk,Sharpe:1992ft}.
\end{enumerate}

In this section we study the pion mass in the limit of small 
quark mass.  Demonstrating consistent chiral behavior for the pion 
mass in the limit $m_f + \mres \rightarrow 0$ is a critical component 
in establishing the ability of the domain wall fermion formalism to 
adequately describe chiral physics.  If we discover that the limit 
$m_f+\mres \rightarrow 0$ is obscured by large $O(a^2)$ effects or 
large violations of chiral symmetry caused by unanticipated propagation 
between the domain walls, little may be gained from this new formalism.  
For the $\rho$ and nucleon masses reported in Section~\ref{sec:sim_results}, 
the masses were shown to be well fit by a linear dependence on the input 
quark mass, $m_f$.  Any possible non-linearities are not resolvable 
within our statistics.  For the pion, the statistical errors for these 
values of $m_f$ are smaller and we have also run simulations at smaller 
values for $m_f$ so we might hope to learn more about this important 
quantity.  We begin by investigating the effects of topological 
near-zero modes on the pion.


\subsection{Topological near-zero mode effects on the pion:
  analytic considerations}

We have seen that topological near-zero modes dominate $\qbq$ for small
$m_f$ and, by continuity, they will also alter the value for $\qbq$
determined with larger quark masses.  Through the Ward-Takahashi
identity, these modes also appear in the pion correlator, $\corrpp$ and
therefore can enter in the determination of the pion mass in a lattice
simulation.  Alternatively, the axial-vector correlator can be used to
measure the pion mass and the zero modes may affect this correlator
differently.  It is vital to understand the role of these topological
near-zero modes, since a study of the chiral limit of $\mpi2$ depends
on an accurate measurement of the mass of the pion state.  In this 
section we will study the way in which topological zero-modes might
be expected to effect pion correlation functions for the continuum
theory using our results as a guide to the study of the domain wall
amplitudes.

Before proceeding, we first establish our notation for
susceptibilities and the integrated Ward-Takahashi identity.
In general we define
\begin{equation}
  \chi_{CD} \equiv {1 \over 12} \sum_x \langle C(x) D(0) \rangle
\end{equation}
where $C$ and $D$ are any two hadronic interpolating fields.  In
particular
\begin{eqnarray}
  \chipp  & \equiv & {1 \over 12}\sum_x \corrpp \\
  \chiaa  & \equiv & {1 \over 12}\sum_x \langle A^a_0(x) \, A^a_0(0) \rangle \\
  \chimpp & \equiv & {1 \over 12}\sum_x \langle \mppi(x) \, \pi^a(0) \rangle
\end{eqnarray}
where no sum over $a$ is intended and the factor of $1/12$ has been 
introduced to maintain consistency with our somewhat unconventional 
normalization for the chiral condensate given in Eq.~\ref{eq:qbd_def}.
Then the Ward-Takahashi identity, Eq.\ \ref{eq:ward_tak_id}, 
with $O(0) = \pi^a(0)$ and summed over $x$ becomes
\begin{equation}
  m_f \chipp + \chimpp = \qbq
  \label{eq:int_ward_tak_id}
\end{equation}
which we will refer to as the integrated Ward-Takahashi identity.

We first consider Eq.\ \ref{eq:int_ward_tak_id} for large $L_s$, where
we should recover the continuum version of the identity.  To simplify
the presentation, we start with the notation of Section
\ref{sec:dwf_banks_casher} for the continuum four-dimensional
anti-hermitian Dirac operator.  We immediately deduce 
from Eq.~\ref{eq:int_ward_tak_id} that a $1/m$ divergence 
in $\qbq$ from topological zero modes dictates a $1/m^2$
divergence in $\chipp$.  In addition, $\chipp$ should have a $1/m$
divergence for large volumes from the pion pole and, as we will see
below, there can also be a $1/m$ pole from topological zero modes.
However, the volume dependence of these various pole terms should be
different.  Pole terms from topological near-zero modes should have a
coefficient which is $O(V^{-1/2})$ for large volumes, while the $1/m$
term from the pion pole should be volume independent,

Thus, we expect
\begin{equation}
 \chipp = \frac{1}{V^{1/2}} \frac{a_{-2}}{m^2}  
          + \frac{a_{-1}}{m} +  O(m^0).
\label{eq:qbq_pole}
\end{equation}
The coefficients $a_{-2}$ and $a_{-1}$ should become volume independent 
in the infinite volume limit.  However, the ``pion pole'' piece, $a_{-1}$,
may contain an additional $1/V^{1/2}$ term arising from zero modes.  
Note, a particular order of limits must be understood when interpreting 
Eq.~\ref{eq:qbq_pole}.  One expects that the usual relation 
$m_\pi^2 \propto m$ will hold only when $m \gg 1/V|\qbq|$
\cite{Leutwyler:1992yt}.  Although this prevents our taking the 
$m \rightarrow 0$ limit of Eq.~\ref{eq:qbq_pole}, it is fully consistent
with the domain $m \propto 1/V^{1/2}$ where the $1/m^2$ term in 
Eq.~\ref{eq:qbq_pole} may be as large as or much larger than the conventional
$1/m$ term coming from the pion.  For domain wall fermions at finite 
$L_s$ these pole terms will be rendered less singular by the presence of 
the $\dmi$ terms in the eigenvalues for $D_H$.

Lattice measurements of the pion mass come from the exponential decay
of a correlator like $\corrpp$ in the limit of large $|x|$.  Having 
examined the zero mode effect in the somewhat simpler susceptibilities, 
we will now investigate the topological zero mode contributions to
two-point functions from their spectral decomposition to understand
how zero modes can distort measurements of the pion mass.
We have
\begin{eqnarray}
  \corrpp & = & \left \langle \tr \, [ \, S^{(4)}_{x,0} \, \gamma_5
		\, S^{(4)}_{0,x} \, \gamma_5 \, ] \, \right \rangle \\
          & = & \left \langle \sum_{\lambda, \lambda^\prime}
		\frac{ \psi_\lambda^\dagger(x) \psi_{\lambda^\prime}(x)
		\; \psi_{\lambda^\prime}^\dagger(0) \psi_\lambda(0) }
		{ (-i\lambda + m ) ( i \lambda^\prime + m ) }
		\right \rangle.
\end{eqnarray}
First we consider the terms in the sum where both $\lambda$ and
$\lambda^\prime$ are zero.  This gives a $1/m^2$ pole in
$\corrpp$, provided the eigenfunctions in the numerator are non-zero
at $x$ and 0.  (Integrating $\corrpp$ over $x$ shows that these 
topological near-zero modes give the $1/m^2$ contribution to $\chipp$.)
The terms in the sum where neither $\lambda$ or $\lambda^\prime$ are
zero should include the small eigenvalues which are responsible for
the Goldstone nature of the pion.  For large $|x|$, the total
contribution to $\corrpp$ from these modes should be proportional to
\begin{equation}
   | \langle 0 | \pi(0) | \pi \rangle |^2 \;
   \frac{e^{- m_\pi |x|} }{m_\pi}
   \sim \frac{1}{\sqrt{m}} \, e^{- m_\pi |x|}.
\end{equation}
(Integrating over $x$ gives another factor of $m_\pi$ in the
denominator, which produces the $1/m$ pion pole in $\chipp$.) Lastly,
the terms with either $\lambda$ or $\lambda^\prime$ zero, but not both,
can be written as
\begin{equation}
  \left \langle \sum_{\lambda > 0, \lambda^\prime = 0}
	\frac{ \psi_\lambda^\dagger(x) \psi_{\lambda^\prime}(x) \;
		\psi_{\lambda^\prime}^\dagger(0) \psi_\lambda(0) }
		{ \lambda^2 + m^2 } \right \rangle
    <   \left \langle \sum_{\lambda \neq 0 }
	  \frac{1} { \lambda^2 + m^2 } \right \rangle
   =  \frac{\qbq_{\rm nz}}{m}
\end{equation}
where $\qbq_{\rm nz}$ is the chiral condensate measured without
zero mode contributions.  Here we have used the symmetries in Eq.~\ref{eq:anti_herm_d4_evalues} to combine the $\pm\lambda$ 
terms in the sum over $\lambda$ and remove the term odd in 
$i\lambda$.   Since $\qbq_{\rm nz}$ should be non-zero
as $m \rightarrow 0 $, we see that the contribution to the
correlator from terms with zero modes in one of the propagators
can produce at most a $1/m$ pole term in $\corrpp$.

Thus, we expect
\begin{equation}
 \corrpp = \frac{1}{V^{1/2}} \left ( \frac{c_{-2}(x,0)}{m^2} +
	    \frac{c_{-1}(x,0)}{m} \right )
   + c_{-1/2} \; \frac{e^{- m_\pi |x|} } {\sqrt{m}} + \cdots
\end{equation}
for small $m$.  The first two terms represent the possible zero
mode contributions.  It is important to note that $c_{-2}(x,0)$
gets contributions from the modulus squared of the zero-mode
eigenfunctions at the points 0 and $x$, while $c_{-1}(x,0)$ does not.
In particular, for a configuration with a single zero mode,
$c_{-2}(x,0)$ is positive definite, being given by
\begin{equation}
  c_{-2}(x,0) = V^{1/2} \left \langle  |\psi_0(x)|^2 \; |\psi_0(0)|^2
		\right \rangle.
\end{equation}
Thus, one could expect $c_{-2}(x,0)$ to be a number of order
the inverse of the mean zero-mode size squared, while $c_{-1}(x,0)$
could be much smaller due to the terms of differing sign appearing
in the sum over eigenmodes.

For large enough $|x|$, only the true pion state should contribute. 
Such large $|x|$ requires a correspondingly large $V$ with the necessary
suppression of zero modes.  However, at a fixed separation $|x|$ in
a finite volume and for simulations with small enough $m$, the physical 
pion contribution to $\corrpp$ can be completely negligible.
For finite $L_s$, the domain wall fermion spectral form, 
Eq.\ \ref{eq:dwf_corrpp_spec}, gives the precise role of
the topological near-zero modes.  The double sum over $\Lambda_H$ and
$\Lambda_H^\prime$ decomposes as we have done above for $D^{(4)}$ and
the dominant contribution of the topological near-zero modes enters as
$1/(\lambda_i^2 + (m_f + \dmi)^2)$ provided $ \langle \Lambda_H | \qw |
\Lambda_H^\prime \rangle$ is well approximated by $O(1)
\delta_{\Lambda_H, \Lambda_H^\prime}$.  Thus, for $\lambda_i$ and $\dmi$
small, there should be a region in $m_f$ where $\corrpp$ displays a
$1/m_f^2$ character.

The pion mass can also be measured from the axial current correlator,
$\corraa$.  The susceptibility for this correlator, $\chiaa$,
is not constrained by the integrated Ward-Takahashi identity
as is $\chipp$.  However, there must be a pion pole contribution in
addition to any zero mode terms.  Therefore
\begin{eqnarray}
  \chiaa & \sim & \frac{\langle 0 | A_0(0) | \pi \rangle |^2 }{\mpi2}
	       + O(m^{-n}) \; {\rm zero \; mode \; poles } \\
	 & = & O(m^0)
	       + O(m^{-n}) \; {\rm zero \; mode \; poles }
\end{eqnarray}
where we have used $\langle 0 | A_0(0) | \pi \rangle \sim m_\pi$.  The
physical pion contribution is independent of $m$ for small $m$, which
is to be compared with the $1/m$ contribution of the physical pion to
$\chipp$.  We now turn to the question of the zero mode contribution.

Once again we consider the $L_s = \infty$ case and use the notation for
$D^{(4)}$.  For the axial vector correlation function, the spectral
decomposition is
\begin{eqnarray}
  \corraa & = & -\left \langle \tr \, [ \, S^{(4)}_{x,0}
		\, \gamma_0 \, \gamma_5
		\, S^{(4)}_{0,x}
		\, \gamma_0 \, \gamma_5 \, ] \, \right \rangle \\
          & = & \left \langle \sum_{\lambda, \lambda^\prime}
		\frac{ \psi_\lambda^\dagger(x) \, \gamma_0 \, 
		\psi_{\lambda^\prime}(x)
		\; \psi_{\lambda^\prime}^\dagger(0) \, \gamma_0 \,
		\psi_\lambda(0) }
		{ (-i\lambda + m ) ( i \lambda^\prime + m ) }
		\right \rangle.
\end{eqnarray}
The terms in the sum where both $\lambda$ and $\lambda^\prime$ are zero
modes vanish here, since the zero modes have a definite chirality and
$\gamma_0$ couples different chiral components.  (On a given
configuration, all the exact zero modes must have the same chirality
since exact zero modes can only occur through the index theorem.)  Thus,
there are no $1/m^2$ terms in $\corraa$.  Note that a $1/m$
contribution to $\corraa$ can appear from terms in the sum with either
$\lambda$ or $\lambda^\prime$ a zero mode (as we saw for $\corrpp$).
The size of such a contribution depends on the matrix element of
$\gamma_0$ between eigenfunctions.

The terms with neither $\lambda$ nor $\lambda^\prime$ a zero mode
give the physical pion contribution, which should have the form
\begin{equation}
  m_\pi^2 \, \frac{e^{-m_\pi |x| }}{m_\pi} \sim \sqrt{m} \,
    e^{-m_\pi |x| }.
\end{equation}
Thus, we expect that
\begin{equation}
  \corraa = \frac{d_{-1}(x,0)}{mV^{1/2}} + d_{1/2} m^{1/2} \,
	    e^{-m_\pi |x| } + \cdots
\end{equation}
with other possible subleading terms from topological zero modes.  As
for the coefficient $c_{-1}(x,0)$ in $\corrpp$, the coefficient
$d_{-1}(x,0)$ above involves matrix elements between different
eigenfunctions and could be quite small from cancellations.  Thus, even
though in $\corraa$, the physical pion contribution can still be
$O(m^{3/2})$ smaller than the zero mode contribution, the effects of
zero modes in this correlator are likely suppressed by the smaller
coefficient $d_{-1}(x,0)$.

To finish our discussion of the topological zero modes in correlators,
we now examine the spectral form of $\corrss$, where the $c$ subscript
means that we only consider the connected part of the correlator.
We find
\begin{equation}
 -\corrss = \left \langle \sum_{\lambda, \lambda^\prime}
	  \frac{ \psi_\lambda^\dagger(x) \, \gamma_5 \,
	    \psi_{\lambda^\prime}(x)
	  \; \psi_{\lambda^\prime}^\dagger(0) \, \gamma_5 \,
	    \psi_\lambda(0) }
	  { ( -i\lambda + m ) ( i \lambda^\prime + m ) } \right \rangle.
\end{equation}
Since zero modes are eigenfunctions of $\gamma_5$, their contribution
to the $1/m^2$ and $1/m$ terms in $\corrpp$ and $-\corrss$ are equal.
Thus, we have
\begin{equation}
 -\corrss = \frac{1}{V^{1/2}} \left ( \frac{c_{-2}(x,0)}{m^2} +
	    \frac{c_{-1}(x,0)}{m} \right )
   + c_{\sigma} \; e^{- m_{\sigma_c} |x|} + \cdots
\end{equation}
for small $m$.  By considering $\corrpp + \corrss$, we obtain
a two point function with no zero mode effects, but which
contains both the physical pion and a heavier state from
$\corrss$.  Thus, to reduce the effects of topological near-zero 
modes in this way requires that one works with correlators where 
the heavy mass $\sigma_c$ state is present.

To summarize this section, we have seen how the topological near-zero
modes for domain wall fermions should enter the correlators which
are used to determine the pion mass.  For $\corrpp$, there must be a
$1/m_f^2$ contribution from near-zero modes, compared with the
$1/\sqrt{m_f}$ contribution expected from the physical pion.  For
$\corraa$, the topological near-zero modes can contribute a term of
order $1/m_f$, while the physical pion should produce a $\sqrt{m_f}$
contribution.  However, the coefficient of the $1/m_f$ term can be
small.  We also have pointed out that the volume 
dependence of the contribution of the topological near-zero modes to
the correlator is different from the contribution due to the modes
responsible for chiral symmetry breaking in QCD so that the zero-mode
effects should vanish as the space-time volume increases.  

The above discussion explicitly addresses the behavior to be
found in a chiral theory.  Thus, it will apply to the domain wall
theory in the limit $L_s\rightarrow\infty$.  We might expect two 
sorts of modified behavior for a theory with finite $L_s$.  First,
the chiral properties of the exact zero modes which 
eliminate the most singular terms from the
$\corraa$ and $\corrpp+\corrss$ will no longer be exact for finite
$L_s$ allowing more singular terms suppressed exponentially in $L_s$
to appear.  Second the zero-mode singularities themselves may be
softened by additional mass contributions to the denominators.
We now turn to the results of our simulations.


\subsection{Topological near-zero mode effects on the pion:
  numerical results}

The first detailed studies of $\mpi2$ as $m_f \rightarrow 0$, done on
$8^3 \times 32$ lattices with $\beta = 5.7$ and a variety of values of
$L_s$, showed that $\mpi2(m_f = 0)$ was not decreasing exponentially to
zero as $L_s \rightarrow \infty$, but rather seemed to be approaching a
constant value of 
$\sim 200$ MeV \cite{Mawhinney:1998ut,Malureanu:1998thesis,Wu:1999cd}.  
The pion mass was extracted from
$\corrpp$ and the resulting $\mpi2$ versus $m_f$ showed noticeable 
curvature for the quark masses used, which were in the range $0.02-0.22$.  
Therefore, the extrapolation to $m_f = 0$ was done using only the three 
lightest quark masses: 0.02, 0.06 and 0.10.  Figure~\ref{fig:mpi2_vs_ls_b5.7} 
updates the earlier graph in \cite{Wu:1999cd} with more data at $L_s = 48$ 
and a new point at $L_s = 64$.  The additional data does show a 
behavior that is more consistent with a monotonic decrease of $\mpi2(m_f = 0)$ 
with increasing $L_s$ that than seen in our earlier study \cite{Wu:1999cd}.
However, the dependence on $L_s$ shown in Fig.~\ref{fig:mpi2_vs_ls_b5.7} 
still cannot be described by a single falling exponential and, for large 
$L_s$ is falling quite slowly.  In this section we will probe this issue 
and others related to the chiral limit, using information from our 
simulations at both $\beta = 5.7$ and 6.0 for many values of $m_f$ 
and $L_s$.

\begin{figure}
\epsfxsize=\hsize
\epsfbox{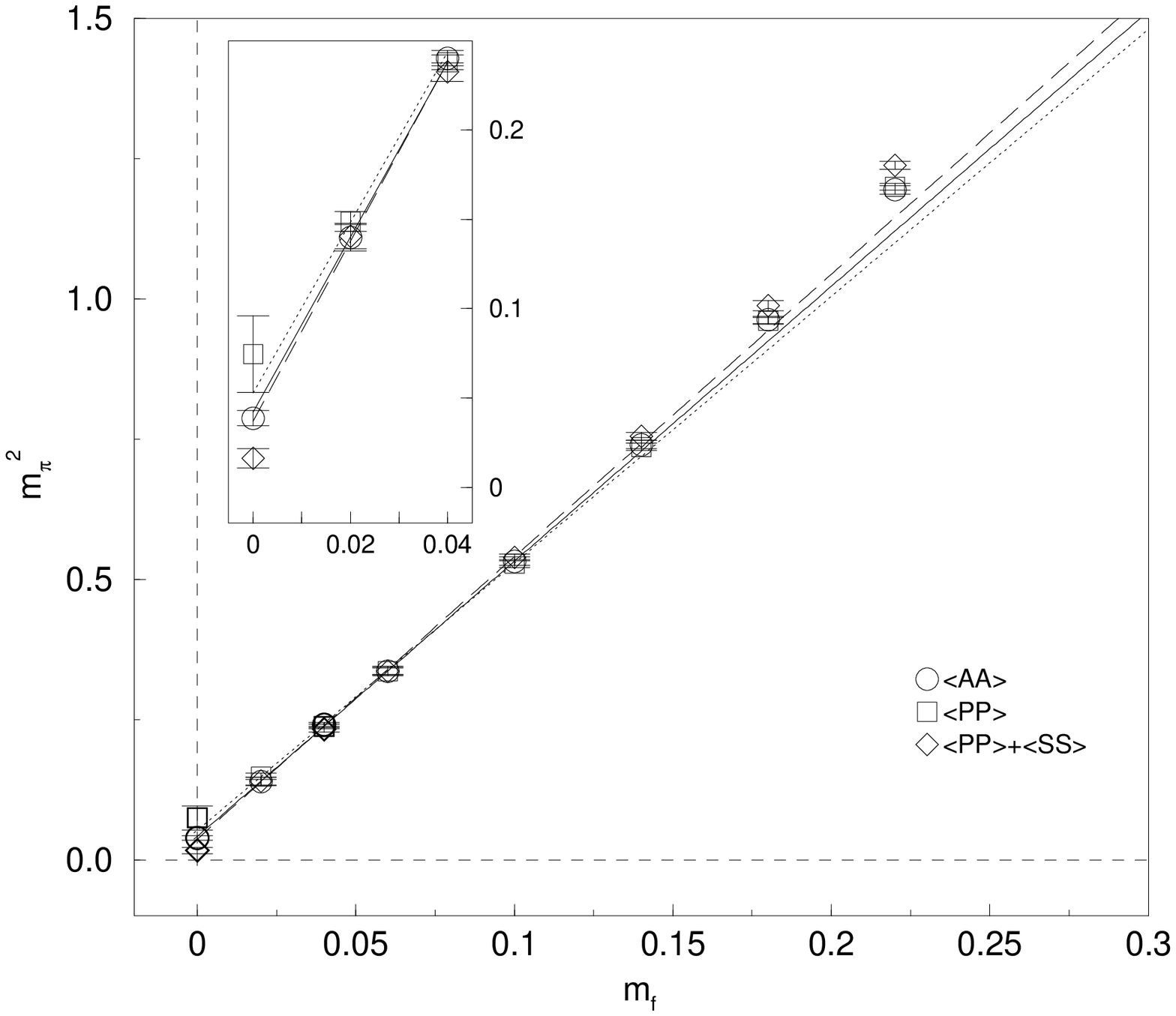}
\caption{The pion mass squared versus $m_f$ from $\corrpp$ ($\Box$),
$\corraa$ ($\circ$) and $\corrpp + \corrss$ ($\Diamond$) for
quenched simulations done on $8^3 \times 32$ lattices at $\beta = 5.7$
with $L_s = 48$.  For $m_f = 0.0$, the correlators all give
different masses due to the differing topological near-zero mode
contributions for each one.  For large enough $x$, all the correlators
should give the same mass.  However, this limit requires a large volume which is
expected to suppress such zero-mode effects.  The dotted line is the fit of Eq.\
\ref{eq:b5_7_8nt32_0.02_0.1_pp_fit}, the solid line is from Eq.\
\ref{eq:b5_7_8nt32_0.02_0.1_aa_fit} and the dashed line is from Eq.\
\ref{eq:b5_7_8nt32_0.02_0.1_pp+ss_fit}.}
\label{fig:mpi2_vs_mf_b5_7_8nt32_ls48}
\end{figure}

Figure \ref{fig:mpi2_vs_mf_b5_7_8nt32_ls48} shows results for $\mpi2$
versus $m_f$ for $8^3 \times 32$ lattices at $\beta = 5.7$ with $L_s =
48$, including results for $m_f = 0.0$.  The pion mass is extracted
from three different correlators: $\corrpp$, $\corraa$ and $\corrpp +
\corrss$.  For $0.02 \le m_f \le 0.15$, the pion masses extracted
from the different correlators are in good agreement.  As $m_f
\rightarrow 0$ the masses begin to disagree, presumably due to the differing
contributions of the topological near-zero modes to each correlator.
Table \ref{tab:mpi_mf_small} gives our fitted pion masses for $m_f <
0.01$.  While the different correlators generally have reasonable
values for $\chi^2/{\rm dof}$ the fitted masses disagree
substantially.  For large enough separation of the interpolating
operators, the three correlators should give the same mass.  
However, we cannot take this large separation limit in our finite
volume.  The results in Table \ref{tab:mpi_mf_small} are the apparent 
masses as determined from fitting to the correlators for finite 
separation of the interpolating operators.

The lines drawn in Figure \ref{fig:mpi2_vs_mf_b5_7_8nt32_ls48} are
from correlated linear fits to $\mpi2$ using $m_f = 0.02$ to 0.1. 
The dotted line is for $\mpi2$ from $\corrpp$, the solid line
for $\corraa$ and the dashed line for $\corrpp+\corrss$.  The fit
results are
\begin{eqnarray}
    \mpi2  =  0.053(8) + 4.76(12) m_f & \;\;\;
        & \chi^2/{\rm dof} = 0.7 \pm 2.5
      \label{eq:b5_7_8nt32_0.02_0.1_pp_fit} \\
    \mpi2  =  0.042(8) + 4.90(9) m_f & \;\;\;
        & \chi^2/{\rm dof} = 0.01 \pm 0.25
      \label{eq:b5_7_8nt32_0.02_0.1_aa_fit} \\
    \mpi2 = 0.037(8) + 5.04(6) m_f & \;\;\;
        & \chi^2/{\rm dof} = 1.7 \pm 2.8
      \label{eq:b5_7_8nt32_0.02_0.1_pp+ss_fit}
\end{eqnarray}      
for $\corrpp$, $\corraa$ and $\corrpp+\corrss$ respectively.  Note for 
large mass, $m_f > 0.1$, $\corrpp+\corrss$ gives a mass that is 
systematically higher than that implied by the other two correlators, likely 
due to contamination from heavy states present in $\corrss$.

Figure \ref{fig:mpi_effm_b5_7_8nt32_ls48_m0.0} shows the pion effective
mass from the three different correlators for $8^3 \times 32$ lattices
at $\beta = 5.7$ with $L_s = 48$ and $m_f = 0.0$.  Reasonable plateaus
are present in $\corraa$ and $\corrpp$, although the value for the
effective mass is markedly different.  The $\corrpp + \corrss$
effective mass becomes very small for intermediate values of $t$.
Figure \ref{fig:mpi_effm_b5_7_8nt32_ls48_m0.04} is a similar plot,
except for $m_f = 0.04$.  Here the effective mass plots show nice
plateaus and consistent results.  This supports the
presence of topological near-zero modes affecting the various
correlators in different ways and provides an example where nice
plateaus do not assure a correct asymptotic result.

As a final step in demonstrating the zero mode effects in the various
correlators, in Figure \ref{fig:corr_evol_b5_7_8nt32_ls48_m0_0} the
evolution of $\corrpp$, $\corrss$ and $\corraa$ is shown for $8^3
\times 32$ lattices at $\beta = 5.7$ with $L_s = 48$ and $m_f = 0.0$.
These correlators are from a point source to a point sink and the zero
spatial momentum component is taken for the sink.  The sink is at a
separation $t=8$ from the source.  The correlators $\corrpp$ and
$-\corrss$ show very large fluctuations, which are common to both
correlators, showing the presence of topological near-zero modes.
These large fluctuations are clearly dominating the ensemble average
for the correlators at this separation, $t=8$.  The $\corraa$
correlator does not show large fluctuations where $\corrpp$ and
$\corrss$ do, making the topological near zero mode effects smaller for
this correlator, as expected from the theoretical discussion of the
previous subsection.  Figure \ref{fig:corr_evol_b5_7_8nt32_ls48_m0_04}
is a similar plot, for the same configurations, except with $m_f =
0.04$.  There is no evidence for a large role being played by the
topological near-zero modes.

Similar results have been obtained for simulations on $16^3 \times 32$
lattices at $\beta = 6.0$ with $L_s = 16$.  These lattices have
essentially the same spatial volume, in physical units, as the previous
$8^3 \times 32$, $\beta = 5.7$ lattices since the lattice spacing is half
that for $\beta = 5.7$.  Figure \ref{fig:mpi2_vs_mf_b6_0_16nt32_ls16}
shows $\mpi2$ for $\corraa$, $\corrpp$ and $\corrpp+\corrss$.  For the
smallest $m_f$ points, 0.0 and 0.001, all three correlators give
different results.  For larger values of $m_f$, the pion mass from
$\corraa$ and $\corrpp$ agree, while the $\corrpp+\corrss$ pion mass is
systematically high, likely due to the contribution of the heavy states
in $\corrss$.  The lines drawn in Figure
\ref{fig:mpi2_vs_mf_b6_0_16nt32_ls16} are from correlated linear fits
to $\mpi2$ using $m_f = 0.01$ to 0.04.  The dotted line is for $\mpi2$
from $\corrpp$, the solid line for $\corraa$ and the dashed line for
$\corrpp+\corrss$.  The fit results are
\begin{eqnarray}
    \mpi2  =  0.0132(20) + 3.07(6) m_f & \;\;\;
        & \chi^2/{\rm dof} = 5.0 \pm 5.0
      \label{eq:b6_0_16nt32_0.02_0.1_pp_fit}  \\
    \mpi2  =  0.0098(20) + 3.14(9) m_f & \;\;\;
        & \chi^2/{\rm dof} = 0.03 \pm 0.30
      \label{eq:b6_0_16nt32_0.02_0.1_aa_fit}  \\
    \mpi2 = 0.0020(26) + 3.56(8) m_f & \;\;\; &
        \chi^2/{\rm dof} = 0.06 \pm 0.51
      \label{eq:b6_0_16nt32_0.02_0.1_pp+ss_fit}
\end{eqnarray}
for $\corrpp$, $\corraa$ and $\corrpp+\corrss$ respectively.

Figure \ref{fig:mpi_effm_b6_0_16nt32_ls16_m0.001} shows effective mass
plots for the pion from the three correlators for $m_f = 0.001$ and
Figure \ref{fig:mpi_effm_b6_0_16nt32_ls16_m0.01} is for $m_f = 0.01$.
Both figures show reasonable plateaus,
even though there are differences in the final fitted masses.  We have
also studied the evolution of the correlators at a fixed $t$ for these
$\beta = 6.0$ lattices and see clear topological near-zero mode effects
as were seen at $\beta = 5.7$.

Thus, investigating the chiral limit of domain wall fermions in quenched
QCD by measuring the pion mass is made difficult by the presence of
topological near-zero modes.  One component in the somewhat large values 
of $m_\pi^2$ plotted in Fig.~\ref{fig:mpi2_vs_ls_b5.7} is the effect 
of topological near zero modes.  As can be seen by a comparison with 
Table~\ref{tab:extrap_832_b5_7} the results we find from the correlator
$\corraa$ for $m_\pi^2(m_f=0)$ are about 1 1/2 standard deviations lower
for $L_s=32$ and $48$.  This is likely a systematic bias caused by
the greater influence of the topological near-zero modes on the 
$\corrpp$ correlator.  Unfortunately, these effects may also enter in 
the other correlators that can give $m_\pi$, at least for the source-sink 
time separations currently accessible.  With this large distortion due to
the topological near-zero modes, there we cannot determine the chiral 
limit by extrapolating to the point where $\mpi2$ vanishes.  Subtler 
finite volume effects and possible quenched chiral logarithms are 
completely overshadowed by the singular nature of the basic quark 
propagators for small $m_f$.

In many ways, the presence of these topological near-zero modes is a
welcome change from other lattice fermion formulations because they are
a vital part of the spectrum of any continuum Dirac operator.  However,
in order to further investigate the chiral limit, they must be removed, or 
at least suppressed.  Without adding the fermionic determinant to the path
integral, we can suppress the effect of topological near-zero modes by
going to large volumes.


\subsection{The pion mass for larger volume}

Having seen clear evidence for topological near-zero modes in the
measurements of the pion mass for lattices with a physical size of
$\sim 2$ Fermi, we have worked on a larger physical volume, $\sim 4$
Fermi, to suppress the effects of these modes.  As we saw in Section
\ref{sec:zero_modes_qbq} from studying $\qbq$, the effects of the
topological near-zero modes were dramatically reduced for larger
volumes.  Here we present results for the pion mass from simulating
with $16^3 \times 32$ lattices at $\beta = 5.7$ and $L_s = 48$.

\begin{figure}
\epsfxsize=\hsize
\epsfbox{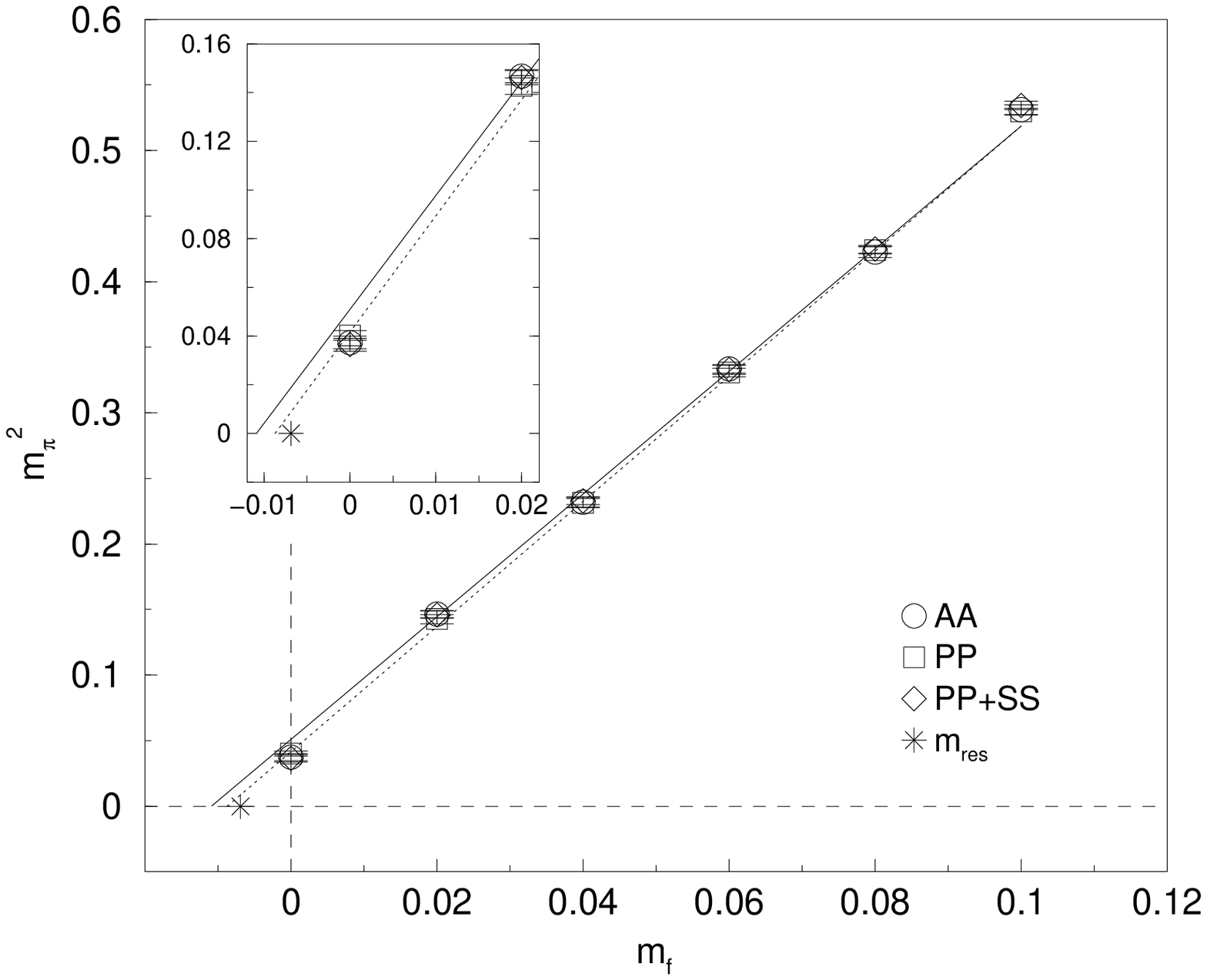}
\caption{The pion mass squared versus $m_f$ from $\corrpp$ ($\Box$),
$\corraa$ ($\circ$) and $\corrpp + \corrss$ ($\Diamond$) for quenched
simulations done on $16^3 \times 32$ lattices at $\beta = 5.7$ with
$L_s = 48$.  The star is the value of $\mres$ as measured from Eq.\
\ref{eq:mres_ratio} and its error bar in the horizontal axis is too
small to show on this scale.  The solid line is the fit to the $\corraa$
correlator for $m_f = 0.02$ to 0.08 given in
Eq.\ \ref{eq:large_vol_0.02_0.08_fit}, while the dotted line is for the
$\corrpp$ correlator for $m_f = 0.0$ to 0.08 as given in
Eq.\ \ref{eq:large_vol_pp_fit}.}
\label{fig:mpi2_vs_mf_b5_7_16nt32_ls48_mres}
\end{figure}

\begin{figure}
\epsfxsize=\hsize
\epsfbox{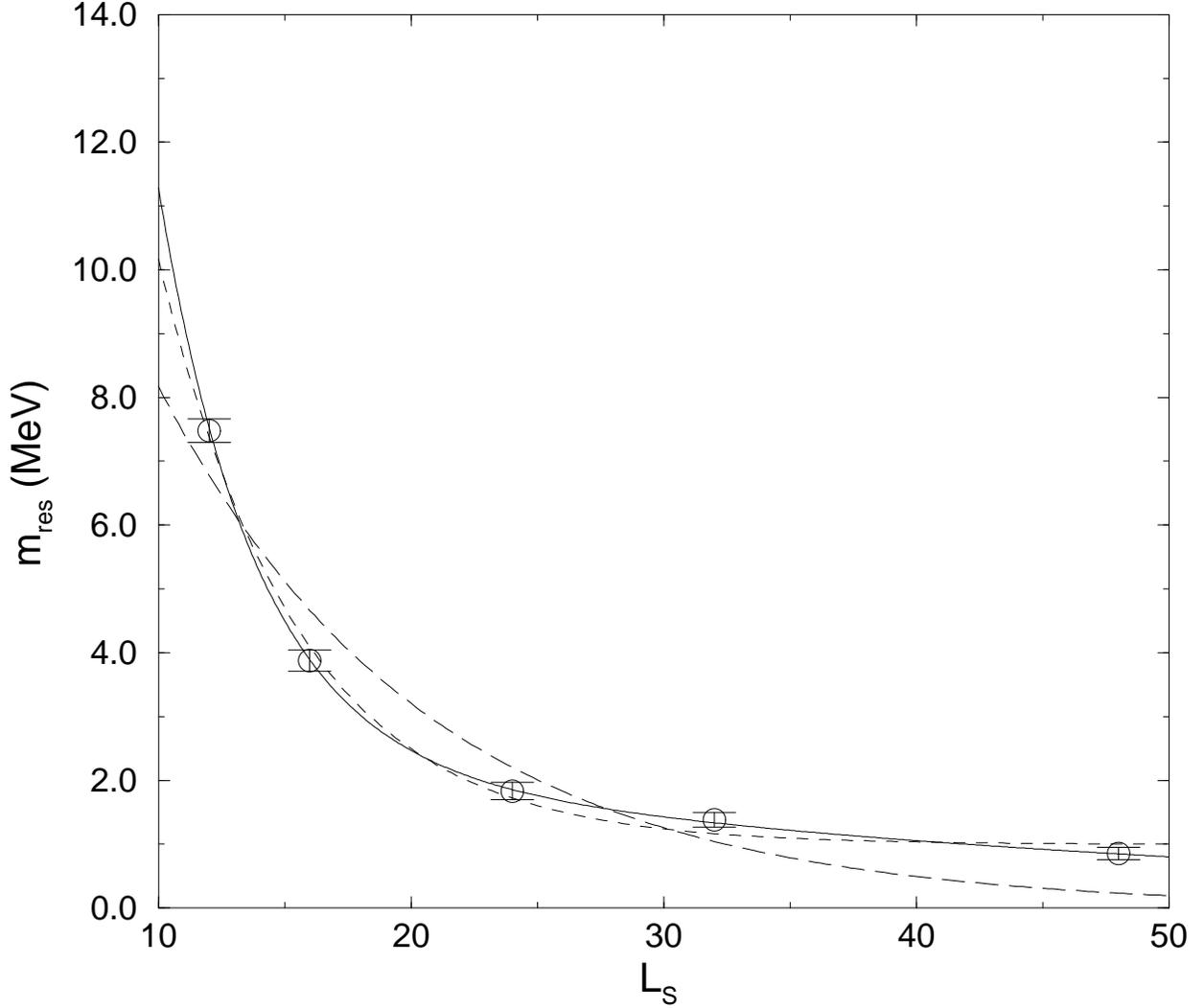}
\caption{The $L_s$ dependence of the residual mass for $16^3 \times
32$ lattices at $\beta = 6.0$.  The long-dashed line is the fit
given in Eq.\ \ref{eq:mres_exp}, the short-dashed line is the fit
from Eq.\ \ref{eq:mres_exp_const} and the solid line is the fit given
in Eq.\ \ref{eq:mres_exp_exp}.  Each of the three fits is made to
all of the $L_s$ points shown.  We have employed an intermediate 
non-perturbative renormalization to convert the plotted values of 
$\mres$ into the $\overline{\rm MS}$ scheme at $\mu = 2$ GeV.}
\label{fig:mres_vs_ls_b6_0}
\end{figure}

\begin{figure}
\epsfxsize=\hsize
\epsfbox{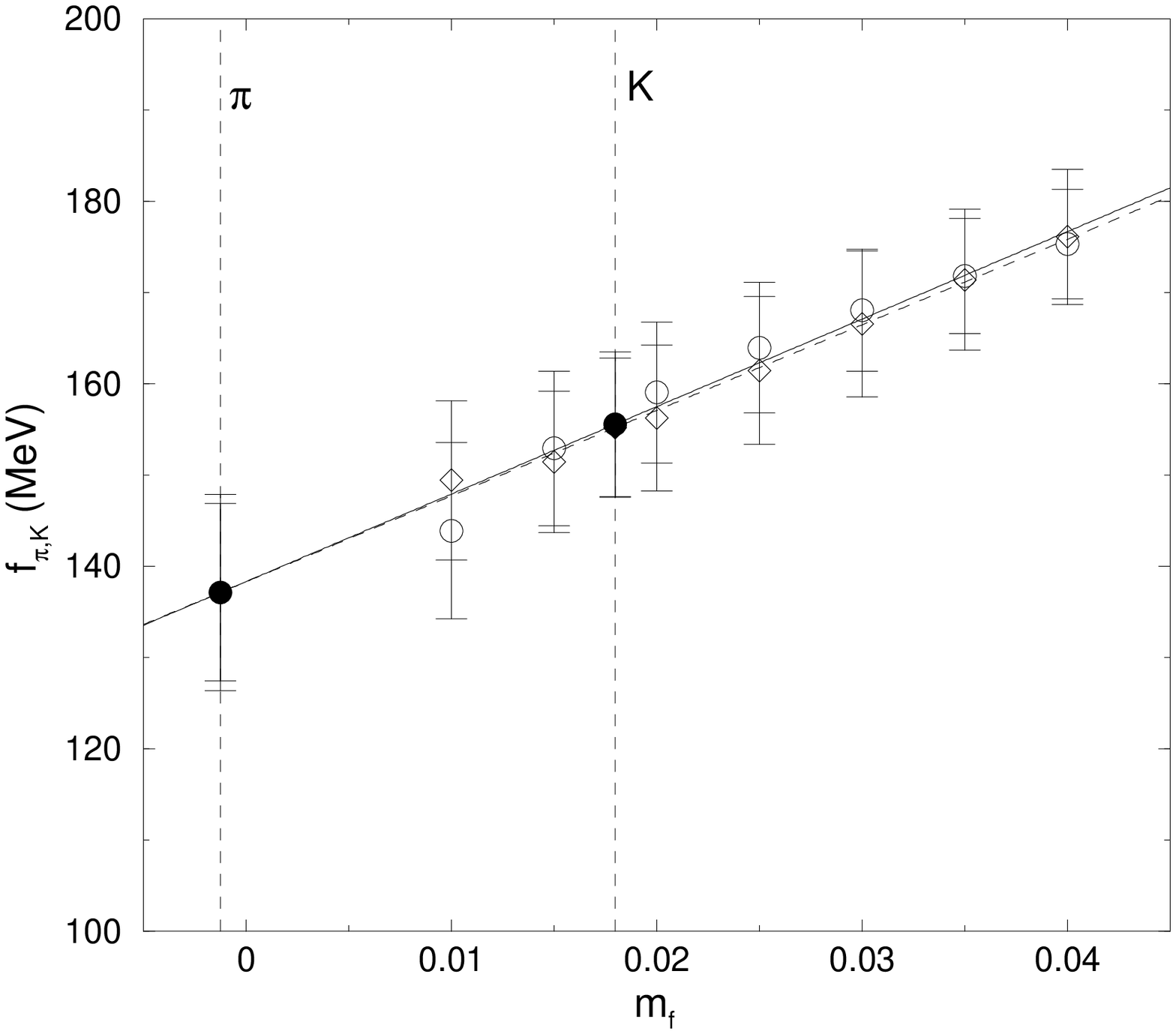}
\caption{Results for $f_\pi$ at $\beta=6.0$ with a $16^3 \times 32$
lattice and $L_s=16$ plotted as a function of $m_f$. The open circles 
are obtained from the $\corraa$ correlator, while the open 
diamonds are obtained from the $\corrpp$ correlator. We also 
show the linear fits which are used to determine our estimate for 
$f_\pi$ and $f_K$. The vertical dashed lines identify the values for 
$m_f$ which locate the chiral limit, $m_f=-\mres$ and give the physical 
ratio for $m_K/m_\rho$.  The solid symbols represent the extrapolations to 
the point $m_f=-\mres$ and interpolations to the kaon mass.}
\label{fig:fpi_b6_0_2parm}
\end{figure}

\begin{figure}
\epsfxsize=\hsize
\epsfbox{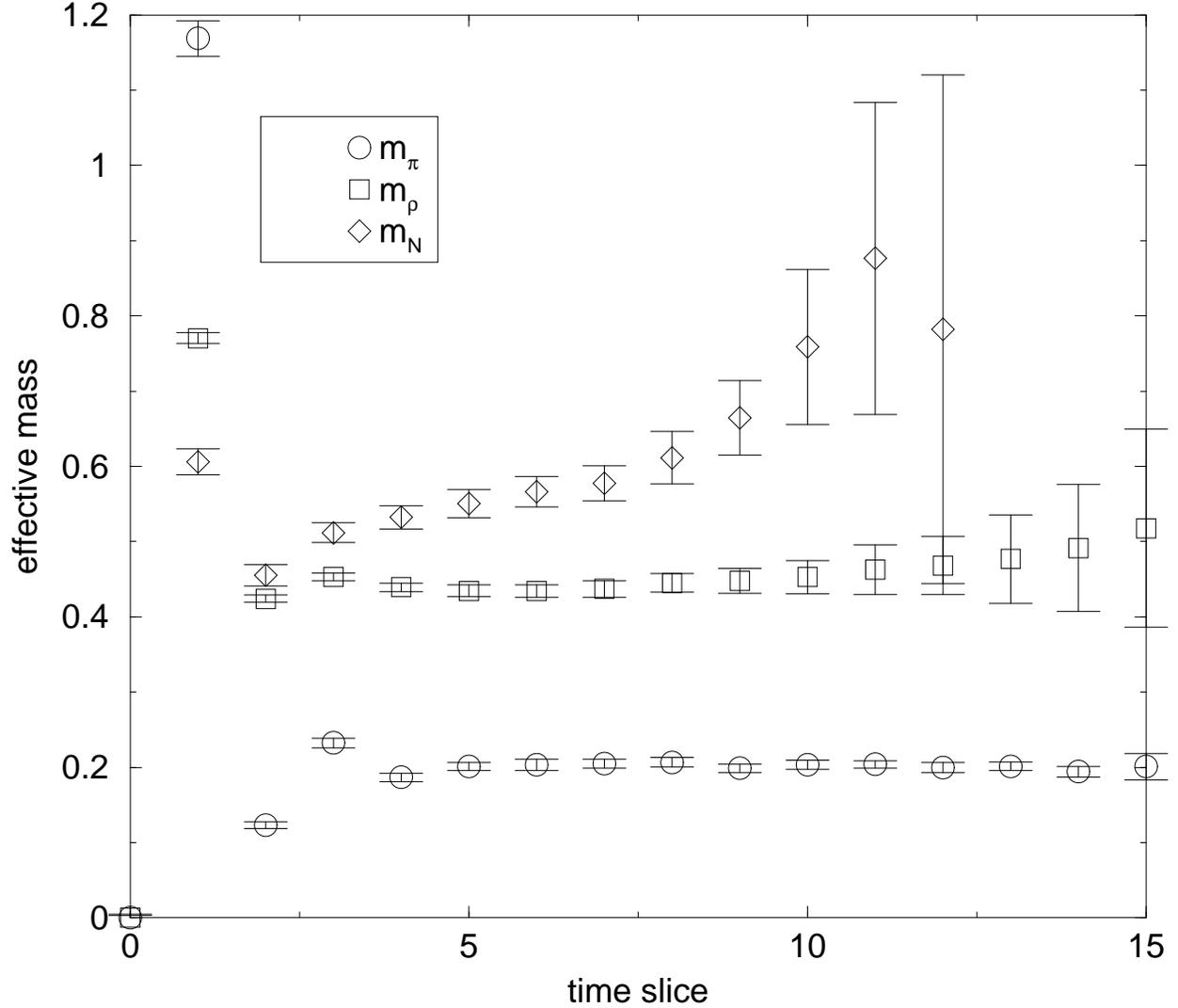}
\caption{Effective mass, $m_{\rm eff}(t)$ is plotted for the 
$16^3 \times 32$, $\beta=6.0$, $L_s=16$, $m_f=0.01$ calculation 
of the $\pi$, $\rho$ and nucleon masses.  While plateau regions for 
$t \ge 5$ are easily identified for $m_\pi$ and $m_\rho$, the nucleon 
fit is less satisfactory.   Although a plateau may be recognized for 
$5 \le t \le 8$, the rapidly growing errors make such an identification 
problematic for this case.  More satisfactory nucleon plateaus are 
seen for the larger values of $m_f$.}
\label{fig:m_eff}
\end{figure}

\begin{figure}
\epsfysize=8.0in
\epsfbox{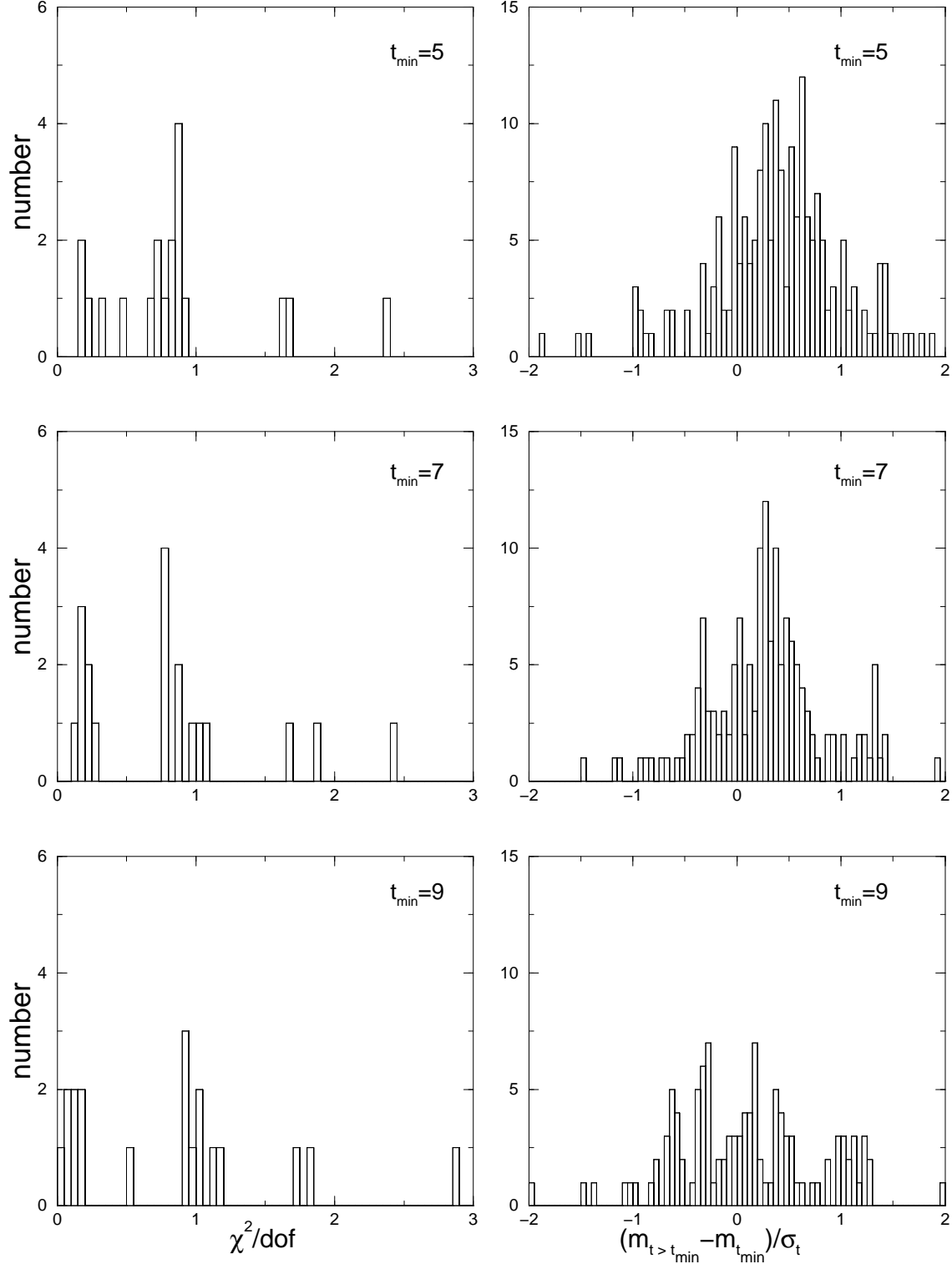}
\caption{Distribution of $\chi^2$ and the mass difference 
$(m_i(t)-m_i(t_{\rm min}))/\sigma_i$ for $t > t_{\rm min}$,
evaluated for three values of $t_{\rm min}$ for the case of 
the $\rho$ mass and a $16^3 \times 32$ lattice, with $\beta=6.0$
and $m_f \ge 0.01$ and $L_s = 12$, 16, 24, 32 and 48.  Both
distributions appear reasonable for each value of $t_{\rm min}$
with only small improvement as $t_{\rm min}$ increases from 5 to
9.  We choose to quote values of $m_\rho$ for all these cases
using the value $t_{\rm min}=7$.}
\label{fig:fit_hist}
\end{figure}

\begin{figure}
\epsfxsize=\hsize
\epsfbox{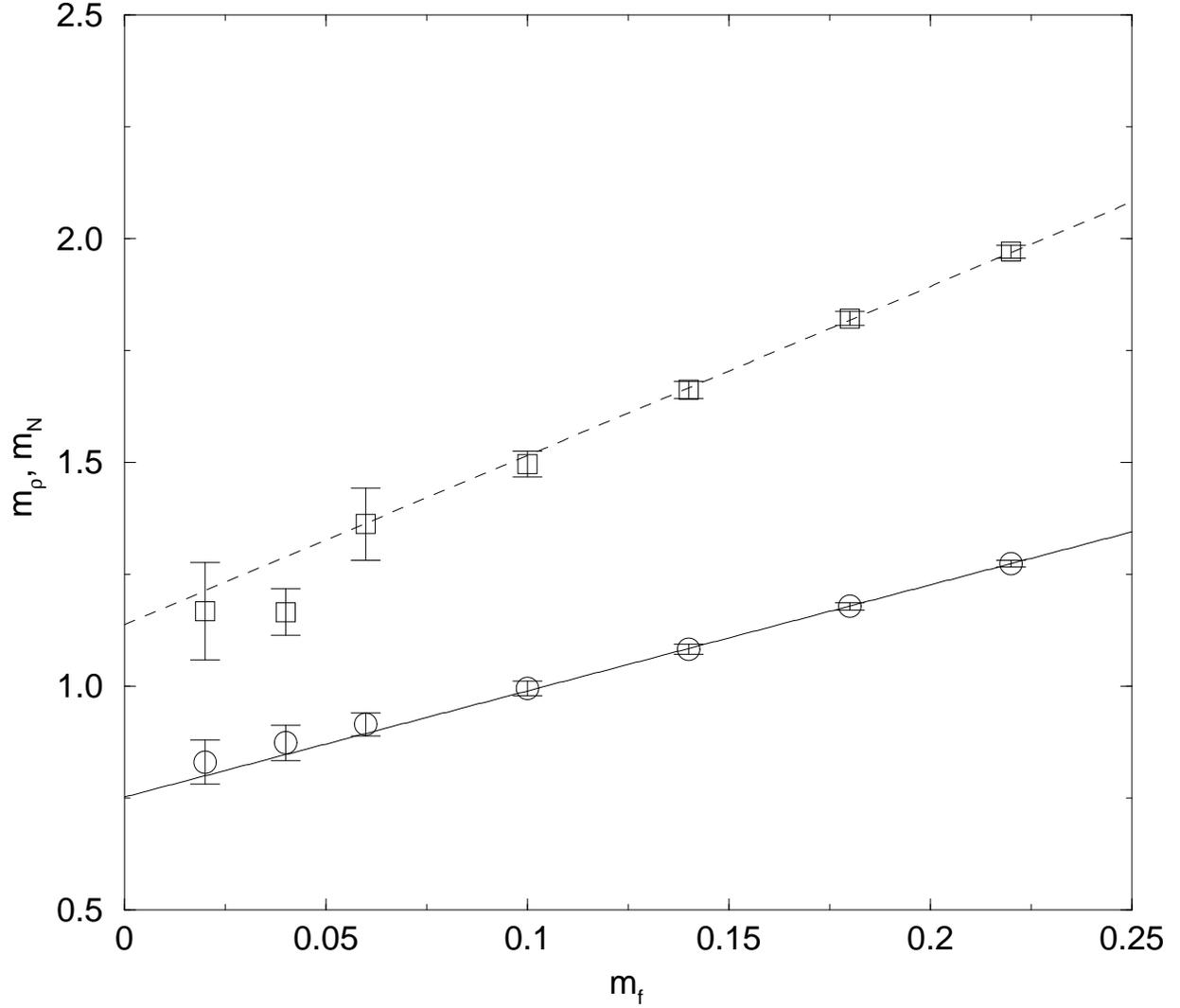}
\caption{The $\rho$ ($\circ$) and nucleon masses ($\Box$) 
plotted as a function of $m_f$ for the case of $\beta=5.7$, $M_5=1.65$, 
$L_s=32$ and a $8^3 \times 32$ lattice.  The lines represent 
least squares fits whose parameters appear in 
Table~\protect{\ref{tab:extrap_832_b5_7}} while the data plotted 
appears in Table~\protect{\ref{tab:mhad_832_b57k165ns32}}.
Note the relatively low value for the $m_f=0.04$ nucleon point results 
from the comparison of two somewhat different data sets.  As mentioned 
in the text, the linear fit was obtain from the 94 configurations 
identified in Table~\protect{\ref{tab:run_parameters_hmc}} while the 
$m_f=0.02$ and 0.04 points plotted also include the further 184 
configurations referenced in Table~\protect{\ref{tab:run_parameters_hb}}.}
\label{fig:m_vs_mf_5.7}
\end{figure}

\begin{figure}
\epsfxsize=\hsize
\epsfbox{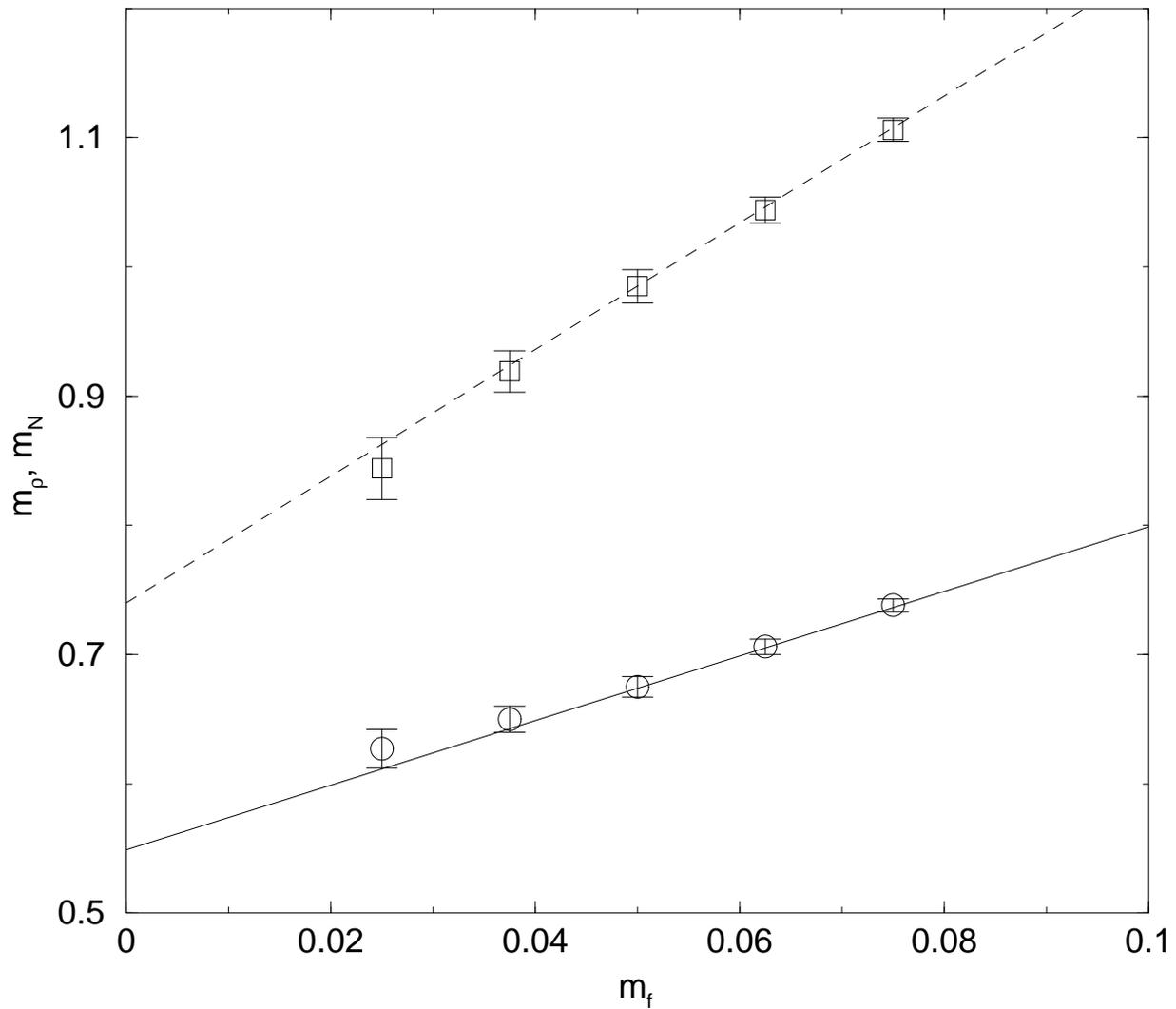}
\caption{The $\rho$ ($\circ$) and nucleon masses ($\Box$) 
plotted as a function of $m_f$ for the case of $\beta=5.85$, $M_5=1.9$, 
$L_s=32$ and a $8^3 \times 32$ lattice.  The lines represent 
least squares fits whose parameters appear in 
Table~\protect{\ref{tab:extrap_1232_b5_85}} while the data 
plotted appears in Table~\protect{\ref{tab:mhad_1232_b585k19ns20}}.}
\label{fig:m_vs_mf_5.85}
\end{figure}

\begin{figure}
\epsfxsize=\hsize
\epsfbox{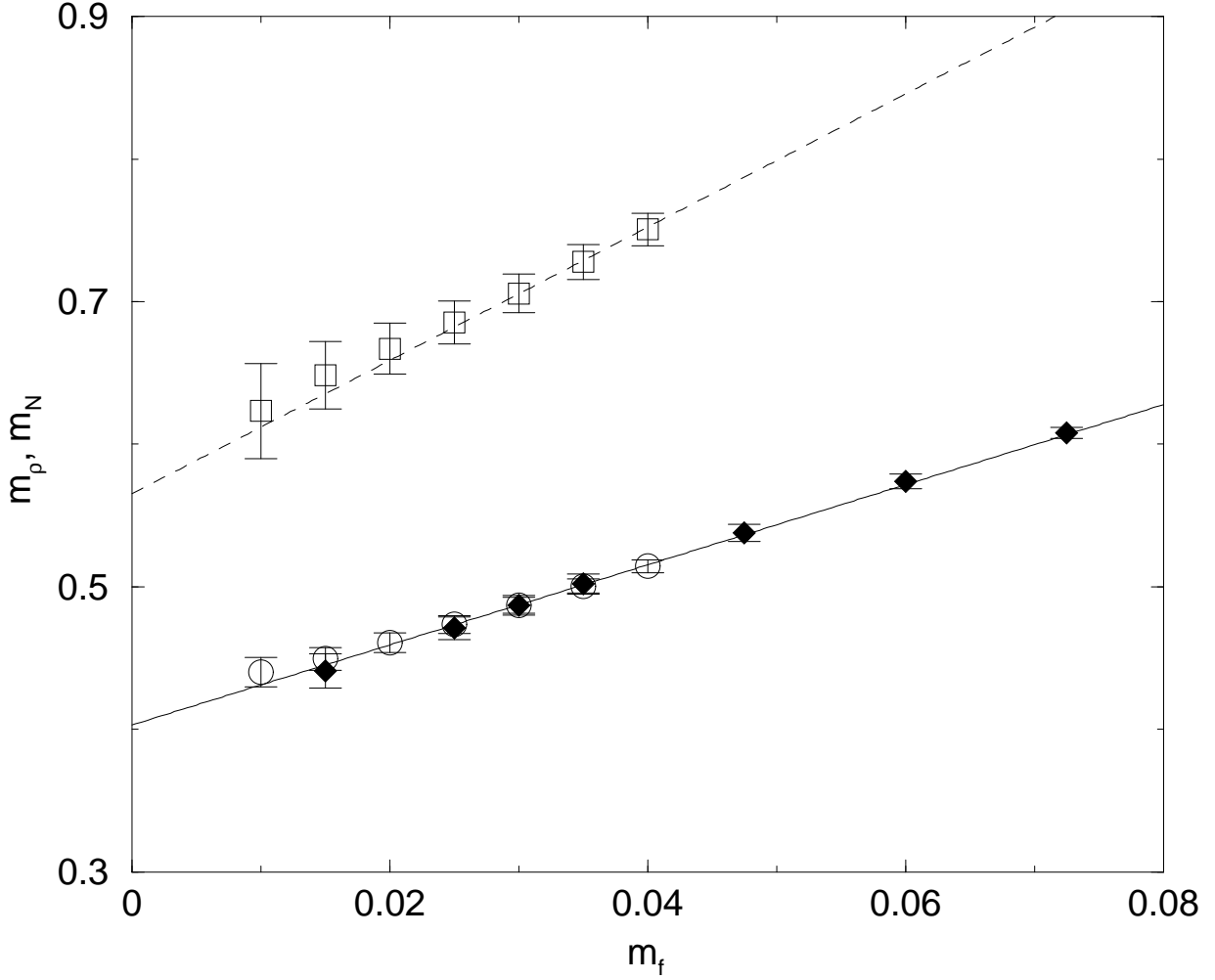}
\caption{The $\rho$ ($\circ$) and nucleon masses ($\Box$)
plotted as a function of $m_f$ for the case of $\beta=6.0$, $M_5=1.8$, 
$L_s=16$ and a $16^3 \times 32$ lattice.  The lines represent 
least squares fits using the parameters appearing in 
Table~\protect{\ref{tab:extrap_1632_b6_0}} while the data plotted 
appears in Table~\protect{\ref{tab:mhad_1632_b60k18ns16}}.  In addition
to these hadron masses computed for the case of equal mass quarks, we
have also plotted the $\rho$ mass for the case of non-degenerate
quarks given in Table~\protect{\ref{tab:mnd_1632_b60k18ns16}}
as a function of the average quark mass, $(m_f(1)+m_f(2))/2$.
These points are plotted as filled diamonds.}
\label{fig:m_vs_mf_6.0}
\end{figure}

\begin{figure}
\epsfxsize=\hsize
\epsfbox{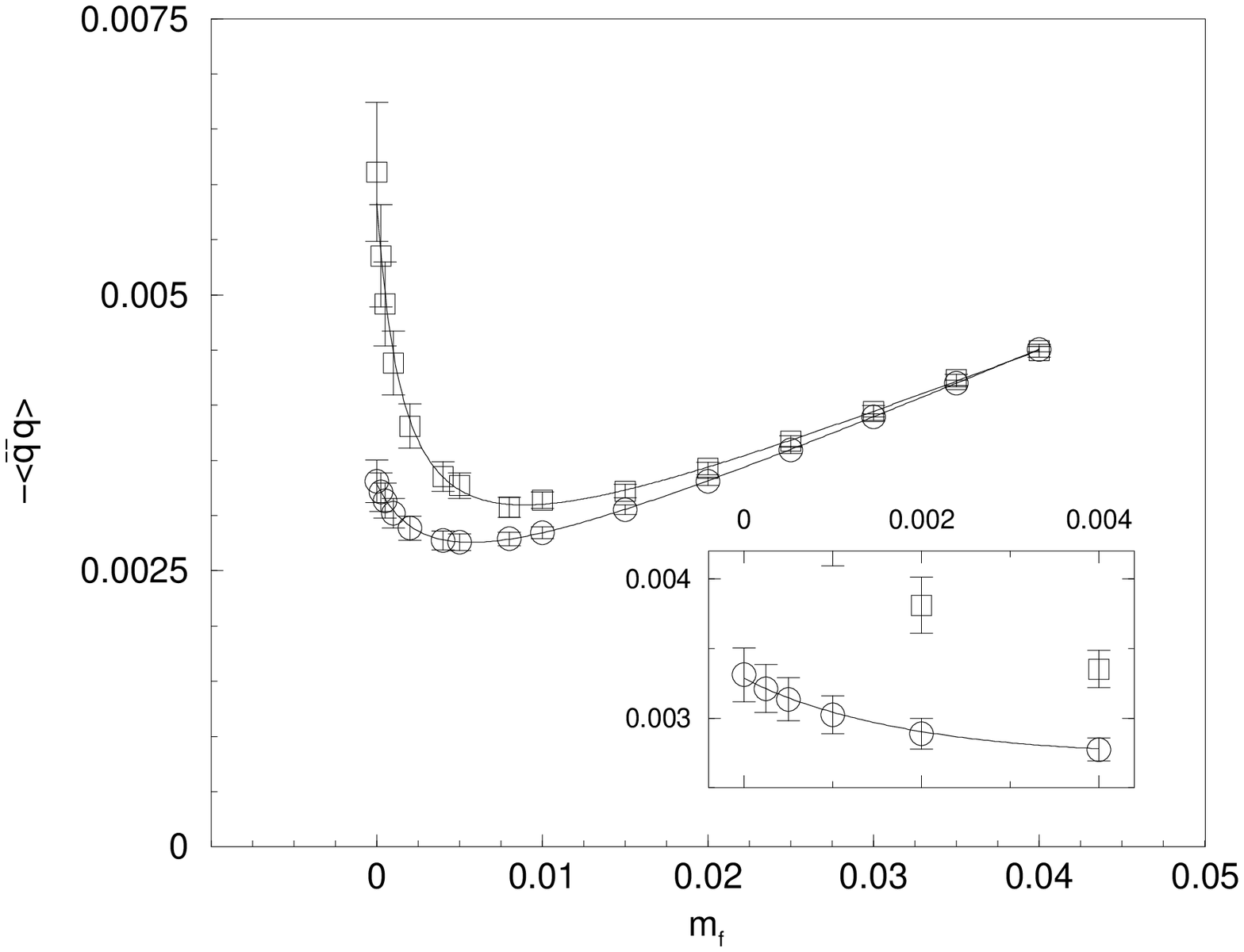}
\caption{$\qbq$ for quenched simulations done on $8^3 \times 32$
lattices at $\beta = 5.7$ for $L_s = 32$ ($\circ$) and $L_s = 48$
($\Box$).  The more pronounced rise as $m_f \rightarrow 0$ for $L_s =
48$ shows that the expected topological near-zero modes have smaller
values for $\lambda_i$ and/or $\dmi$ for this larger $L_s$.}
\label{fig:qbq_b5_7_ls32_48}
\end{figure}

\begin{figure}
\epsfxsize=\hsize
\epsfbox{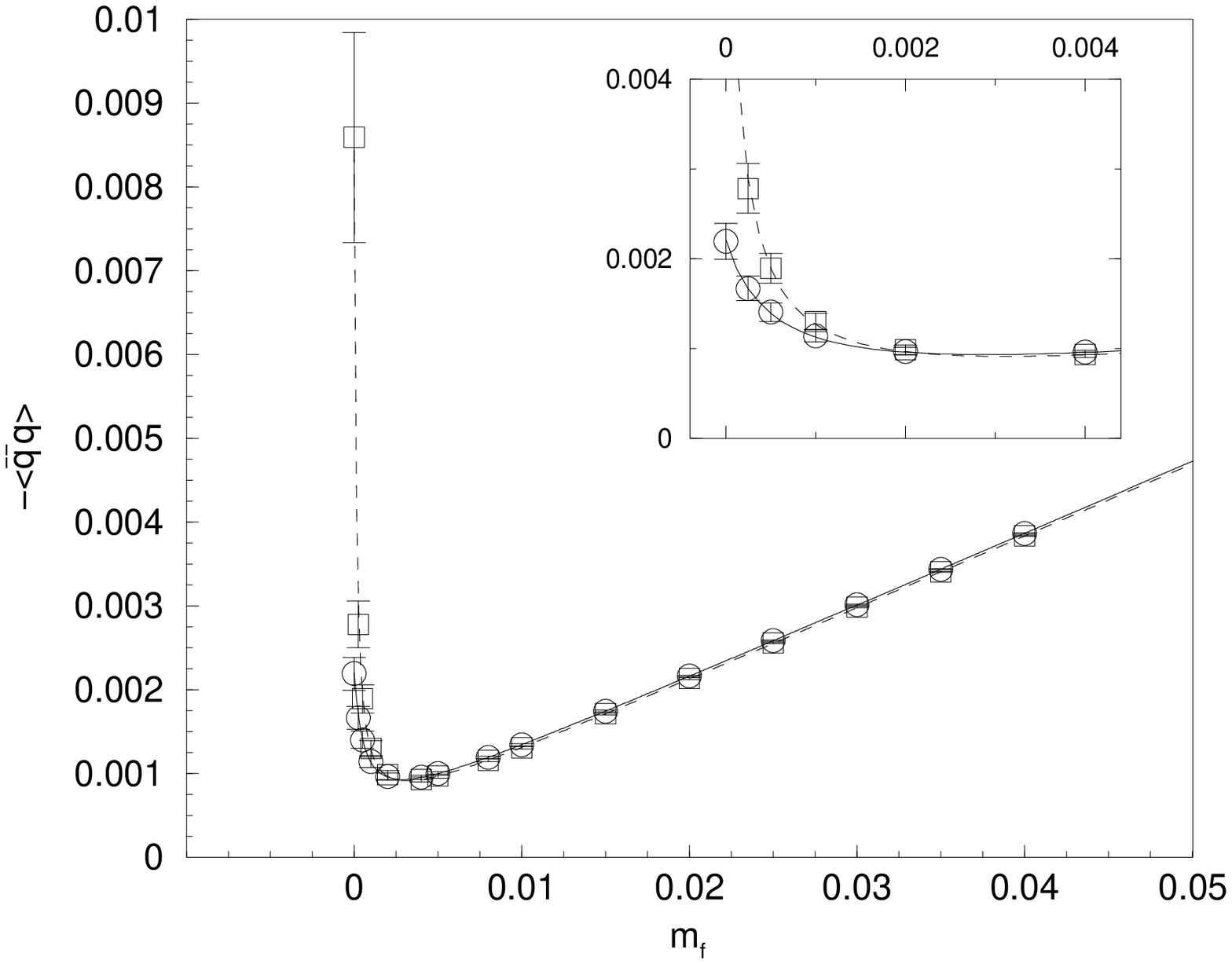}
\caption{$\qbq$ for quenched simulations done on $16^3 \times 32$
lattices at $\beta = 6.0$ for $L_s = 16$ ($\circ$) and $L_s = 24$
($\Box$).  The more pronounced rise as $m_f \rightarrow 0$ for
$L_s = 24$ shows that the expected topological near-zero modes have
smaller values for $\lambda_i$ and/or $\dmi$ for this larger $L_s$.}
\label{fig:qbq_b6_0_ls16_24}
\end{figure}

\begin{figure}
\epsfxsize=\hsize
\epsfbox{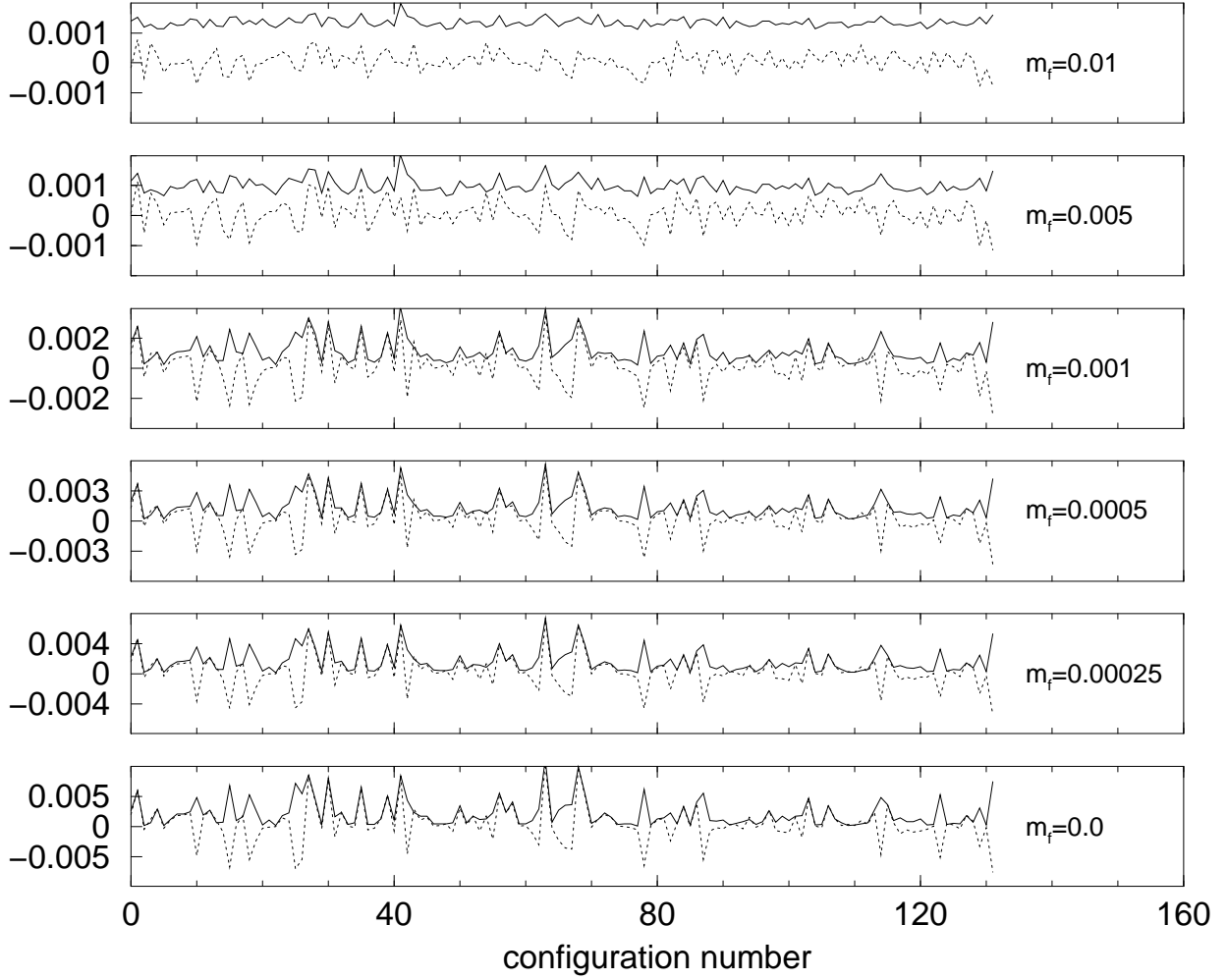}
\caption{Evolutions of $-\qbq$ (solid line) and $-\langle \overline{q}
\gamma_5 q \rangle$ (dotted line) for $16^3 \times 32$ lattices at
$\beta = 6.0$ with $L_s = 16$.  For smaller values of $m_f$ the
evolutions show pronounced fluctuations which have opposite sign
for $\qbq$ and $\langle \overline{q} \gamma_5 q \rangle$, indicating
the presence of eigenfunctions of $D_H$ at this $L_s$ which are very
good approximations to the exact topological zero modes expected
at $L_s \rightarrow \infty$.  Note that the vertical scale
{\em increases} for the smaller values of $m_f$.}
\label{fig:qbq_qbg5q_evol_b6_0}
\end{figure}

\begin{figure}
\epsfxsize=\hsize
\epsfbox{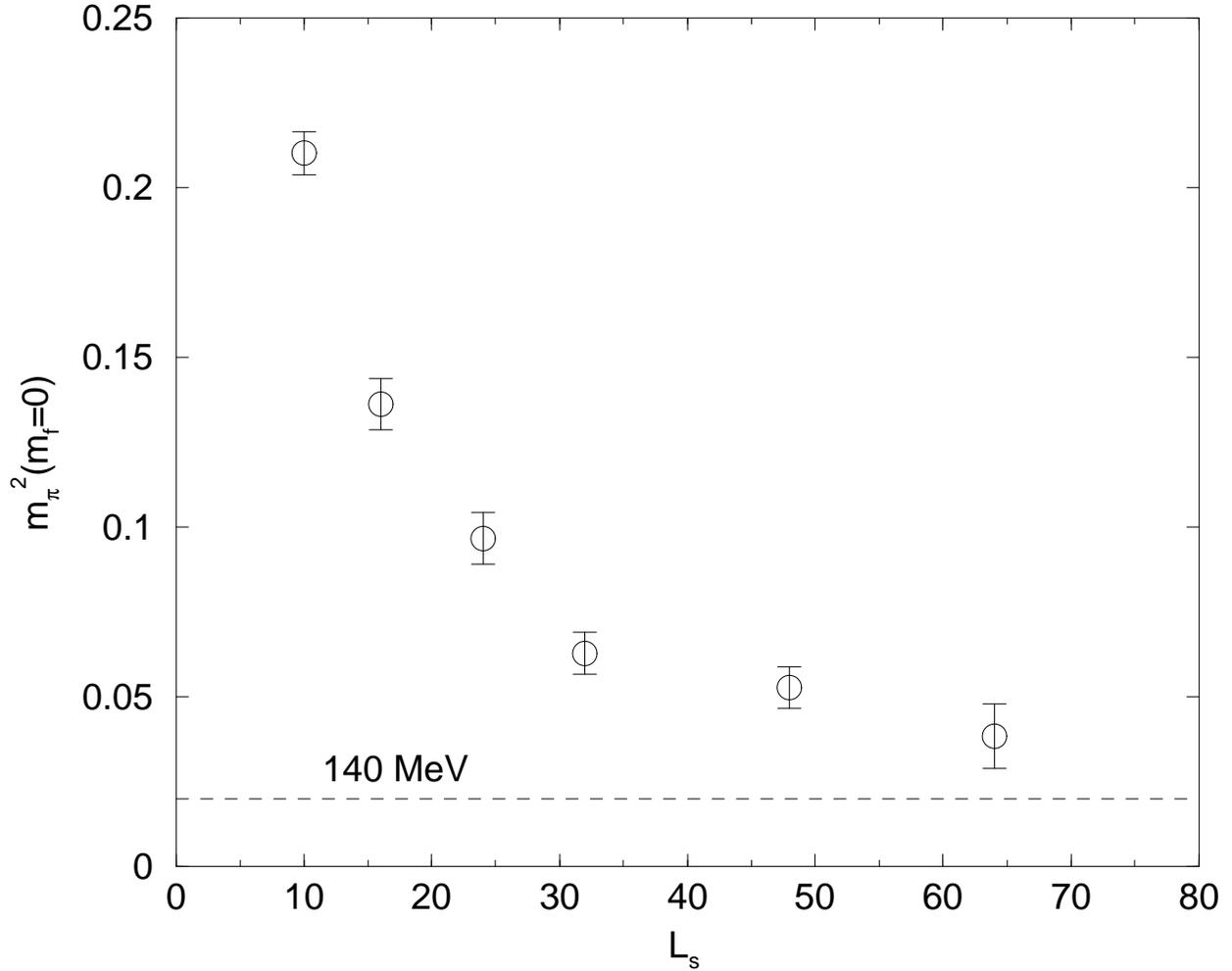}
\caption{The quantity $\mpi2(m_f=0)$ from $\corrpp$ for quenched 
simulations done on $8^3 \times 32$ lattices at $\beta = 5.7$ 
versus $L_s$.  This graph is an updated version of an earlier result 
based on part of the data presented here.  
While a slow decrease in $\mpi2(m_f=0)$ as $L_s$ increases from 32
to 64 is now visible, the effect is much less dramatic than the 
drop seen in the more accurate values of $\mres$, which decrease
from 0.0105(2) to 0.0071(4) as $L_s$ increases from 32 to 48.
This contrast presumably results from the effects of both zero-modes
and non-linearity for $m_\pi^2(m_f)$ in the quenched approximation.}
\label{fig:mpi2_vs_ls_b5.7}
\end{figure}

\begin{figure}
\epsfxsize=\hsize
\epsfbox{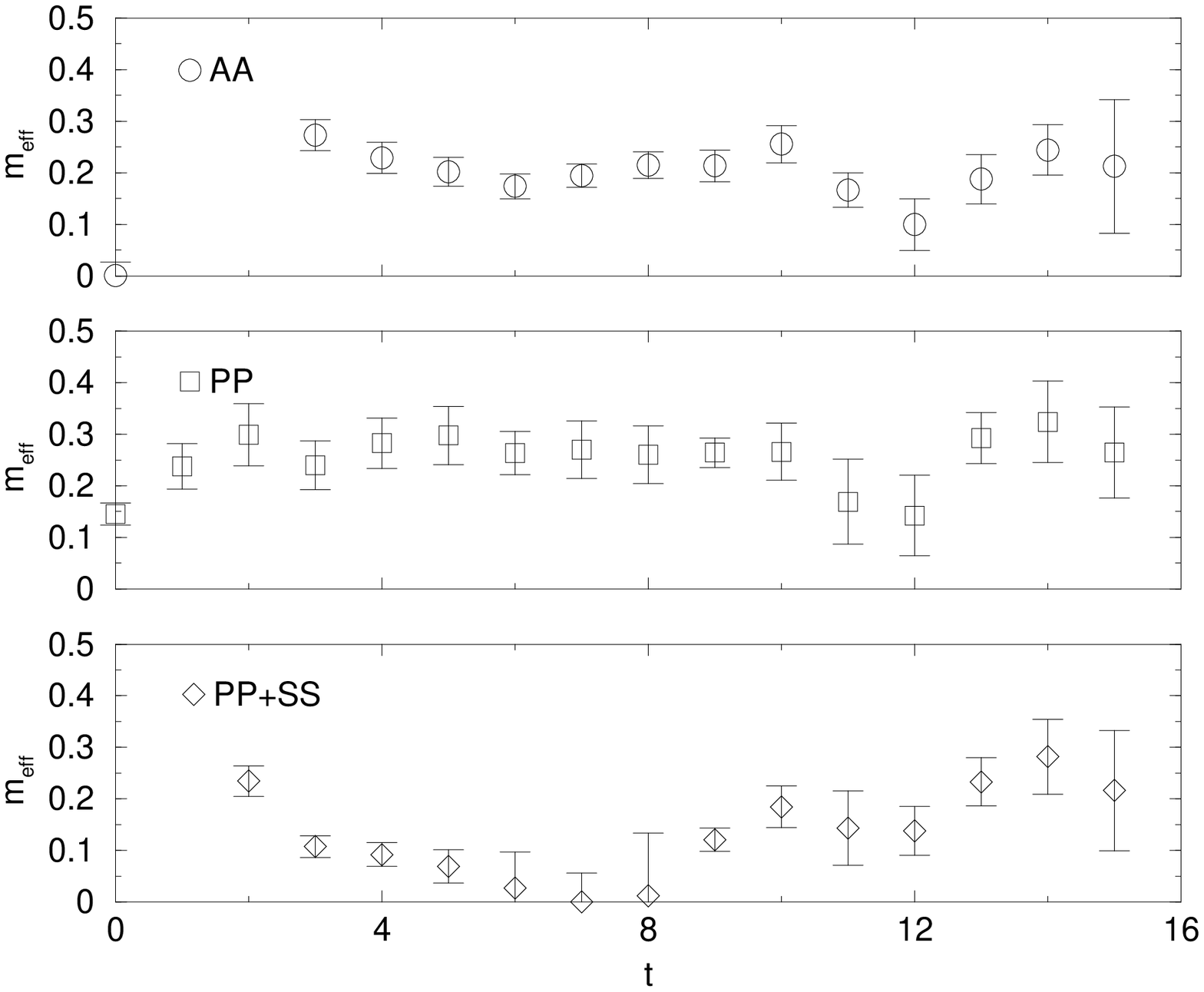}
\caption{The pion effective mass as a function of the source-sink
separation, $t$, for $8^3 \times 32$ lattices at $\beta = 5.7$ with
$L_s = 48$ and $m_f = 0.0$.  The upper panel is from $\corraa$
($\circ$), the middle from $\corrpp$ ($\Box$) and the lower
from $\corrpp+\corrss$ ($\Diamond$).  The $m_f = 0.0$ points in Figure
\ref{fig:mpi2_vs_mf_b5_7_8nt32_ls48} come from fitting from $t = 7$
to $t=16$.}
\label{fig:mpi_effm_b5_7_8nt32_ls48_m0.0}
\end{figure}

\begin{figure}
\epsfxsize=\hsize
\epsfbox{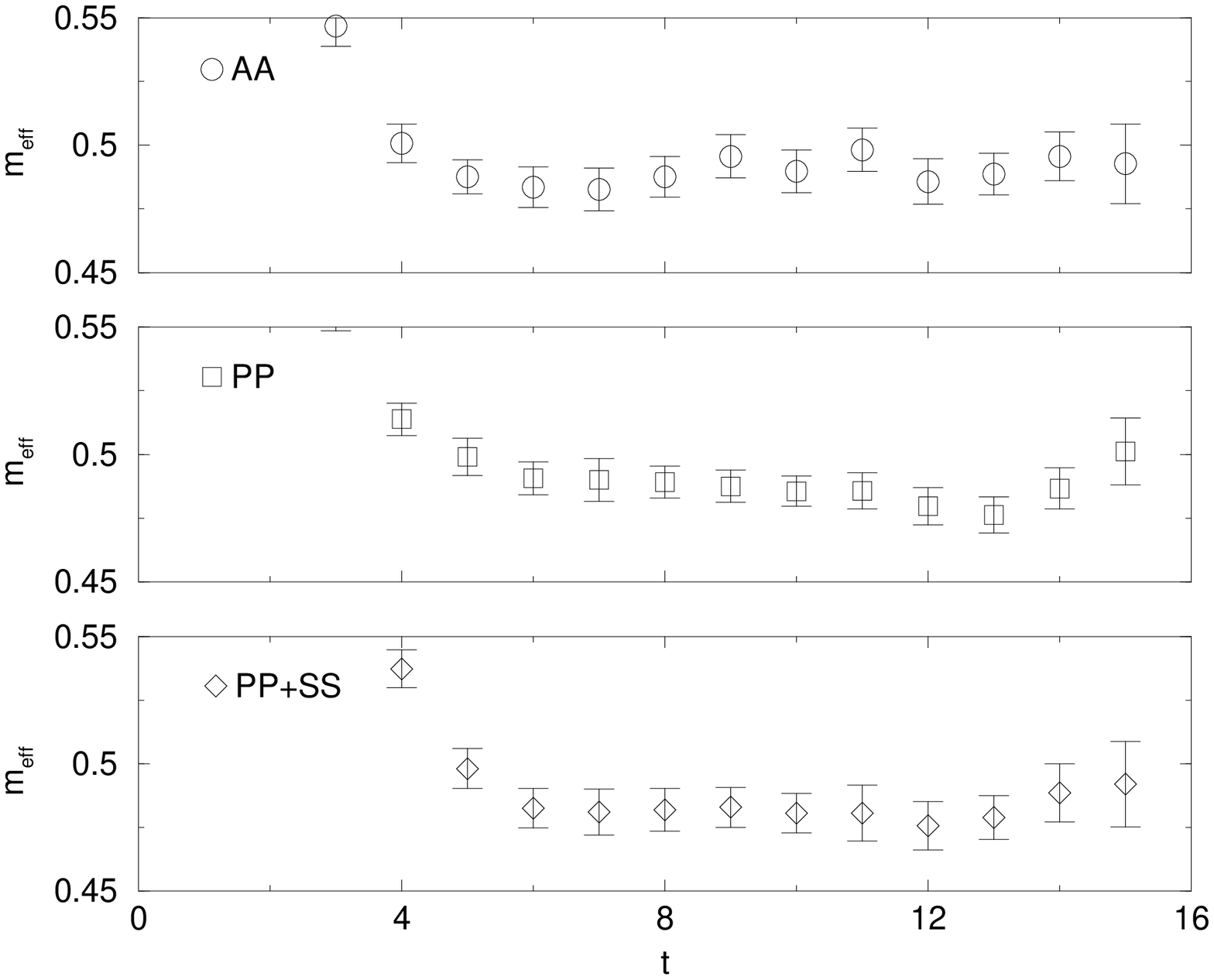}
\caption{The pion effective mass as a function of the source-sink
separation, $t$, for $8^3 \times 32$ lattices at $\beta = 5.7$ with
$L_s = 48$ and $m_f = 0.04$.  The upper panel is from $\corraa$
($\circ$), the middle from $\corrpp$ ($\Box$) and the lower from
$\corrpp+\corrss$ ($\Diamond$).  The effective masses from the
different correlators are quite consistent.}
\label{fig:mpi_effm_b5_7_8nt32_ls48_m0.04}
\end{figure}

\begin{figure}
\epsfxsize=\hsize
\epsfbox{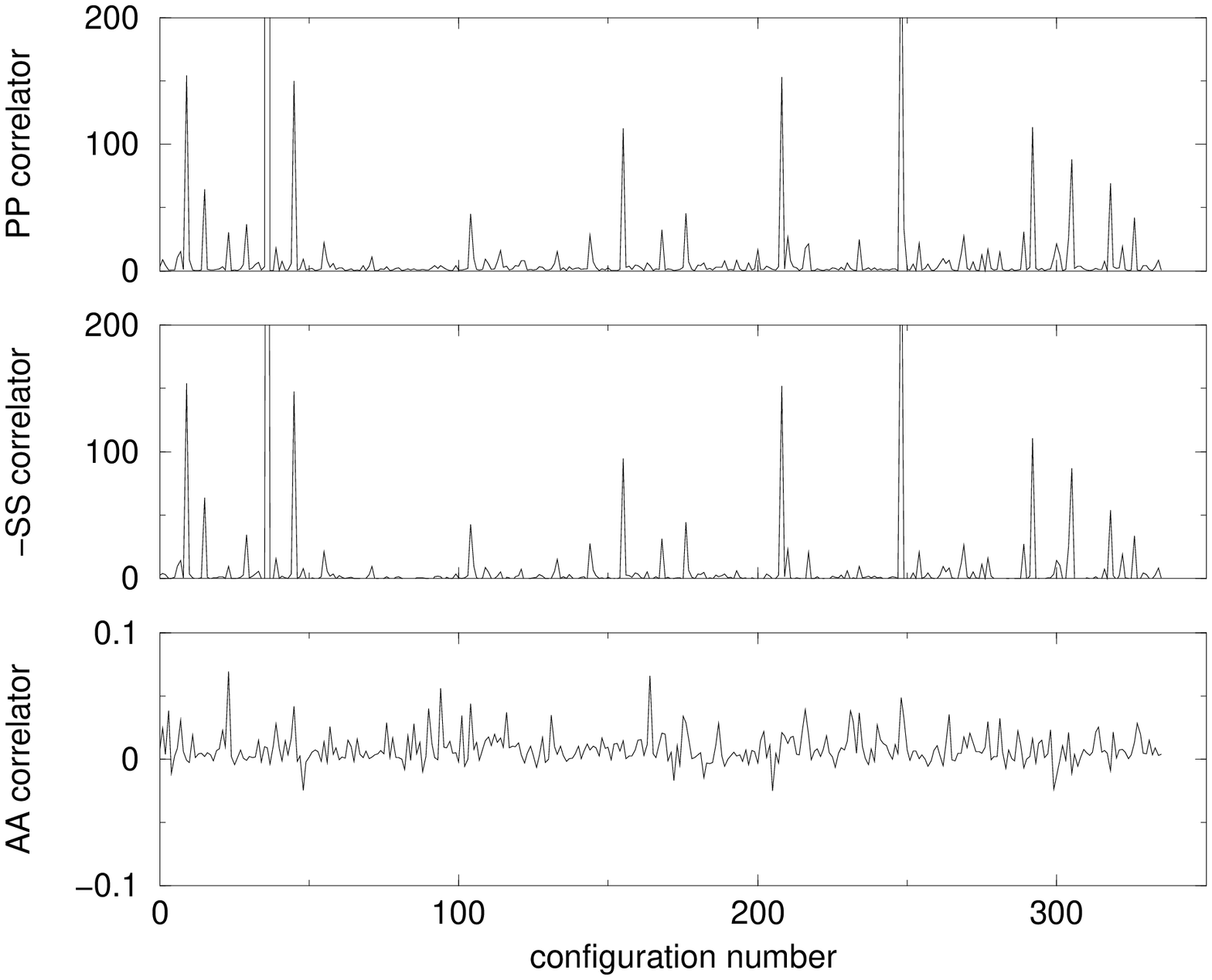}
\caption{The evolution of point source correlators at $t=8$ for $8^3
\times 32$ lattices at $\beta = 5.7$ with $L_s = 48$ and $m_f = 0.0$
The upper panel is $\corrpp$, the middle $-\corrss$ and the lower
$\corraa$.  The large fluctuations that are common to $\corrpp$ and
$-\corrss$ are due to zero modes and show that they dominate the
ensemble average for the correlator at this $t$.}
\label{fig:corr_evol_b5_7_8nt32_ls48_m0_0}
\end{figure}

\begin{figure}
\epsfxsize=\hsize
\epsfbox{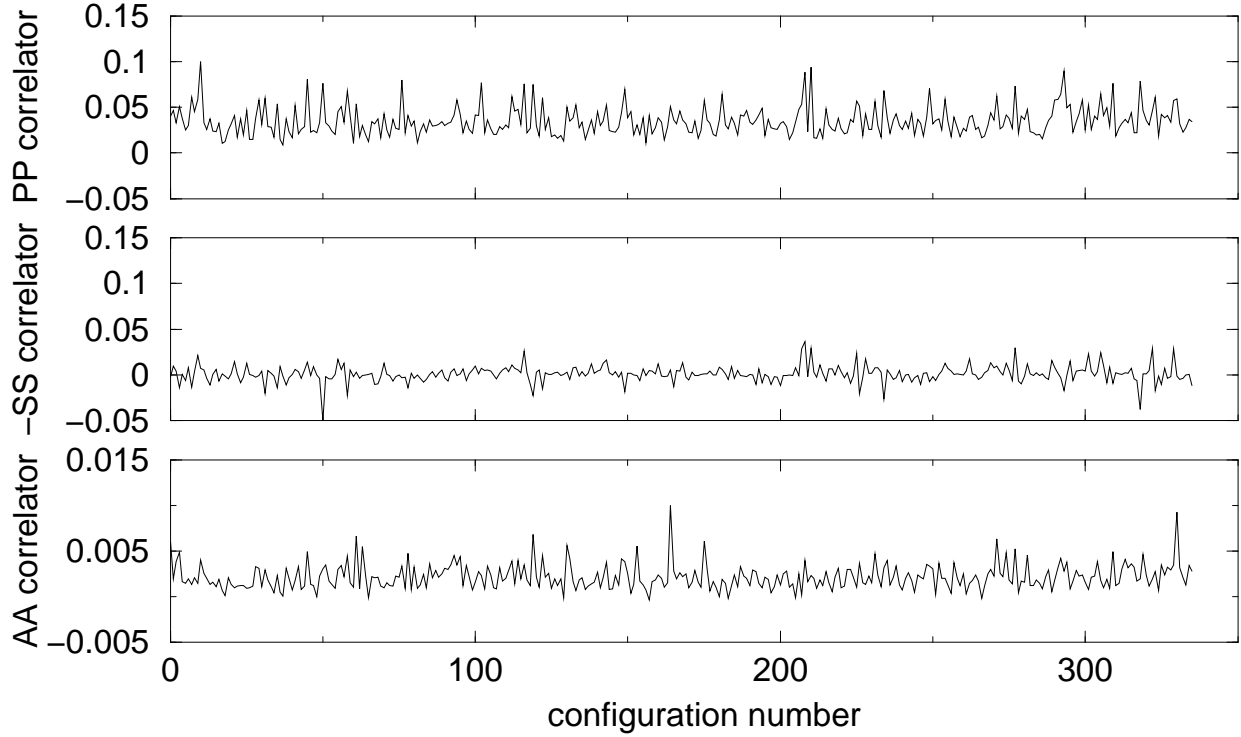}
\caption{The evolution of point source correlators at $t=8$ 
for $8^3 \times 32$ lattices at $\beta = 5.7$ with
$L_s = 48$ and $m_f = 0.04$  The upper panel is $\corrpp$, the
middle $-\corrss$ and the lower $\corraa$.
Zero mode effects seem entirely absent from these evolution plots.}
\label{fig:corr_evol_b5_7_8nt32_ls48_m0_04}
\end{figure}

\begin{figure}
\epsfxsize=\hsize
\epsfbox{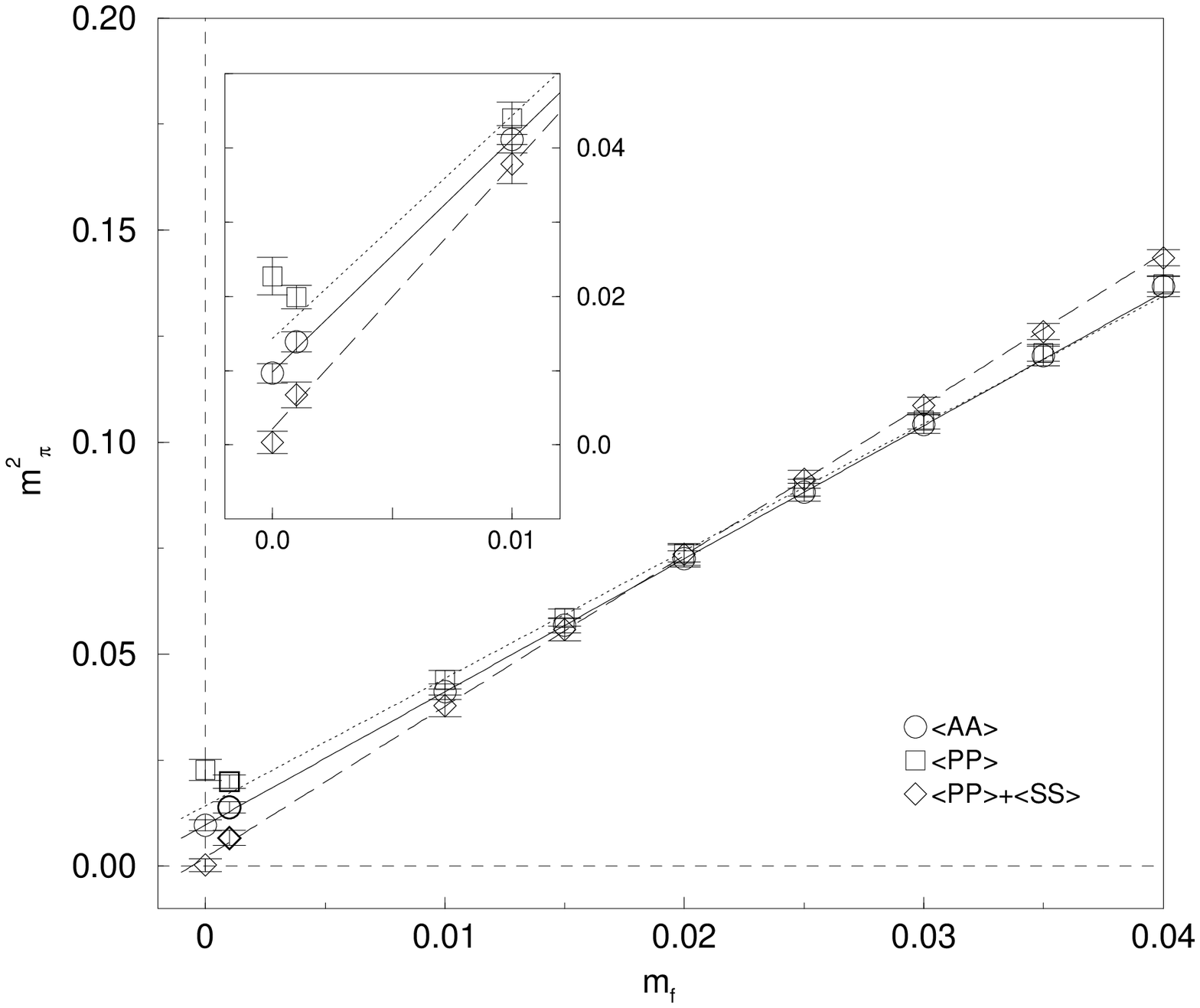}
\caption{The pion mass squared versus $m_f$ from $\corrpp$ ($\Box$),
$\corraa$ ($\circ$) and $\corrpp + \corrss$ ($\Diamond$) for
quenched simulations done on $16^3 \times 32$ lattices at $\beta = 6.0$
with $L_s = 16$.  For $m_f = 0.0$ and 0.001, the correlators all give
different masses due to the differing topological near-zero mode
contributions for each one.  For larger $m_f$, the pion mass
determination from $\corrpp+\corrss$ is likely contaminated by the
heavy mass states in the $\corrss$.  The dotted line is the fit of Eq.\
\ref{eq:b6_0_16nt32_0.02_0.1_pp_fit}, the solid line is from Eq.\
\ref{eq:b6_0_16nt32_0.02_0.1_aa_fit} and the dashed line is from Eq.\
\ref{eq:b6_0_16nt32_0.02_0.1_pp+ss_fit}.}
\label{fig:mpi2_vs_mf_b6_0_16nt32_ls16}
\end{figure}

\begin{figure}
\epsfxsize=\hsize
\epsfbox{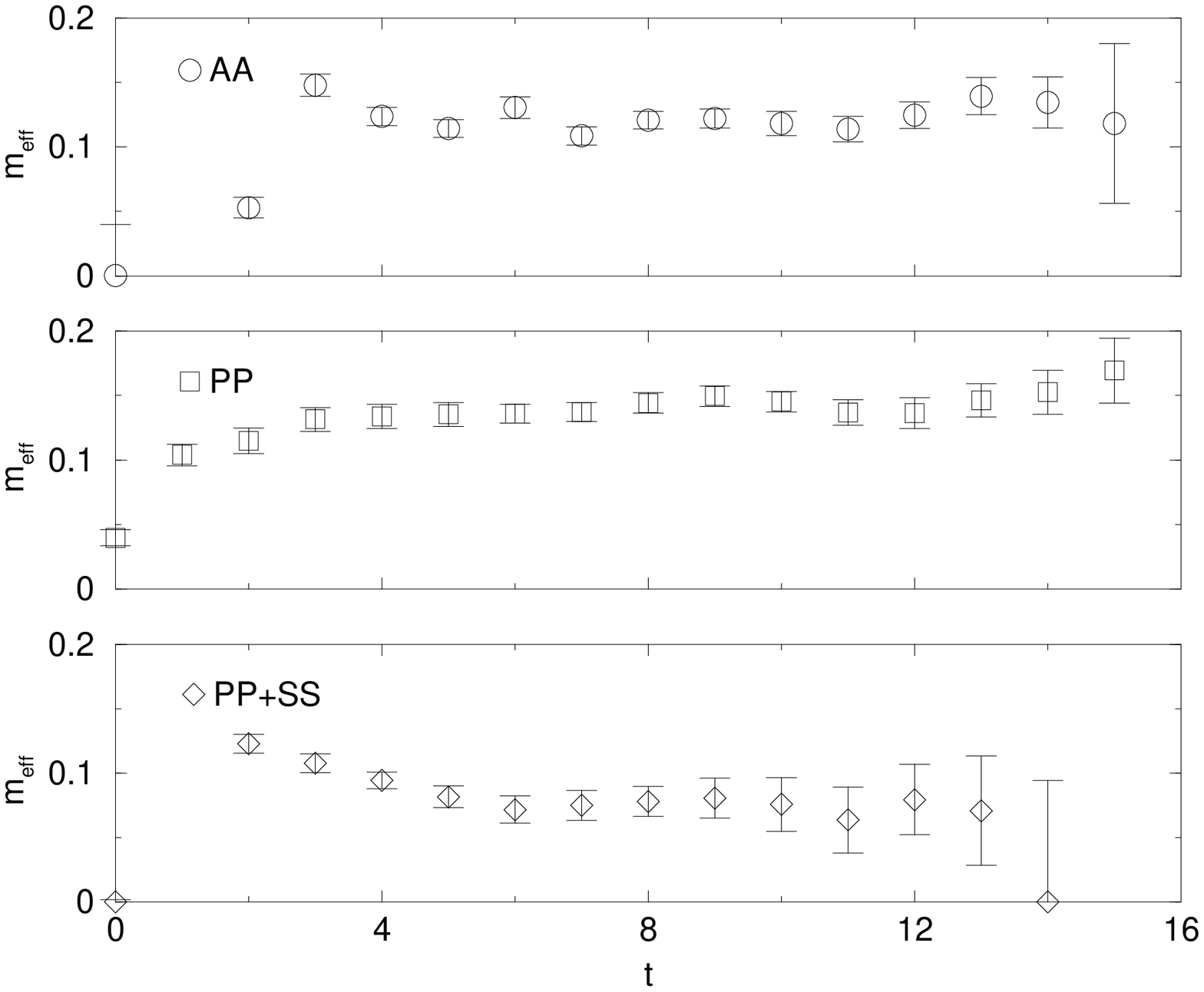}
\caption{The pion effective mass as a function of the source-sink
separation, $t$, for $16^3 \times 32$ lattices at $\beta = 6.0$ with
$L_s = 16$ and $m_f = 0.001$.  The upper panel is from $\corraa$
($\circ)$, the middle from $\corrpp$ ($\Box$) and the lower from
$\corrpp+\corrss$ ($\Diamond$).}
\label{fig:mpi_effm_b6_0_16nt32_ls16_m0.001}
\end{figure}

\begin{figure}
\epsfxsize=\hsize
\epsfbox{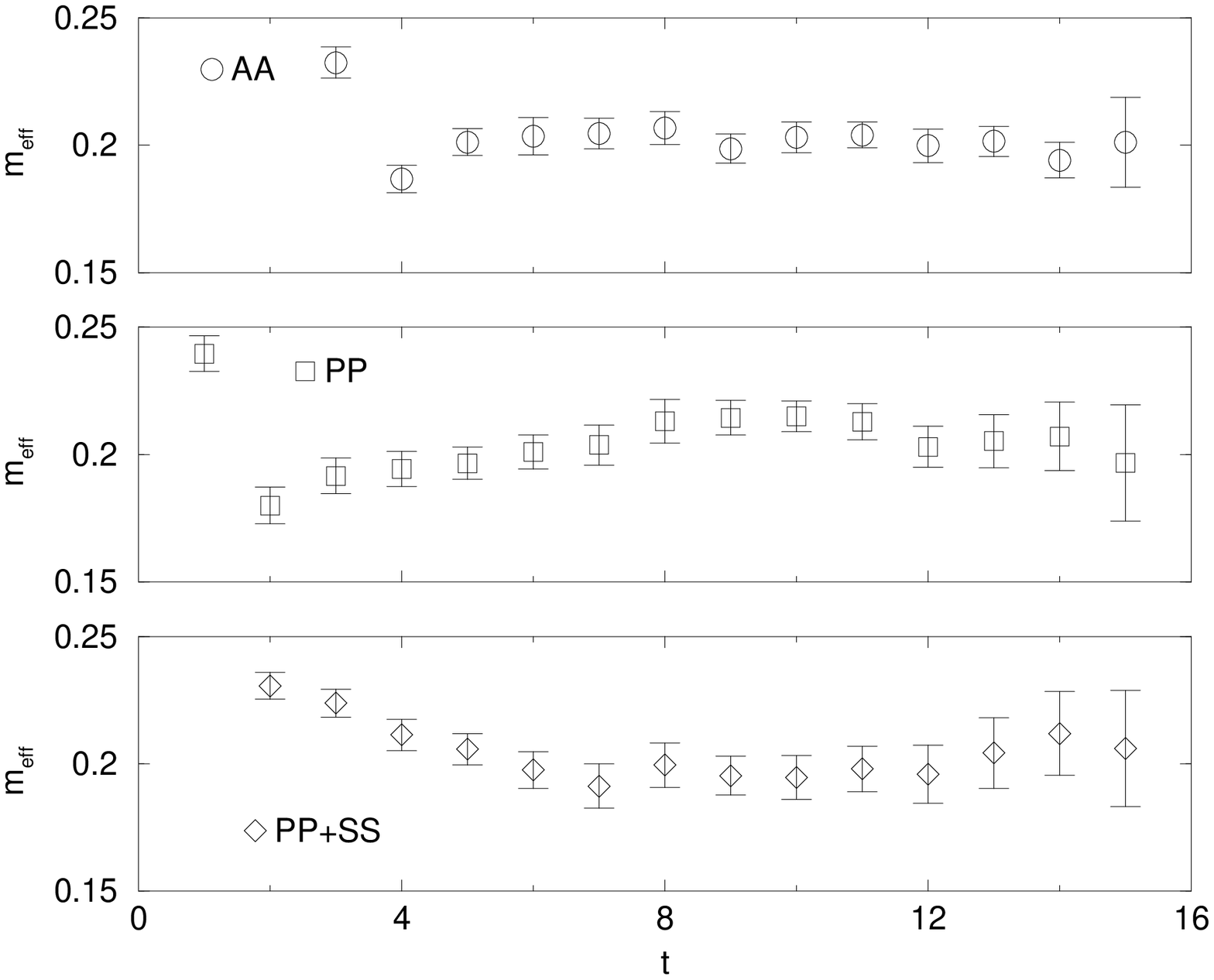}
\caption{The pion effective mass as a function of the source-sink
separation, $t$, for $16^3 \times 32$ lattices at $\beta = 6.0$ with
$L_s = 16$ and $m_f = 0.01$.  The upper panel is from $\corraa$
($\circ$), the middle from $\corrpp$ ($\Box$) and the lower from
$\corrpp+\corrss$ ($\Diamond$).}
\label{fig:mpi_effm_b6_0_16nt32_ls16_m0.01}
\end{figure}

\begin{figure}
\epsfxsize=\hsize
\epsfbox{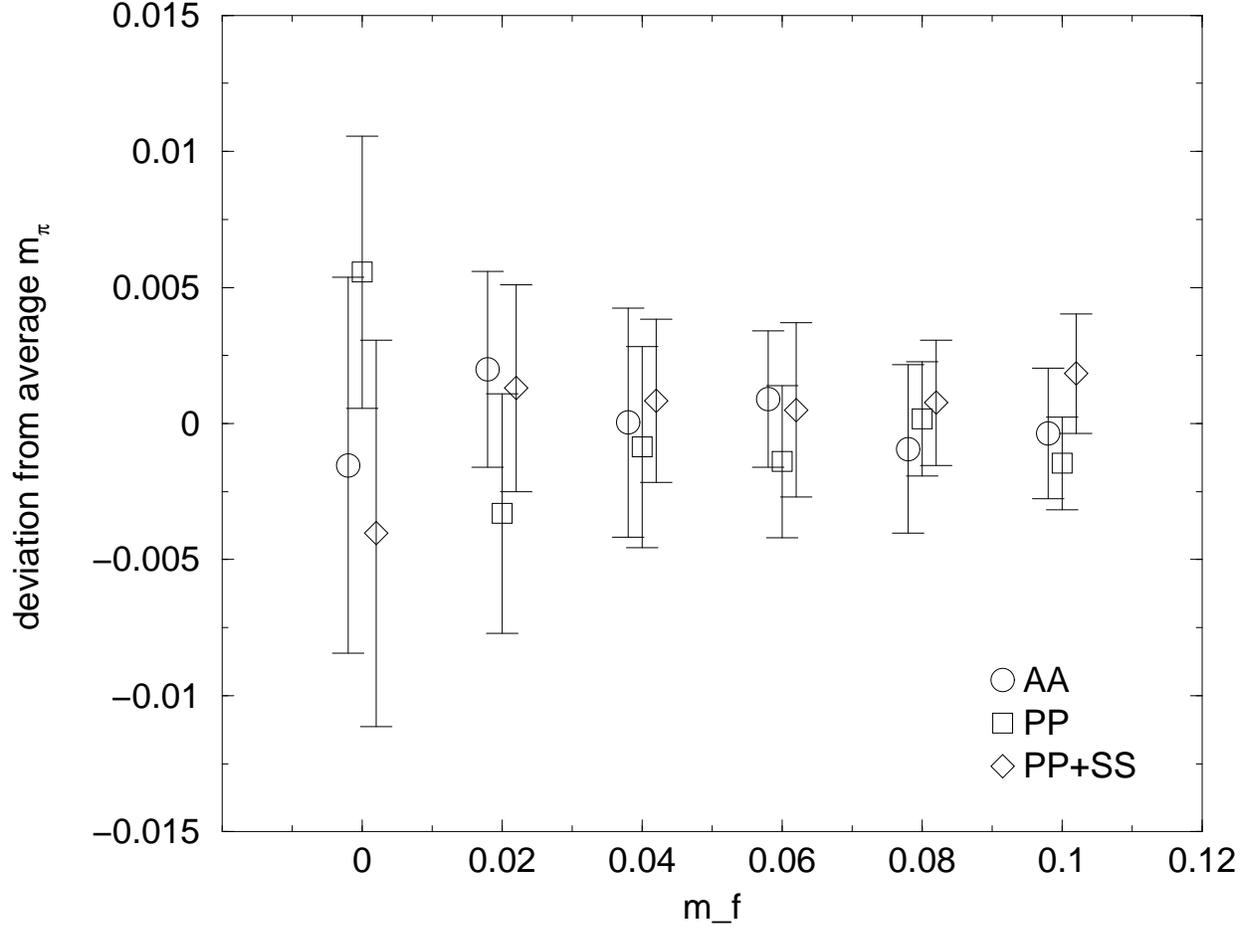}
\caption{For each $m_f$, the average value of $m_\pi$ is calculated for
the three correlators and the graph above shows the deviation of each
correlator from the average.  For each $m_f$, the result from $\corraa$
is shifted slightly to the right and the result from $\corrpp+\corrss$
to the left for clarity.  No systematic deviation is visible from the
data.}
\label{fig:mpi_vs_mf_b5_7_16nt32_ls48_dev}
\end{figure}

Figure \ref{fig:mpi2_vs_mf_b5_7_16nt32_ls48_mres} shows $\mpi2$ plotted
against $m_f$ for these runs.  In contrast to the smaller volume $8^3
\times 32$ result shown in Figure \ref{fig:mpi2_vs_mf_b5_7_8nt32_ls48},
all three correlators now give the same results for the pion mass, within
statistics (Table \ref{tab:mpi_mf_small}).  The larger volume has
clearly reduced the effects of the zero modes.  Further evidence of the
consistency of the mass from the three correlators is shown in Figure
\ref{fig:mpi_vs_mf_b5_7_16nt32_ls48_dev}.  Here, for each $m_f$, the
average value of $m_\pi$ is calculated and then the deviation from that
average, for each correlator, is plotted.  For each $m_f$, $\corraa$ is
offset to the left and the $\corrpp+\corrss$ to the right for clarity.

\begin{figure}
\epsfxsize=\hsize
\epsfbox{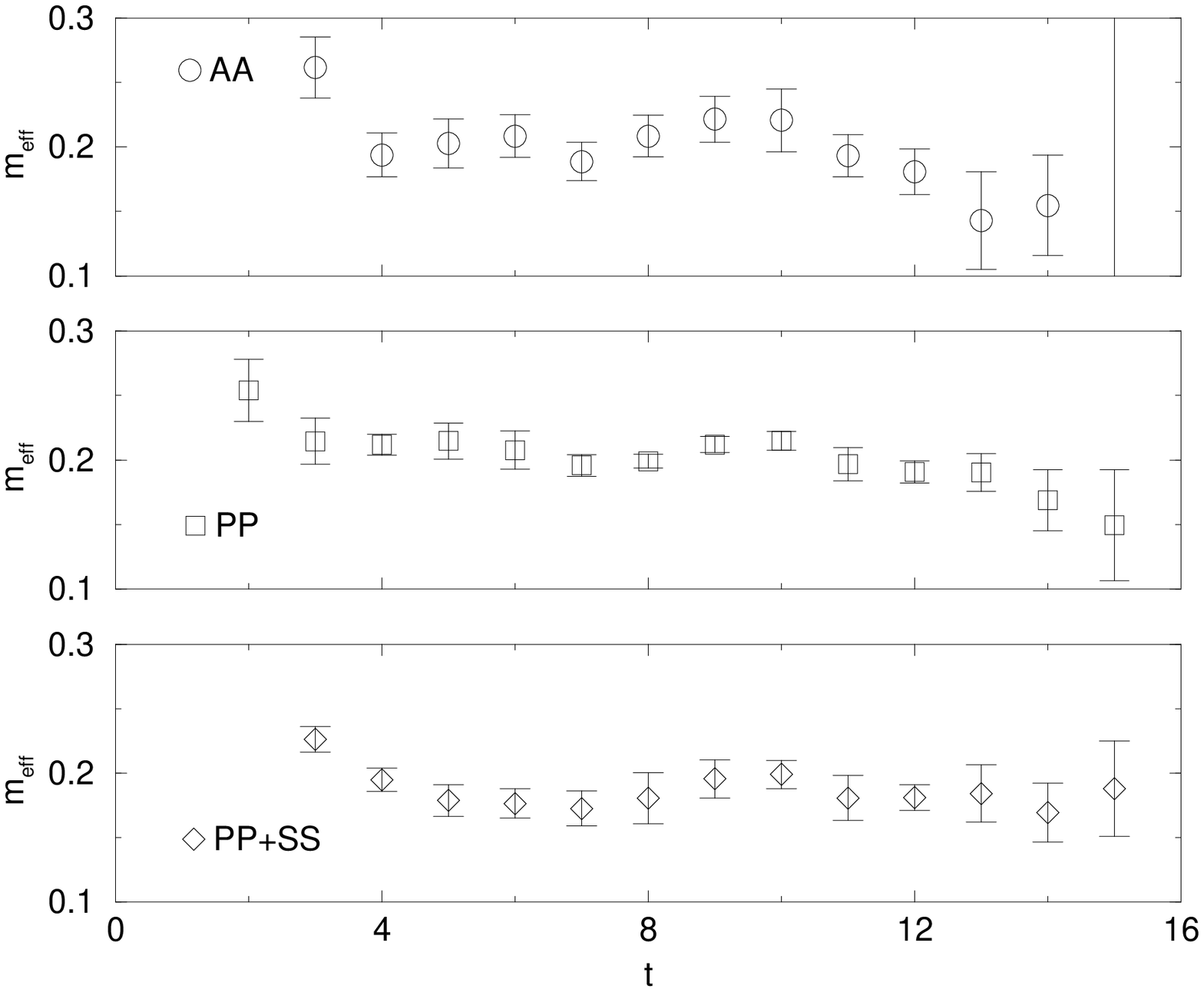}
\caption{The pion effective mass as a function of the source-sink
separation, $t$, for $16^3 \times 32$ lattices at $\beta = 5.7$ with
$L_s = 48$ and $m_f = 0.0$.  The upper panel is from $\corraa$
($\circ$), the middle from $\corrpp$ ($\Box$) and the lower from
$\corrpp+\corrss$ ($\Diamond$).  All three correlators give reasonable
effective masses and the fitted masses agree.}
\label{fig:mpi_effm_b5_7_16nt32_ls48_m0.0}
\end{figure}

\begin{figure}
\epsfxsize=\hsize
\epsfbox{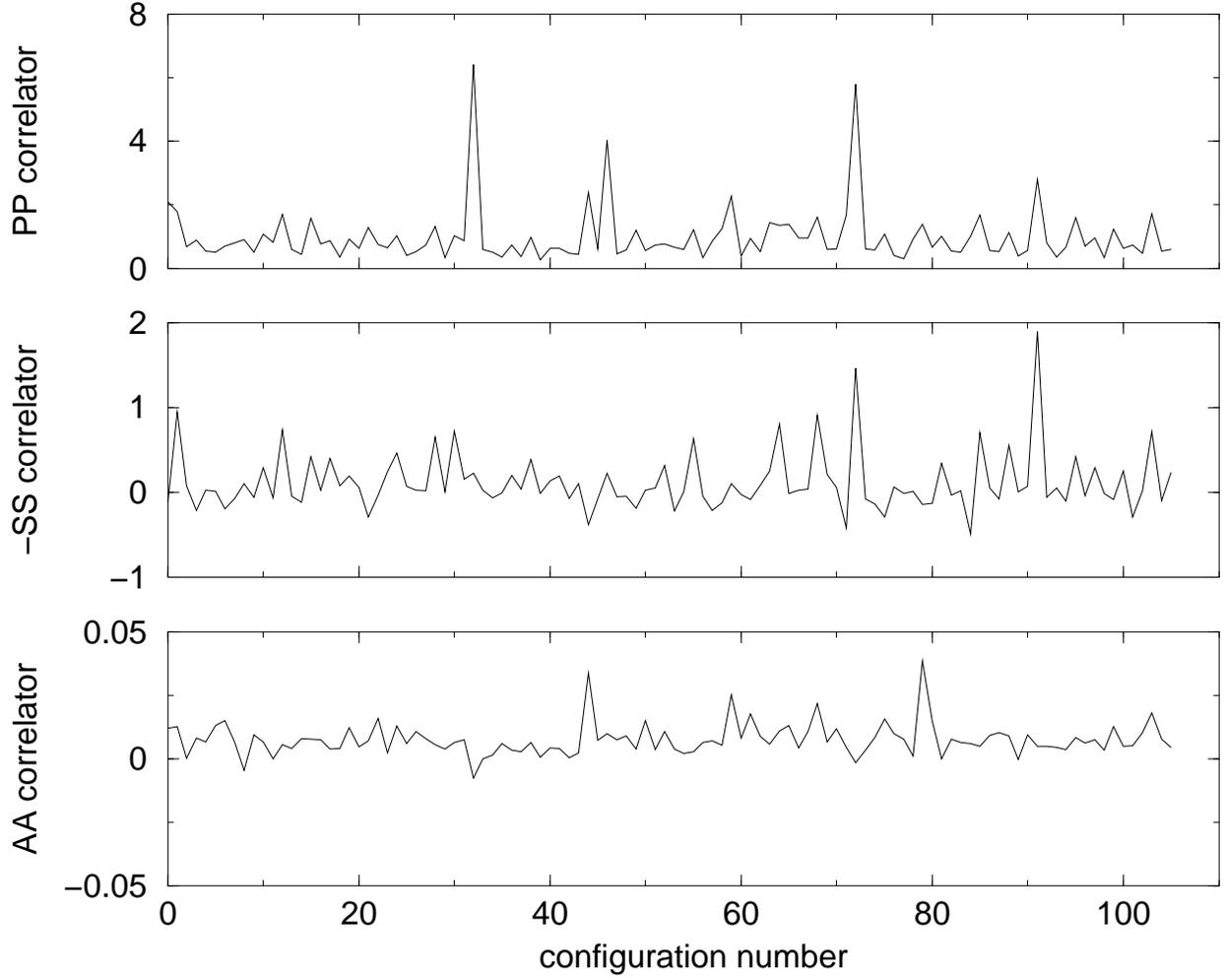}
\caption{The evolution of point source correlators at $t=8$ 
for $16^3 \times 32$ lattices at $\beta = 5.7$ with
$L_s = 48$ and $m_f = 0.0$  The upper panel is $\corrpp$, the
middle $-\corrss$ and the lower $\corraa$ (AA).
There are few, if any, contributions from topological near-zero modes.} 
\label{fig:corr_evol_b5_7_16nt32_ls48_m0_0}
\end{figure}

Figure \ref{fig:mpi_effm_b5_7_16nt32_ls48_m0.0} shows the effective
mass from each of the three correlators for $m_f=0$.  In contrast to the smaller
volume case, the effective masses have quite similar values and
lead to the same fitted mass, within errors.  As a last comparison
with the small volume, Figure \ref{fig:corr_evol_b5_7_16nt32_ls48_m0_0}
shows the three correlators at a time separation of $t=8$ as a
function of configuration number.  Little if any effect of
topological near-zero modes is seen.  Thus, we conclude that this
larger volume has suppressed these effects as expected.

\begin{figure}
\epsfxsize=\hsize
\epsfbox{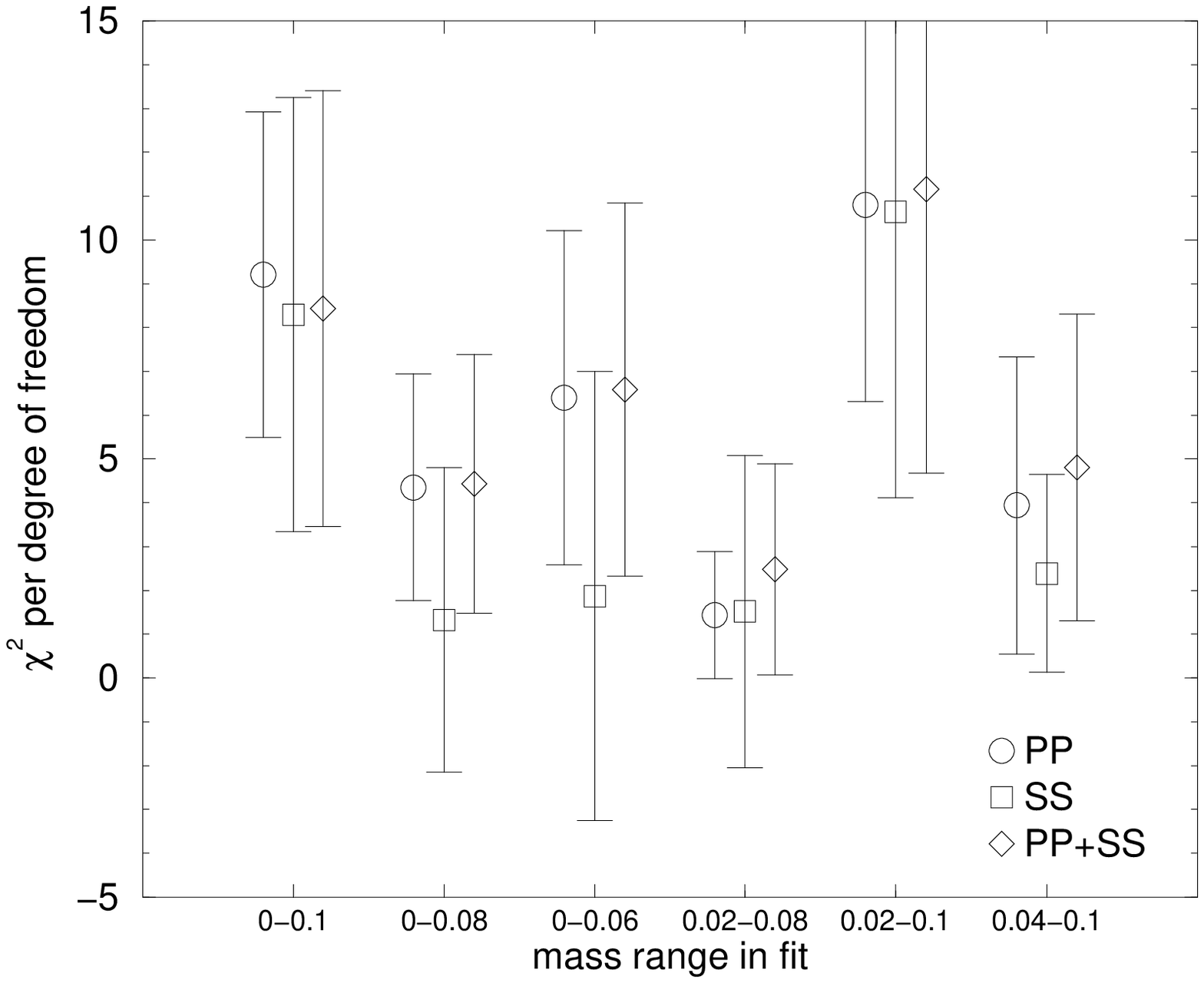}
\caption{The $\chi^2$ per degree of freedom for linear fits of $\mpi2$
versus $m_f$ for $16^3 \times 32$ lattices at $\beta = 5.7$ with $L_s =
48$ from $\corraa$ ($\circ$), $\corrpp$ ($\Box$) and $\corrpp+
\corrss$ ($\Diamond$).  Only the range $m_f = 0.02$ to 0.08 gives a fit
with an acceptable value for $\chi^2$ per degree of freedom.  After
using the large volume to eliminate zero modes, and presumably also
finite volume effects, we have evidence for a non-linear dependence of
$\mpi2$ on $m_f$.}

\label{fig:mpi2_vs_mf_b5_7_16nt32_ls48_chisq}
\end{figure}

Having established that a consistent pion mass can be determined from
our fitting range, we discuss the result of linear fits of $\mpi2$ as
a function of $m_f$.  We have done correlated linear fits of 
$\mpi2$ to $m_f$ for each of the correlators, using a variety of 
different ranges for $m_f$ in the fit.  The resulting $\chi^2$ per degree 
of freedom is shown in Figure \ref{fig:mpi2_vs_mf_b5_7_16nt32_ls48_chisq}, including the jackknife error on the $\chi^2$.  (The plotted error bars are 
the $\pm 1\sigma$ errors from the jackknife procedure and do not mean that
$\chi^2$ can become negative.)   The pion propagator for $m_f = 0.0$ and
0.04 was measured on the same set of configurations, with some of the
$m_f = 0.08$ propagators also measured on those configurations.  The
$m_f = 0.02$, 0.06 and 0.10 points were all measured on the same
configurations, along with the remaining 0.08 propagators.  Thus, these
points are less correlated in $m_f$ than the corresponding measurements
on the smaller volumes.

Now let us discuss the quality of these fits.  Given the significant
upward curvature of $m_\pi^2$ for $m_f \ge 0.1$, seen for example
in Figure~\ref{fig:mpi2_vs_mf_b5_7_8nt32_ls48}, we limit the mass
range to $m_f \le 0.08$.  If we do not include the lightest masses
and fit the points with $0.02 \le m_f \le 0.08$, as shown in
Figure~\ref{fig:mpi2_vs_mf_b5_7_16nt32_ls48_chisq} we obtain acceptable 
values for $\chi^2$ per degree of freedom for all three correlators.  
Specifically using the mass range $m_f = 0.02$ to 0.08, the fits to $\mpi2$
from the correlators $\corrpp$, $\corraa$ and $\corrpp+\corrss$
are
\begin{equation}
  \begin{array}{cc}
    \mpi2  =  0.044(5) + 4.75(5) m_f \;\;\;
  & \chi^2/{\rm dof} = 1.4 \pm 3.6 \\
    \mpi2  =  0.051(3) + 4.68(4) m_f \;\;\;
  & \chi^2/{\rm dof} = 1.4 \pm 1.4  \\
    \mpi2  =  0.049(3) + 4.70(5) m_f \;\;\;
  & \chi^2/{\rm dof} = 2.5 \pm 2.4. \\
  \end{array}
  \label{eq:large_vol_0.02_0.08_fit}
\end{equation}

However, given our confidence that this larger $16^3$ volume 
permits the reliable calculation of the pion mass for smaller 
values of $m_\pi$ we can also attempt a linear fit in the entire range 
$0.0 \le m_f \le 0.08$.  For this mass range, we find
\begin{equation}
  \begin{array}{ccl}
  m_\pi^2 = 0.042(3) + 4.77(3) m_f \;\;\;
& \chi^2/{\rm dof} = 1.3 \pm 3.5 \\
  m_\pi^2 = 0.044(3) + 4.75(5) m_f \;\;\;
& \chi^2/{\rm dof} = 4.3 \pm 2.6 \\
  m_\pi^2 = 0.042(4) + 4.82(6) m_f \;\;\;
& \chi^2/{\rm dof} = 4.4 \pm 3.0   
  \end{array}
  \label{eq:large_vol_pp_fit}
\end{equation}
for the correlators $\corrpp$, $\corraa$ and $\corrpp+\corrss$ 
respectively.  The $\corraa$ and $\corrpp+\corrss$ fits 
suggest that $m_\pi^2$ is not linear in this mass range.
While the $\corrpp$ fit is acceptable, as can be seen from 
a careful examination of 
Figure~\ref{fig:mpi2_vs_mf_b5_7_16nt32_ls48_mres}, this acceptable
fit comes because the $m_f=0.02$ point lies somewhat below while the
$m_f=0.0$ lies somewhat above the masses obtained from the other
two correlators.  Since the smaller volume studies suggest that
the $\corrpp$ correlator is most sensitive to zero modes and such
an upturn for small mass is the effect of zero modes seen at smaller
volume, this could easily be a remaining zero mode distortion.

It is difficult to draw a firm conclusion from the relatively large
correlated $\chi^2$/dof presented in Eq.~\ref{eq:large_vol_pp_fit}.
As is indicated by the errors shown, these $\chi^2$/dof are not
reliably known.  However, the comparison of the $\chi^2$/dof between
Eqs.~\ref{eq:large_vol_0.02_0.08_fit} and \ref{eq:large_vol_pp_fit}
may be more meaningful.  We attribute significant weight to the fact 
that the lightest $m_f=0$ point lies below the value predicted by a 
linear extrapolation from larger masses as can be easily seen in 
Figure~\ref{fig:mpi2_vs_mf_b5_7_16nt32_ls48_mres}.
 
We conclude that a linear fit does not well represent our data over the 
full mass range $m_f = 0.0$ to 0.1.  Of course, non-linearities for larger
masses can come from a variety of sources including terms from the
naive analytic expansion in powers of $m_f$.  However, for small $m_f$,
linearity is expected for large volumes in full QCD.  In contrast,
in the quenched approximation the absence of the fermion determinant 
may result in complex and more singular infrared behavior.  For example, 
it has been argued that a quenched chiral logarithm can appear in $\mpi2$ 
versus $m_f$ for quenched QCD \cite{Sharpe:1990me,Bernard:1992mk,Sharpe:1992ft}.  
The results just presented may be evidence for some non-linear behavior
of this sort.

Because of the poor linear fits found for small $m_f$, our data does not 
allow a determination of the location of the chiral limit for quenched
domain wall fermions by a simple extrapolation of $\mpi2$.
Even with the suppression of topological near-zero mode effects that
has been achieved by going to larger volume, further theoretical
input may needed if we are to deduce $\mres$ from these measurements
of $m_\pi^2$.  In the next section we will discuss our determination 
of the location of the chiral limit using other techniques and then 
return to the question of the behavior of $\mpi2$ with $m_f$.


\section{The residual mass}
\label{sec:m_res}


\subsection{Determining the residual mass}

\begin{figure}
\epsfxsize=\hsize
\epsfbox{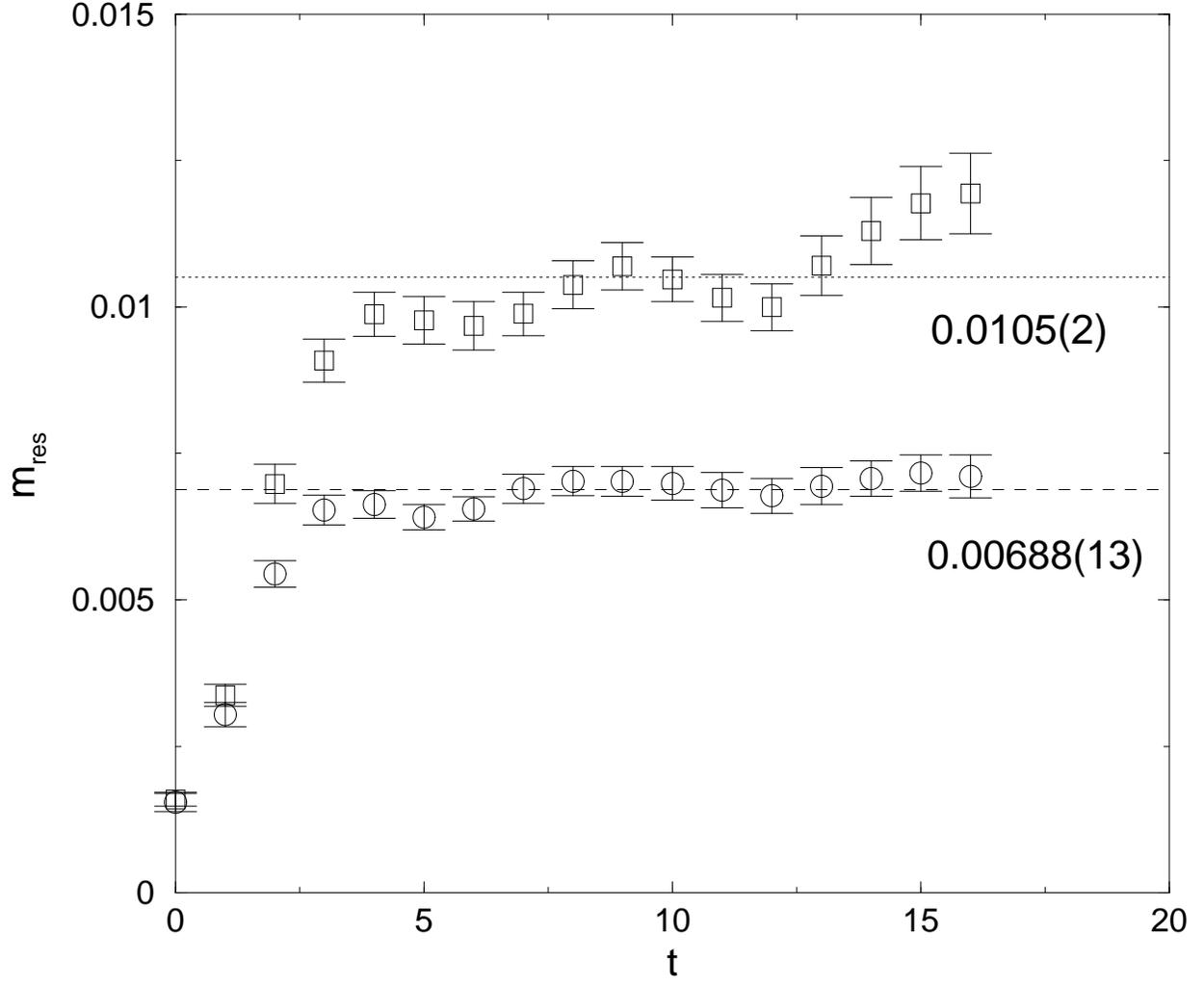}
\caption{The residual mass for $8^3 \times 32$ lattices with
$L_s=32, 48$ and $m_f=0.04$ at $\beta=5.7$. 
The labels for the horizontal lines are 
the averages over the range $4 \leq t \leq 16$ with jackknife errors. }
\label{fig:mres_plateau}
\end{figure}

In this section, we discuss our determination of $\mres$ using the 
low-momentum identity in Eq.\ \ref{eq:mres}.  This can done by
calculating the ratio
\begin{equation}
  R(t) = 
  \frac{\langle \sum_{\vec{x}} J^a_{5q} (\vec{x}, t)
  \pi^a(0) \rangle } 
  {\langle \sum_{\vec{x}} J^a_5(\vec{x}, t) 
  \pi^a(0) \rangle } 
  \label{eq:mres_ratio}
\end{equation}
as a function of $t$ (no sum on $a$), where $\pi^a(0)$ is a source 
evaluated at $t=0$ but possibly extended in spatial position.  This 
ratio was first used to determine $\mres$ in Ref.~\cite{Blum:1998ud} 
and later in Refs.~\cite{Aoki:1999uv,Aoki:2000pc}.  Our results are consistent 
with this earlier work, but a much more detailed study is undertaken 
here.  For $t$ outside some short-distance region, $t \ge t_{\rm min}$, 
$R(t)$ should be simply equal to $\mres$.  Using $R(t)$ for 
very large $t$ gives $\mres$ as the coupling of the pion 
to the mid-point pseudoscalar density divided by its coupling 
to the wall pseudoscalar density.  Of course, $\mres$ is 
an additive contribution to the effective quark mass at low
energies which effects all low-energy physics, not just the pion.  To
understand how large $t$ must be, Figure \ref{fig:mres_plateau} shows a
typical good plateau and a poor one.  Results are shown for $8^3 \times
32$ lattices with $m_f=0.04$ and $\beta=5.7$ for $L_s = 32$ and 48.
The good plateau is obtained from 335 configurations for $L_s=48$,
while the poor plateau is obtained from 184 configurations for
$L_s=32$.  The fewer measurements for $L_s = 32$ likely is the cause
for the upturn in the data at large $t$ and adding more configurations
at this $L_s$ should improve the signal.

From observing the onset of the plateaus in our data, we calculate
$\mres$ from the ratio in Eq.\ \ref{eq:mres_ratio} using the range 
$4 \leq t \leq 16$ for $\beta=5.7$, $6 \leq t \leq 26$ for $\beta=5.85$,
and $2 \leq t \leq 16$ for $\beta=6.0$.  The jackknife method is used 
to measure the statistical uncertainty and our $\mres$ results at 
$\beta=5.7, 5.85$ and $6.0$ are listed in Tables~\ref{tab:m_res_b6_0} 
and \ref{tab:m_res_b5_7}.  For most data sets, nice plateaus can be 
seen over the selected range, while for the few others with the poor 
plateaus, using a different range could change the results by $< 5\%$.  
We have also measured $\mres$ for different values of $m_f$ for 
$\beta=5.7$ on $16^3 \times 32$ lattices with $L_s = 48$.  
Table~\ref{tab:m_res_b5_7} gives the results and shows that the 
residual mass has little dependence on the input quark mass, 
reflecting the expected universal character of $\mres$.  Our $\beta=6.0$ 
results for $\mres$ appear to be a consistent extension of the values 
plotted in Figure~5 of Ref.~\cite{Aoki:2000pc} for $L_s= 4$, 6 and 10.

The $L_s$ dependence of $\mres$ is of vital importance to numerical
simulations with domain wall fermions.  Without the effects of 
topological near-zero modes, quenched chiral logs and finite volume, 
$\mpi2(m_f = 0)$ should be proportional to $\mres$ and should vanish 
with $\mres$ as $L_s \rightarrow \infty$.  However, in Section
\ref{sec:chiral_m_pi} we discussed how topological near-zero mode
effects alter $\corrpp$ and can distort the value of $\mpi2(m_f = 0)$ 
for large $L_s$ shown in Figure~\ref{fig:mpi2_vs_ls_b5.7}.  By 
measuring the ratio in Eq.\ \ref{eq:mres_ratio}, we can determine 
$\mres$ for non-zero $m_f$ and suppress all these effects which make 
the $m_f \rightarrow 0$ limit problematic.  This allows us to study 
the $L_s$ dependence of $\mres$, to which we now turn.

From the two values of $L_s$ shown in Figure \ref{fig:mres_plateau}, we see
that the residual mass for $8^3 \times 32$ lattices at $\beta=5.7$
falls from $0.0105(2)$ to $0.00688(13)$ as $L_s$ is increased from $32$
to $48$.  This is in sharp contrast to the almost identical results for
$\mpi2(m_f=0)$ at these two values for $L_s$ (Figure
\ref{fig:mpi2_vs_ls_b5.7}).  The overlap of the surface states is
significantly suppressed, as expected, even at this relatively strong
coupling.  We have not pursued the asymptotic behavior for large $L_s$
at $\beta = 5.7$, due to the large values for $L_s$ required, but instead
have studied this question for $\beta = 6.0$.

Figure \ref{fig:mres_vs_ls_b6_0} shows a similar study of the $L_s$
dependence of the residual mass for $16^3 \times 32$ lattices with
$m_f=0.02$ and $\beta = 6.0$.  The number of configurations used is
modest for the larger values of $L_s$.  We have used the factor 
$Z_S({\rm \overline{MS}, 2GeV}) = 0.619(25)$ obtained by a combination 
of non-perturbative renormalization and standard perturbation theory 
\cite{Dawson:1999yx,Npr:2000zz} to convert the plotted values of 
$\mres$ into MeV.  The value of $\mres$ is
decreasing with $L_s$ for all values of $L_s$, but is poorly fit by a
simple exponential.  In particular, an exponential fit using all
values of $L_s$ gives
\begin{equation}
  \mres = 0.0068(4) \exp(-0.094(4) L_s), \ \ \ \chi^2/{\rm dof}(3)=32
  \label{eq:mres_exp}
\end{equation}
which clearly does not match the measured values.  Adding a constant
to the fit gives
\begin{equation}
  \mres = 0.00032(3) + 0.018(3)\ {\rm exp} (-0.181(13) L_s), \ \ \
  \chi^2/{\rm dof}(2)=4.1
  \label{eq:mres_exp_const}
\end{equation}
where again all values for $L_s$ were used.  Even if this is the
correct asymptotic form, the value of $\mres$ for $L_s \rightarrow
\infty$ is very small, 1 MeV.

We have also tried fitting the largest three $L_s$ points to a simple
exponential and find
\begin{equation}
  \mres = 0.0012(2)\ {\rm exp}( -0.032(6) L_s), \ \ \
  \chi^2/{\rm dof}(1)=0.074.
  \label{eq:mres_exp_large_ls}
\end{equation}
Our data is consistent with the residual mixing vanishing exponentially
as $L_s \rightarrow \infty$, but the 0.032 coefficient in the exponent of 
Eq.~\ref{eq:mres_exp_large_ls} is quite small.   Of course, we can easily 
obtain an excellent fit to our five points if we include a second 
exponential.  For example, as shown in the figure, the five points fit 
well to two-exponential function
\begin{equation}
  \mres = 0.038(16)\ {\rm exp}(-0.26(4) L_s)
        + 0.0010(3)\ {\rm exp}(-0.027(7) L_s), \ \ \
  \chi^2/{\rm dof}(1)=0.1. 
  \label{eq:mres_exp_exp}
\end{equation}

Our measurements do not demonstrate a precise asymptotic form for
$\mres$ as a function of $L_s$.  However, we do see $\mres$ 
decreasing for large $L_s$ until, for $L_s \ge 24$, 
it has become so small as to be essentially negligible for current 
numerical work.  For $\beta = 5.7$ at $L_s$ = 48, $\mres$ is 
$0.074(5)$ in units of the strange quark mass, while
for $\beta = 6.0$ at $L_s = 16$ it is $0.033(3) m_s$.  
In the latter case, where we know the renormalization factors, 
$\mres$ in the $\overline{\rm MS}$ scheme at 2 GeV is 3.87(16) MeV.
Thus, even though more simulations will be needed to get the
precise asymptotic form, we find domain wall fermions having
the expected chiral properties for large $L_s$, even for lattice
spacings of around 1 Fermi.

In the next subsection, we will use the values of $\mres$ that have
just been determined to investigate further the $m_f$ dependence of
$m_\pi^2$, looking in particular at possible non-linear behavior as
$m_f + \mres \rightarrow 0$.  Here we would like to discuss a
simpler consistency check on the values of $\mres$ just obtained.
For the $\pi$, $\rho$ and nucleon we have established good linear 
$m_f$ behavior for larger values of $m_f$ with slopes and intercepts
given in Tables~\ref{tab:extrap_832_b5_7} and 
\ref{tab:extrap_1632_b5_7}.  If the only effect on these masses
of changing $L_s$ is to change the effective quark mass through the 
corresponding change in $\mres$, then we should be able to relate the 
the differences in the intercepts given in these tables to the 
product of the corresponding slope times the change in $\mres$ 
given in Tables ~\ref{tab:m_res_b6_0} and \ref{tab:m_res_b5_7}.  

While this comparison shows no inconsistencies, the errors in the 
intercepts are typically too large to permit a detailed confirmation.
For example, the difference in intercepts for $m_\pi^2$ at 
$\beta=6.0$ between $L_s=16$ and 24 is 0.0004(30) while the
difference predicted from the slope and the measured change
in $\mres$ is 0.0020(2).  The best test of this sort can be
made using the actual value for $m_\pi$ determined at 
$\beta=5.7$ and $m_f=0.04$ for $L_s=32$ and 48.  Here the 
difference of the masses squared is 0.012(6) while the prediction 
from the slope and change in $\mres$ is 0.0176(12).  Thus, we
can demonstrate consistency with the expected behavior but cannot 
make a definitive test.


\subsection{The residual mass and $\mpi2$ versus $m_f$}

The definition of $\mres$ and its measurement mean that we have
determined the value of $m_f$ for which the pion should become 
massless if the domain wall method is successfully representing
the chiral limit of the underlying theory.   We can now return to the
question of the dependence of $\mpi2$ on $m_f$, starting with the $16^3
\times 32 $ simulations at $\beta = 5.7$ and $L_s = 48$.  Recalling
Figure \ref{fig:mpi2_vs_mf_b5_7_16nt32_ls48_mres}, we found that the
larger volume gave consistent pion mass measurements from the three
correlators, but $\mpi2$ was not well fit as a linear function of $m_f$
for two of the correlators if the $m_f = 0.0$ point was included.  In
Figure \ref{fig:mpi2_vs_mf_b5_7_16nt32_ls48_mres}, we have included the
value of $\mres$ (the starred point) as measured from Eq.\
\ref{eq:mres_ratio}. (Its error bar on the horizontal axis is a
vertical line on this scale.) The solid line is the fit to the $\corraa$
correlator for $m_f = 0.02$ to 0.08 given in
Eq.\ \ref{eq:large_vol_0.02_0.08_fit} while the dotted line is for the
$\corrpp$ correlator for $m_f = 0.0$ to 0.08 as given in
Eq.\ \ref{eq:large_vol_pp_fit}.  Thus, we see that linear fits poorly
represent the data when the $m_f = 0.0$ point is included for the $\corraa$
and $\corrpp+\corrss$ case and fail for all three correlators when the 
pion mass is required to vanish at $m_f = -\mres$.

We can make this conclusion more quantitative by comparing our
accurate value for $\mres = 0.00688(13)$ at $L_s=48$ determined on 
an $8^3\times 32$ lattice with the naive linear extrapolation of
$m_\pi^2(m_f)$ to the point $m_\pi^2=0$.  Using the most reliable 
linear fits obtained by excluding the $m_f=0$ point in 
Eq.~\ref{eq:large_vol_0.02_0.08_fit} we obtain the $x$-intercept
values shown in Table~\ref{tab:q_chiral_log}: -0.0092(12),
-0.0108(7) and -0.0104(7) for the $\corrpp$, $\corraa$ and 
$\corrpp+\corrss$ correlators respectively.  These differ from this 
value of $\mres$ by $\approx 50\%$ and 2, 5 and 6 standard
deviations respectively.  We conclude that there is a significant
discrepancy between the $m_f$-dependence of these $m_\pi$ results
and the hypothesis that $m_\pi^2(-\mres)=0$.  However, notice that 
if the $m_f=0$ points are included in the linear fits, and the 
less accurate $\mres$ from the same volume is used, this 
discrepancy can be reduced.  For example, a linear fit to the data 
from the $\corrpp$ correlator in Eq.~\ref{eq:large_vol_pp_fit} has 
an intercept at -0.0088(5) while $-\mres = -0.0072(9)$ on the 
same volume.  We believe that such an interpretation should be
discounted as failing to exploit all the available information.

Is this significant discrepancy caused by essential 
non-linearities in the quenched approximation or by a breakdown 
of the domain wall method, for example, large $O(a^2)$ effects?  
We can address this question by making a similar comparison
for $\beta=6.0$ where $O(a^2)$ effects should be significantly 
reduced.  Since we have not investigated a large volume 
at this weaker coupling, we propose to examine the $\corraa$
correlator because reduced zero-mode effects were seen for this
correlator in our $\beta=5.7$ studies.  Using the three lightest
masses we find $x$-intercepts of -0.0031(7) and -0.0030(9) for 
the $L_s=16$ and 24 cases respectively.  Again, these are
dramatically farther from the origin than the corresponding values
of $\mres=0.00124(5)$ and 0.00059(4).  These are each three standard
deviation effects.  However, they are obtained on independent
configurations and together can be viewed as a 6 standard 
deviation discrepancy.  Thus, if possible finite-volume difficulties 
are ignored, we have again strong evidence for a discrepancy.  Rather
than decreasing by a factor of four as would be expected from
an $O(a^2)$ error, this fractional discrepancy is substantially
larger in this $\beta=6.0$ comparison.  Thus, it is natural to conclude
that domain wall fermions are accurately representing the chiral
behavior of quenched QCD.

At the beginning of Section \ref{sec:chiral_m_pi} we listed possible
systematic effects influencing the chiral limit for $m_\pi$.  With a
measurement of $\mres$ we have quantified the role of finite $L_s$ and
with the larger volume used for $\beta = 5.7$ we have reduced, if not
eliminated the topological near-zero modes.  We should also have
minimized other finite volume distortions of the density of
eigenvalues, which also influence the pion mass.  Finally with the
comparison above, we have examined the possibility of $O(a^2)$ errors.
Thus, we now address the question of quenched chiral logarithms.  
Predictions of this particular pathology of quenched simulations 
were made some time ago. There is certainly much data indicating 
possible support for the predictions, but there is disagreement 
about its conclusiveness, see for example 
refs.~\cite{Kim:1995tg,Duncan:1996ma,Bernard:1998db,Aoki:1999yr,Bardeen:2000cz,Gockeler:2000pg}.
Since many other effects must be removed before these subtle logarithms
are convincingly seen, it is a challenging numerical issue.

The natural first place to look for quenched chiral logarithm effects
is in $m_\pi$, but this is difficult for Wilson fermions, where
the chiral point is not crisply defined for finite lattice spacing.
For staggered fermions, where the chiral limit occurs when the
input quark mass is zero, the issue is complicated by the presence
of only a single Goldstone pion.  In some respects, domain wall
fermions are an ideal place to look for these effects, except
that the statistical resolution needed is difficult to achieve
with the additional computational load of the fifth dimension.
In addition, the topological near-zero modes are a much larger
quenched pathology at moderate volumes.

\begin{figure}
\epsfxsize=\hsize
\epsfbox{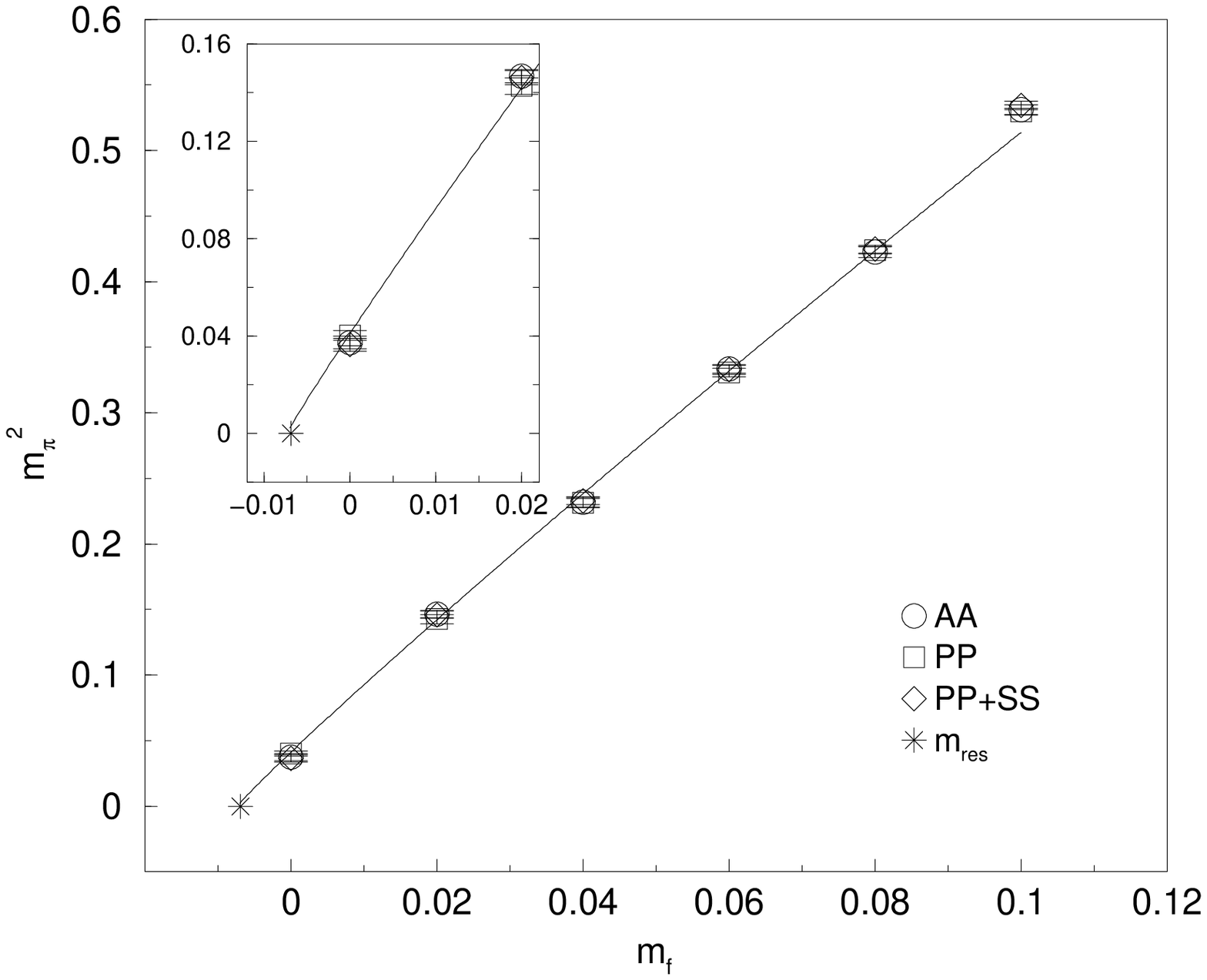}
\caption{The pion mass squared versus $m_f$ from $\corrpp$ ($\Box$),
$\corraa$ ($\circ$) and $\corrpp + \corrss$ ($\Diamond$) for quenched
simulations done on $16^3 \times 32$ lattices at $\beta = 5.7$ with
$L_s = 48$.  The star is the value of $\mres$ as measured from Eq.\
\ref{eq:mres_ratio} and its error bar in the horizontal axis is too
small to show on this scale.  The solid line is a fit of the $\corraa$
correlator for $m_f = 0.0$ to 0.08 to the quenched chiral logarithm
form given in Eq.\ \ref{eq:q_chiral_log}.  This fit gives the pion mass
vanishing in very good agreement with the value of $\mres$ determined
from Eq.\ \ref{eq:mres_ratio}.}
\label{fig:mpi2_vs_mf_b5_7_16nt32_ls48_log}
\end{figure}

As one way of probing the non-linearity in $\mpi2$ versus $m_f$,
we have fitted our data for $\mpi2$ for $16^3 \times 32$ lattices at
$\beta = 5.7$ and $L_s = 48$ to the form
\cite{Sharpe:1990me,Bernard:1992mk,Sharpe:1992ft}
\begin{equation}
  \mpi2 = a_0 ( m_f + a_1 ) ( 1 + a_2 \ln(m_f + a_1 ) )
  \label{eq:q_chiral_log}
\end{equation}
and the results are given in Table~\ref{tab:q_chiral_log}.  The fit
yields a value for the residual mass (the parameter $a_1$ above) and
the results are quite close to those measured from the ratio of
Eq.\ \ref{eq:mres_ratio}.  Figure
\ref{fig:mpi2_vs_mf_b5_7_16nt32_ls48_log} shows the result from fitting
$\corraa$ for $m_f = 0.0$ to 0.08 to the quenched chiral logarithm form
given in Eq.\ \ref{eq:q_chiral_log}.  We have excluded the larger
values of $m_f$ from our fits, since higher order terms are needed in
Eq.\ \ref{eq:q_chiral_log} to accommodate the upward curvature of our
$\mpi2$ data.  While the $\chi^2$/dof for the logarithmic fit is only 
marginally better than those obtained for the simple linear fits described 
earlier in Eq.~\ref{eq:large_vol_pp_fit} for this same mass range, the 
ability of the logarithmic fit to predict the appropriate $\mres$ 
value is significant.

For the simulations at smaller physical volumes, $8^3 \times 32$ at
$\beta = 5.7$ and $16^3 \times 32$ at $\beta = 6.0$, the values for
$\mres$ measured from Eq.\ \ref{eq:mres_ratio} are generally smaller
than the $x$ intercepts for the linear fits shown in Figures
\ref{fig:mpi2_vs_mf_b5_7_8nt32_ls48} and
\ref{fig:mpi2_vs_mf_b6_0_16nt32_ls16}.  This indicates curvature in the
direction given by a chiral logarithm, but the other phenomena that may 
be affecting these chiral limits make quantitative analysis ambiguous.
We note that $\mpi2$ from $\corrpp+\corrss$ seems to
smoothly curve towards the value of $\mres$ from the previous
subsection.  However, we are not sufficiently certain of the absence
of zero mode effects in the $\corrpp+\corrss$ correlator to describe
a logarithmic fit to these cases.

This nice agreement between the values of $\mres$ determined from the 
location of the $m_\pi^2=0$ point in these fits and that computed by 
other means earlier in the paper implies consistency between
our results and the logarithmic form of Eq.~\ref{eq:q_chiral_log}.
Of course, other non-linear terms could be used to explain this 
curvature and, given our statistics, would provide an equally 
consistent description of our data.  

However, our most important
conclusion is not related to quenched chiral logarithms, but rather
to having seen all the expected properties for the chiral limit
with domain wall fermions.  Once the topological near-zero mode
effects are reduced or eliminated, consistent pion masses can
be measured.  A precise measurement of $\mres$ is consistent
with our $\mpi2$ versus $m_f$ dependence if, for example, a 
chiral logarithm term is included.  In short, domain wall fermions 
are showing sensible chiral properties, even on lattices with a 
lattice spacing of $\sim 1$ Fermi.

We have chosen not to pursue an additional method of determining
$\mres$ that has been proposed in two of our previous publications 
\cite{Fleming:1999eq,Chen:2000zu}.  In that method, one examines the 
integrated Ward-Takahashi identity in Eq.~\ref{eq:int_ward_tak_id} and uses 
the location of the pion pole in $\chipp$ to determine $\mres$.  While 
this technique should be reliable for dynamical fermion calculations, 
{\it e.g.} as used in Ref~\cite{Chen:2000zu}, it does not explicitly 
allow for the effects of topological near-zero modes or possible 
non-linear behavior of $m_\pi^2(m_f)$ that we have found to be important 
in the quenched approximation.  Thus, even though this method gave a 
result for $\beta=5.7$ quite close \cite{Fleming:1999eq} to the $L_s=48$ 
value $\mres=0.00688(13)$ presented in this paper, more analysis is needed
to adequately justify its use in this quenched case.

\subsection{Eigenvalue properties and $\mres$}

A comparison of the approximate form of the Banks-Casher relation
for domain wall fermions given in Eq.~\ref{eq:dwf_banks_casher}
with the usual 4-dimensional expression in Eq.~\ref{eq:banks_casher}
suggests a close relationship between the parameter $\delta m_i$
deduced from the $i^{th}$ eigenvalue $\Lambda_{H,i}$ of $D_H$ and
the residual mass $\mres$.  In this section we will explore this 
relation further making use of an exploratory study of the 
low-lying spectrum of $D_H$ \cite{eigen:2000aa}.

These eigenvalues were calculated for 32 configurations obtained
at $\beta=6.0$ on a $16^4$ lattice with $L_s=16$ listed in Table~\ref{tab:run_parameters_hb} and beginning with an equilibrated
configuration from an earlier run.  We used the 
Kalkreuter-Simma \cite{Kalkreuter:1996mm} method to find the 
19 lowest eigenvalues on each configuration\footnote{We thank Robert
Edwards whose program formed the basis of the code used in this
part of the calculation.}.   We apply this method 
to the positive matrix $D_H^2$, and then determine the eigenvectors 
and eigenvalues of $D_H$ by a final explicit diagonalization
of $D_H$ in the subspace of the eigenvectors of $D_H^2$ just 
determined.  The details of our application of this method and a 
more complete description of these results will be presented in 
a later publication \cite{eigen:2000aa}.

While this method determines both the eigenvalues and eigenvectors, 
we have chosen to examine only the $s$-dependent, four-dimensional 
inner products:
\begin{equation}
\Gamma_{R/L}(s)_{i,j} = \sum_x \Psi^\dagger(x,s)_{\Lambda_{H,i}} P_{R/L}
                           \Psi(x,s)_{\Lambda_{H,j}},
\end{equation}
where the indices $i,j$ run over all of the 19 eigenvalues while $P_R$
and $P_L$ are the left and right spin projection operators defined
above Eq.~\ref{eq:axial_cc}. In order to be able to make use of the 
mass dependence of the eigenvalues, we have repeated the calculation 
of $\Lambda_i(m_f)$ and $\Gamma_{R/L}(s,m_f)_{i,j}$ five times on each 
configuration for the five different mass values $m_f=0.0$, 0.0025, 0.005, 
0.0075 and 0.001.

Here we will describe some of the overall features of this calculation
and then examine more closely the relation between the parameters 
$\delta m_i$ and the value of $\mres$ determined earlier in this paper.
First we examine the diagonal elements of the matrix $\Gamma(s)$
\begin{equation}
{\cal N}(s)_i = \Gamma_R(s)_{i,i} + \Gamma_L(s)_{i,i}.
\label{eq:norm_s}
\end{equation}
This is the contribution to the norm of the 5-dimensional wave
function from the 4-dimensional hyperplane with a specific value 
of $s$.  For these low lying eigenvalues, we expect that this norm should
be concentrated on the $s=0$ and $s=L_s-1$ walls, which we find to be true 
to good accuracy.  For the entire group of $32 \times 19\ \times 5 = 3,040$
eigenvectors computed, the ratio of the sum of the norm on the two walls 
to the minimum value of this norm between the walls was always greater than 34, 
${\cal N}(0)+{\cal N}(L_s-1) > 34 \cdot {\cal N}(s_{\rm min})$.  The median
value for this ratio was 744.  Thus, the general framework upon which
the domain wall formalism rests appears approximately valid.

As a test of our method for determining the eigenvectors, we evaluate
the left- and right-hand side of the symmetry relation, 
Eq.~\ref{eq:sym_constraint}, between pairs of eigenvectors on a given 
configuration.  The resulting equality:
\begin{equation}
(\Lambda_{H,i}+\Lambda_{H,j})\langle\Lambda_{H,i}|\Gamma_5|\Lambda_{H,j}\rangle 
         = \langle\Lambda_{H,i}|\Bigl(2 m_f\qw + 2\qmp \Bigr)|\Lambda_{H,j}\rangle
\label{eq:ks_test}
\end{equation}
provides a good test of our diagonalization procedure.  The vectors
$|\Lambda_{H,j}\rangle$ needed to evaluate this expression are eigenvectors
of the Dirac operator $D_H$, not $D_H^2$.  We determine the eigenvectors of
$D_H$ by diagonalization within the 19-dimensional subspace found by applying
the Kalkreuter-Simma method to $D_H^2$.  In the event that the $19^{th}$ and
$20^{th}$ eigenvalues of $D_H^2$ are nearly degenerate (not entirely unlikely 
given the expectation that the eigenvalues of $D_H$ occur in $\pm \Lambda_H$
pairs), this truncated, 19-dimensional subspace will not be spanned by 
eigenvectors of $D_H$.   It will contain 18 valid eigenvectors and a $19^{th}$
vector, orthogonal to the rest but not an eigenvector of $D_H$.  This
``spurious'' eigenvector can be reliably removed since it will give an 
``eigenvalue'' whose square does not agree with any found for $D_H^2$.  
We remove such eigenvectors from our test of Eq.~\ref{eq:ks_test} and, 
for uniformity, the $19^{th}$ eigenvector in the case that no spurious 
eigenvector occurs.  There are then $32 \times 5 = 160$ instances where we 
can check $18^2$ independent elements of Eq.~\ref{eq:ks_test}.  We
find that 95\% of these 51,840 comparisons have a fractional error below 
5\%.  The few cases with significantly worse agreement, result from infrequent 
near degeneracies which challenge the Rayleigh-Ritz method on which 
the Kalkreuter-Simma algorithm is based.

For most configurations there are easily identified zero modes.  Typically
the few lowest eigenvalues have eigenvectors all of which are bound 
to the same wall, either $s=0$ or $s=L_s-1$.  The corresponding matrix 
elements $\langle\Lambda_{H,i}|\Gamma_5|\Lambda_{H,i}\rangle$ all 
have the same sign and are within a few percent of 1, showing precisely 
the structure expected in a four-dimensional theory as summarized in
Eq.~\ref{eq:anti_herm_d4_evalues}.  

\begin{figure}
\epsfxsize=\hsize
\epsfbox{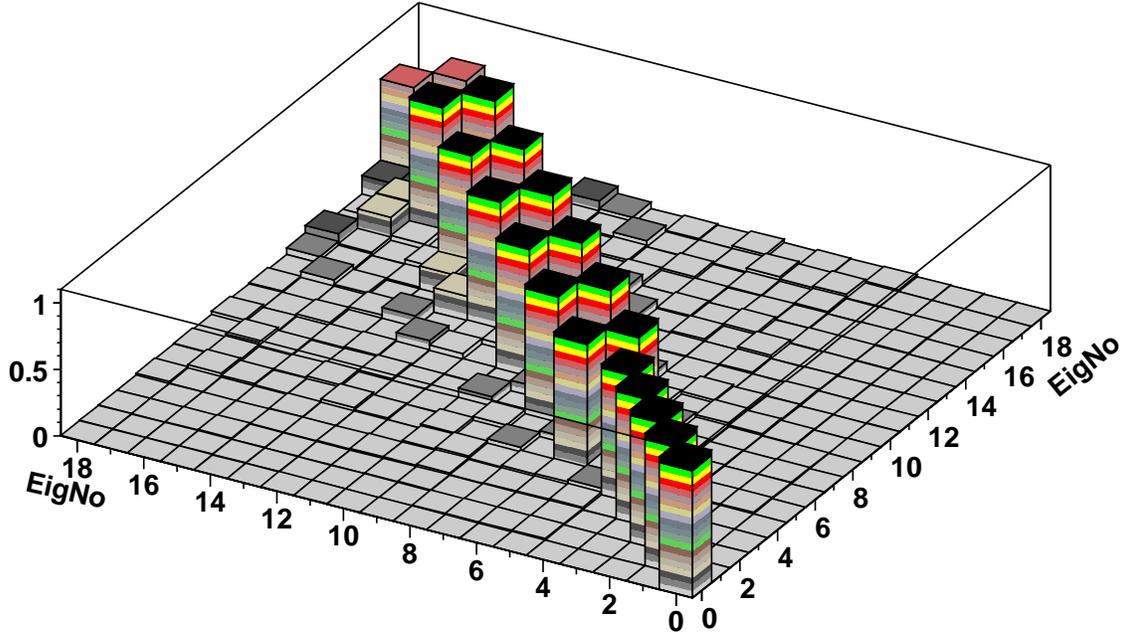}
\caption{A three dimensional ``Lego'' plot showing the matrix elements
of $\Gamma^5$ between all nineteen eigenvectors found for one of the
better configurations in our sample of 32, evaluated at $m_f=0$.  
The height of the box located by horizontal coordinates $(i,j)$ represents 
the magnitude of the matrix element 
$\langle\Lambda_{H,i} | \Gamma^5 |\Lambda_{H,j}\rangle$.  The five zero modes, 
all nearly eigenvectors of $\Gamma^5$ with eigenvalue +1, are easily 
identified.  The remaining seven pairs are also very evident 
corresponding to the expected $|\Lambda_{H}\rangle$ and 
$\Gamma^5|\Lambda_{H}\rangle = |-\Lambda_{H}\rangle$ eigenstates.}
\label{fig:lego_3_134_19}
\end{figure}

\begin{figure}
\epsfxsize=\hsize
\epsfbox{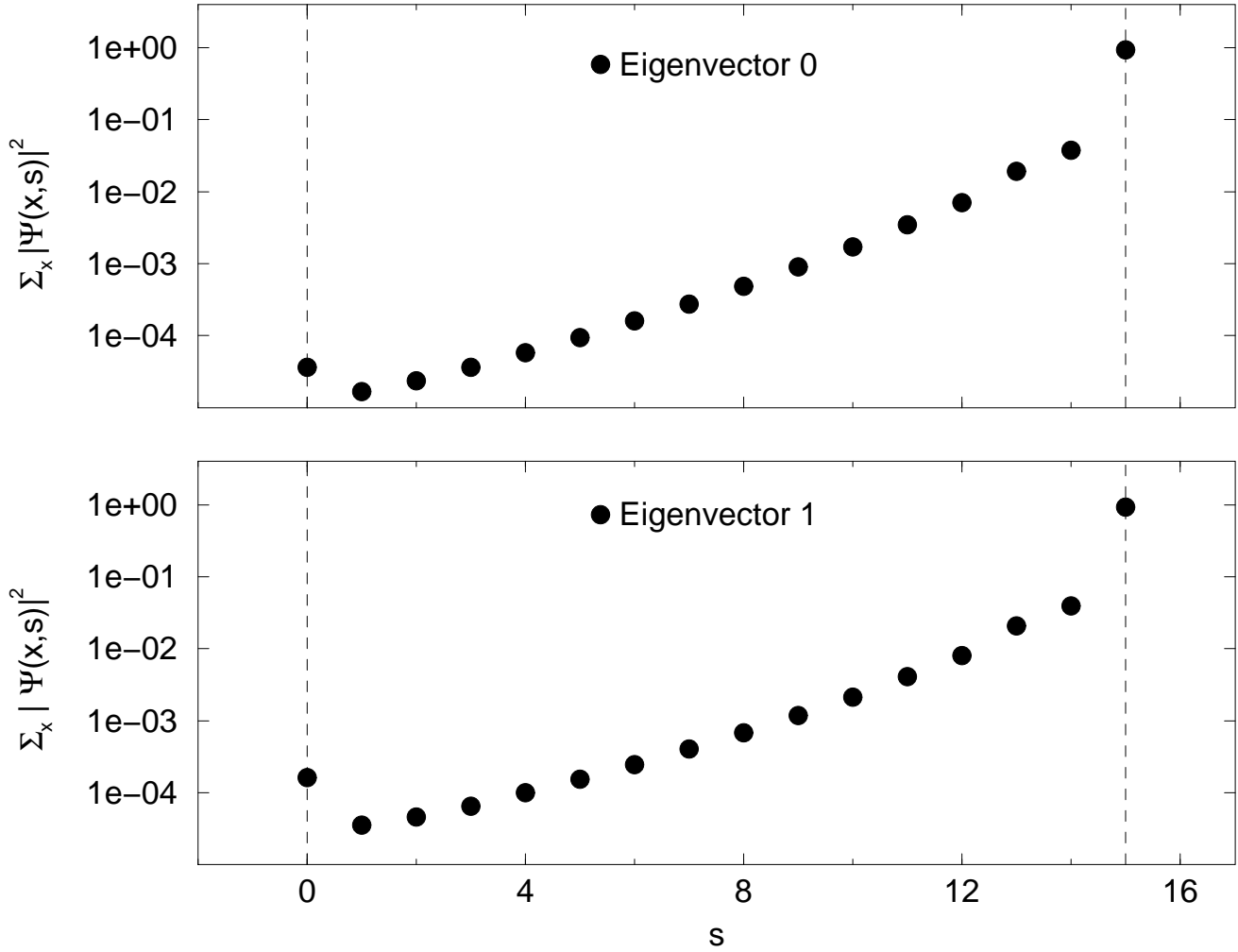}
\caption{The distribution of the 4-dimensional norm, ${\cal N}(s)$
for the first two zero modes shown in Figure~\ref{fig:lego_3_134_19}
as a function of $s$.  Note both states are tightly bound to the $s=L_s-1$
wall, as are the other three zero modes states.}
\label{fig:evector_0_1_vs_s_b6_0_16nt16}
\end{figure}

\begin{figure}
\epsfxsize=\hsize
\epsfbox{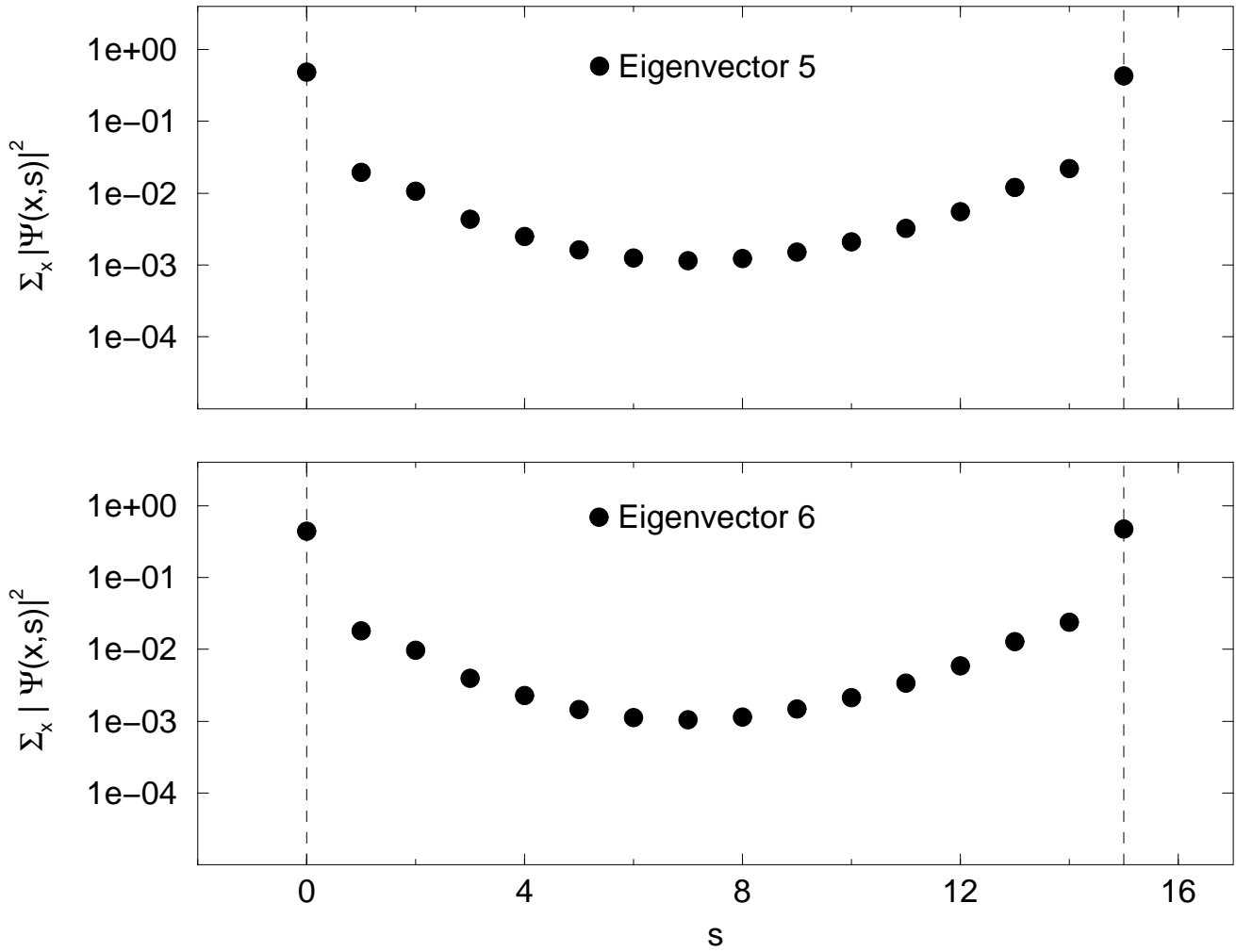}
\caption{The distribution of the 4-dimensional norm, ${\cal N}(s)$
for the first pair of non-zero modes shown in 
Figure~\ref{fig:lego_3_134_19} as a function of $s$.}
\label{fig:evector_5_6_vs_s_b6_0_16nt16}
\end{figure}

The potential of the domain wall method is nicely displayed by examining 
the properties of one of our better configurations.   In 
Fig~\ref{fig:lego_3_134_19}, we show the magnitude of the elements of the matrix
$\langle\Lambda_{H,i}|\Gamma_5|\Lambda_{H,j}\rangle$ in a three-dimensional
plot.  Note the five zero-modes in this configuration are easily recognized.
Each has a diagonal matrix element of $\Gamma^5$ within 1.5\% of 1 and
matrix elements with other vectors all of magnitude below 0.06.  The values
of $\lambda_i$ for these five eigenvalues all lie in magnitude below
$3.6 \times 10^{-4}$ while the remaining paired eigenvalues lie between
0.028 and 0.093.  In Figures~\ref{fig:evector_0_1_vs_s_b6_0_16nt16} and
\ref{fig:evector_5_6_vs_s_b6_0_16nt16} we show the $s$-dependence of the 
first two zero-modes and the first pair of non-zero eigenvectors, numbers
5 and 6.  One sees precisely the expected behavior.  Both zero-modes are 
bound to the same wall (as are the other three zero modes) while the 
two paired non-zero modes are nearly symmetrical between right and 
left.  This is clearly identified as a configuration with topological 
charge $\nu=+5$.

Of direct interest in this section is the mass dependence of 
$\Lambda(m_f)_{H,i}$ and a quadratic fit of the sort proposed in
Eq.~\ref{eq:dherm_lambda_vs_mf}.  For the small masses we have used, this
quadratic form provides an excellent fit, after some re-sorting of 
eigenvalues is performed to account for infrequent level crossings
as $m_f$ is varied.   In order to avoid the possibility that these
level crossings may have pushed a needed eigenvalue up to beyond 
number 19, we have excluded those quadratic fits which contain the 
largest eigenvalue at $m_f=0$ for each of the 32 configurations.  
The resulting root-mean-square of the fractional differences 
between the left- and right-hand sides of Eq.~\ref{eq:dherm_lambda_vs_mf} 
is very small.  The average root-mean-square of the fractional 
difference is $1.3\times 10^{-4}$ while the largest value is 
$4.3\times 10^{-3}$.

\begin{figure}
\epsfxsize=\hsize
\epsfbox{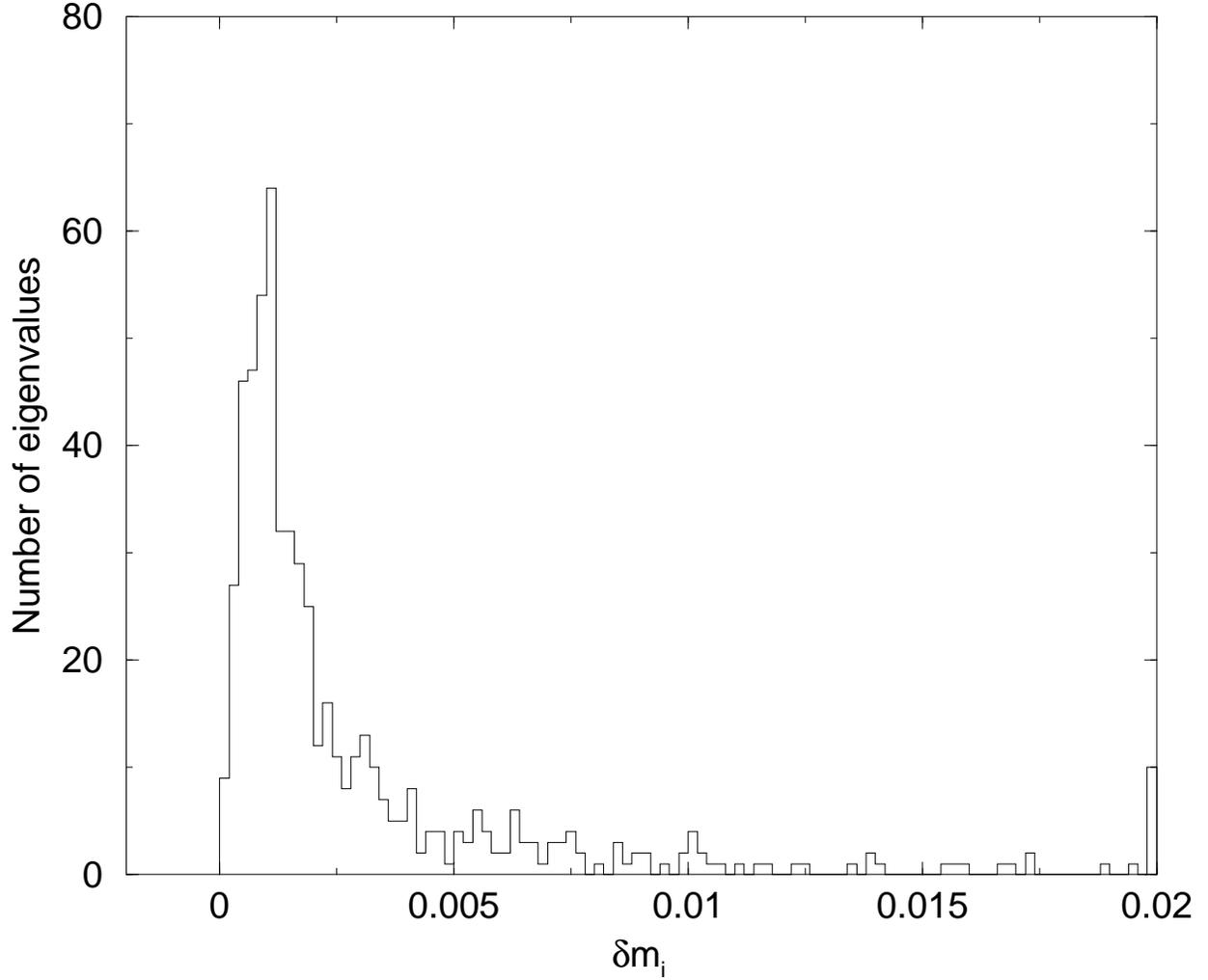}
\caption{The distribution of values of the quadratic fit parameter 
$\delta m_i$ defined in Eq.~\ref{eq:qbq_phenom}.  These parameters were
determined from a total of 576 eigenvalues obtained from 32 configurations
computed with $\beta=6.0$, $16^4$ and $L_s=16$.  The peak of the 
distribution lies remarkably close to the value we find for the residual 
mass, $\mres=0.00124(5)$.}
\label{fig:deltam_hist}
\end{figure}

In Figure~\ref{fig:deltam_hist} we present a histogram of the 
distribution of fit parameters for the $18 \times 32 = 576$, 
$\delta m_i$ values that we obtain.  The majority of values 
are quite small, very much on the order of $\mres$.  While 
a few larger values of $\delta m_i$ are seen (the largest 
is 0.0660), the median of the distribution is 
$\overline{\delta m} = 0.00147$ which is remarkably close to the value 
of $\mres=0.00124$ found earlier for this value of $\beta$ and $L_s$.

The 4-dimensional expression for $\langle \overline{\psi} \psi \rangle$ 
in Eq.~\ref{eq:banks_casher} and the 5-dimensional result in 
Eq.~\ref{eq:dwf_banks_casher} as a function of $m = Z m_f$ must 
agree in the continuum limit after a rescaling and overall subtraction.
This must be true even if $\mres$ is held fixed in physical units
as $ a \rightarrow 0$.  Therefore, in the limit of zero lattice 
spacing, the histogram shown in Figure~\ref{fig:deltam_hist} must 
approach a delta function so that $\delta m_i$ has the unique value 
$\mres$.  Thus, we might interpret the width of the distribution 
in Figure~\ref{fig:deltam_hist} as a result of $O(a^2)$ effects.  
The large size of the fluctuations relative to the central value 
is presumably a result of the small central value produced by our 
quite large separation of 16 between the walls.


\section{Hadronic Observables}
\label{sec:hadron}

We can now use the results of the previous sections to compute 
a variety of hadronic properties.  In this section we will
discuss two topics: the evaluation of the pion decay constant
$f_\pi$ and the scaling properties of the nucleon to $\rho$ mass
ratio.  The first topic is of greatest interest since we can 
compute the pion decay constant using two independent methods, 
one of which depends directly on the residual mass determined in 
Section~\ref{sec:m_res}.  The close agreement between these
two approaches provides a very important consistency check of the
analysis and results presented in this paper.


\subsection{Calculation of $f_\pi$}

In the conventional continuum formulation, the pion decay constant
$f_\pi$ is defined through the equation
\begin{equation}
\Big\langle 0 \Big| i\overline\psi\gamma^\mu_M\gamma_5 t^a \psi
\Big| \pi^b(\vec p)\Big\rangle 
~ \equiv ~ f_\pi 
{p^\mu \delta^{a,b}\over \sqrt{2E_\pi(\vec p)}}
\label{eq:f_pi_cont}
\end{equation}
where the fields $\psi$ and $\overline\psi$ are interpreted as conventional,
Hilbert space quark operators and the pion state obeys the
non-covariant normalization $\langle \pi(\vec p\,')|\pi(\vec p)\rangle
= \delta^3(\vec p -\vec p\,')$.  To be concrete we adopt the Minkowski
metric $g^{\mu\nu}$ with signature $(-1,+1,+1,+1)$ and a Minkowski
gamma matrix convention in which $\gamma^0_M$ is anti-hermitian and
$\{\gamma^\mu_M, \gamma^\nu_M\}=2g^{\mu\nu}$.  With this normalization,
$f_\pi \approx 130$ MeV.

Following the usual methods of lattice gauge theory, we evaluate 
matrix elements of the two-quark operator appearing in 
Eq.~\ref{eq:f_pi_cont} with the Euclidean time dependence resulting 
from use of the evolution operator $e^{-H t}$ where $H$ is the QCD 
Hamiltonian.  Thus, we choose to evaluate
\begin{equation}
f_\pi^2{m_\pi\over 2}e^{-m_\pi t} = \lim_{T\rightarrow\infty}
 {Tr\Big\{e^{-H(T-t)}\;\int d^3 x\;i\bar\psi\gamma^0_M\gamma_5 
        t^a \psi(\vec x,t)\;
  e^{-Ht}\;i\bar\psi\gamma^0_M\gamma_5 t^a \psi(\vec 0,0)\Big\} 
               \over Tr\big\{e^{-HT}\big\} }
\label{eq:euclid_cont}
\end{equation}
where no sum over the flavor index $a$ is intended and the time $t$ is
assumed sufficiently large that only the pion intermediate state
contributes.

The continuum operators in Eq.~\ref{eq:euclid_cont} are easily
represented as lattice, Euclidean-space expressions once the usual
transition to a Euclidean-space path integral has been performed.  In
particular, the operators $\psi(\vec x)$ and $\psibar(\vec x) =
\psi^\dagger(\vec x)\gamma^0$ are replaced by the Grassmann variables
$q(\vec x,t)$ and $\qbar(\vec x,t)$ respectively.  Thus, we extract
$f_\pi^2$ from the usual Euclidean correlation function:

\begin{equation}
{f_\pi^2 \over Z_A^2} {m_\pi\over 2}e^{-m_\pi t} 
=\Big\langle\int d^3 x\;\qbar \gamma^0\gamma^5 t^a q(\vec x, t)
                      \;\qbar \gamma^0\gamma^5 t^a q(\vec 0, 0) 
                                     \Big\rangle
\label{eq:euclid_lat_AA}
\end{equation}
where now Euclidean gamma matrices appear, obeying $\{\gamma^\mu,
\gamma^\nu\}=2\delta^{\mu\nu}$.  Here we have introduced the Grassmann
variables $q$ and $\qbar$ defined earlier in this paper so the axial
current appearing in Eq.~\ref{eq:euclid_lat_AA} is explicitly
constructed from the five-dimensional quark fields $\Psi$ and $\Psibar$
restricted to the $s=0$ and $s=L_s-1$ walls.  This ``local'' current,
$A_\mu^a$ is not conserved in the full five-dimensional theory so the
factor $Z_A$ appearing on the left hand side of
Eq.~\ref{eq:euclid_lat_AA} is needed to make a connection to the
continuum axial current.

The conserved current ${\cal A}^a_\mu$ defined in
Eq.\ \ref{eq:axial_cc} must approach the corresponding, partially
conserved continuum current with unit normalization, when the continuum
limit is taken.  Thus, to order $a^2$, the low energy matrix elements
of ${\cal A}_\mu^a$ and $A_\mu^a$ must be proportional:  ${\cal
A}_\mu^a = Z_A A_\mu^a$.  While we have computed $f_\pi$ using the
local current $A_\mu^a$ we have also compared that current to the
partially conserved domain wall axial current ${\cal A}_\mu^a$,
allowing an accurate determination $Z_A$.

In addition to the procedure just described, there is a second,
independent method that we have used to compute $f_\pi$.  Here we use
the Ward-Takahashi identity to relate ${\cal A}_\mu^a$ and the
pseudo-scalar density $J^a_5$:
\begin{equation}
  \Delta_\mu {\cal {A}}_\mu^a(x) \approx 2(m_f + m_{\rm res}){J}^a_5(x)
\end{equation}
an expression valid for low energy matrix amplitudes .  In particular, 
we have replaced the
usual midpoint term in the exact identity of Eq.~\ref{eq:axial_cc_diverg} 
by its low energy limit:  $2 \mres J^a_5(x)$.  Thus, we can also obtain
$f_\pi$ from the correlation function:
\begin{equation}
-{f_\pi^2 \over (m_f + m_{\rm res})^2 } {m_\pi^3 \over 8}e^{-m_\pi t} 
   = \Big\langle\int d^3 x\;\qbar\gamma^5 t^a q(\vec x, t)
                          \;\qbar\gamma^5 t^a q(\vec 0, 0)
                            \Big\rangle
\label{eq:euclid_lat_PP}
\end{equation}
where again no sum over the flavor index $a$ is intended.  This formula
involves no renormalization factors but requires knowledge of the
residual mass $m_{\rm res}$ induced by mixing between the walls.  Thus,
a comparison of the values for $f_\pi$ obtained from
Eqs.~\ref{eq:euclid_lat_AA} and \ref{eq:euclid_lat_PP} provides a
critical test of the analysis presented in this paper.

We will now discuss these two calculations of $f_\pi$ in detail.  To 
measure the value for the renormalization factor $Z_A$, we compare
the amplitudes of two-point functions $C(t)$ and $L(t)$ defined as
\begin{eqnarray}
C(t+1/2) &=& \sum_{\vec x} \langle {\cal A}^a_0(\vec x ,t)
         \; \pi^a(\vec{0}, 0) \rangle \nonumber \\ 
L(t)     &=& \sum_{\vec x} \langle A^a_0 (\vec x,t) 
         \; \pi^a(\vec{0}, 0) \rangle.
\label{eq:Ct_Lt}
\end{eqnarray}

The $1/2$ in the argument of $C(t+1/2)$ in Eq.~\ref{eq:Ct_Lt} comes
from the fact the conserved axial current ${\cal A}^a_{\mu}(x)$ is not the
current at lattice site $x$ but instead the current carried by the link
between $x$ and $x+\hat{\mu}$.  We take appropriate arithmetic averages
to solve the problem that $C(t+1/2)$ and $L(t)$ are not at the same
location.  To avoid as much systematic error as possible, we define
$Z_A(t)$ as
\begin{equation}
Z_A(t) = \frac{1}{2} 
        \left\{ \frac{C(t+1/2)+C(t-1/2)}{2\ L(t)} + 
        \frac{2\ C(t+1/2)}{L(t)+L(t+1)} \right\}.
\label{eq:Za_def}
\end{equation}
For $t \gg a^{-1}$, $C(t)/L(t)$ behaves like a constant which
can be identified with $Z_A$.   Both terms in Eq.~\ref{eq:Za_def}
estimate this value without $O(a)$ error.  The average of these two,
incorporated in Eq.~\ref{eq:Za_def}, further eliminates a portion
of the $O(a^2)$ error.

\begin{figure}
\epsfxsize=\hsize
\epsfbox{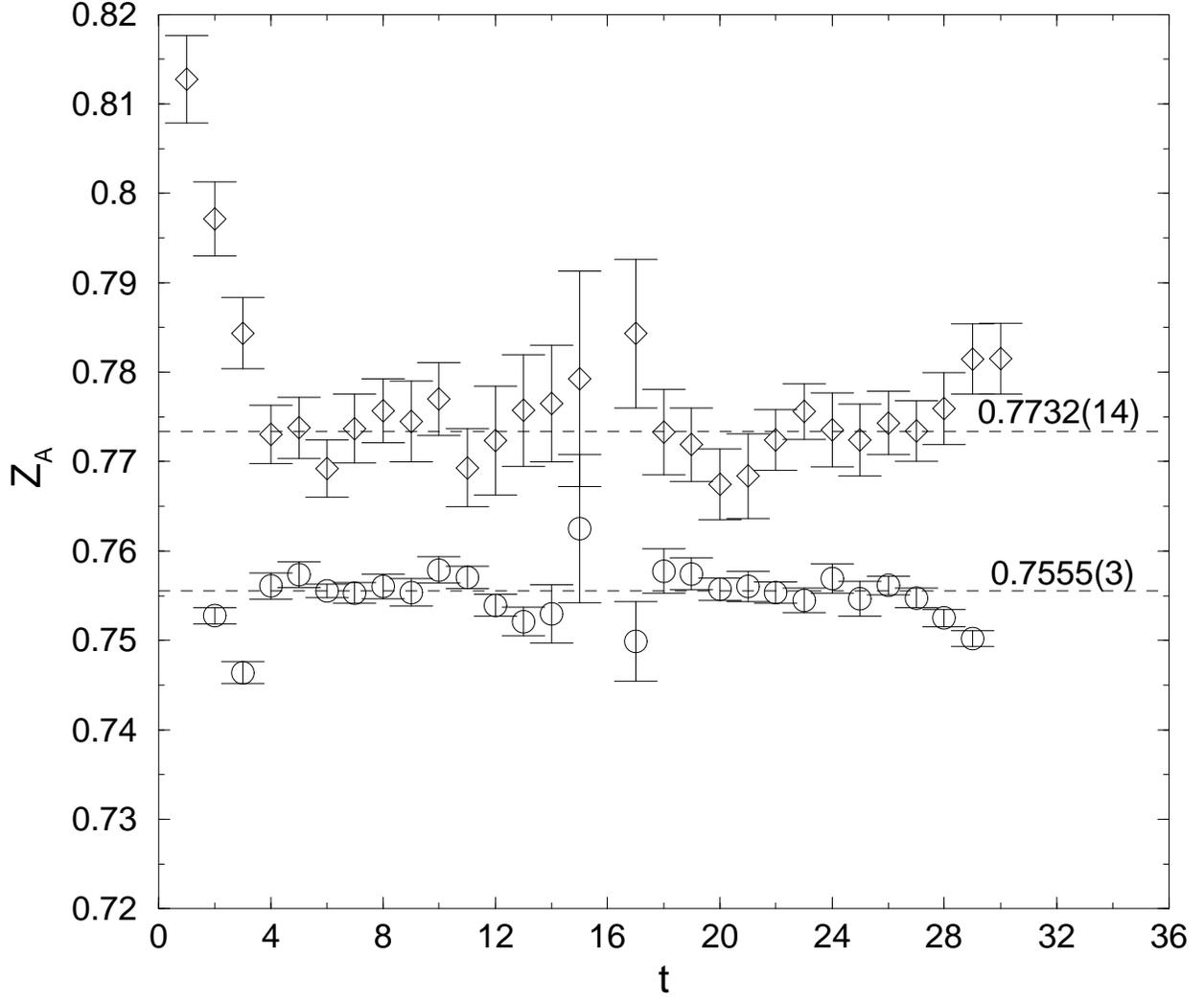}
\caption{The renormalization constant $Z_A$ obtained for a $16^3 \times
32$ lattice with $L_s$=16 and $\beta=6.0$ ($\circ$) and that for a $8^3
\times 32$ lattice with $L_s$=48 and $\beta=5.7$ ($\Diamond$).  The
labels for the horizontal lines are the averages, with jackknife
errors, over the ranges $4 \leq t \leq 14$ and $18 \leq t \leq 28$.}
\label{fig:za}
\end{figure}

Figure \ref{fig:za} shows the ratio $Z_A(t)$ defined in
Eq.~\ref{eq:Za_def} for both a $16^3 \times 32$ lattice with $L_s=16$,
and $\beta=6.0$ as well as the same quantity for a $8^3 \times 32$
lattice with $L_s=48$ and $\beta=5.7$.  We determine the value for the
renormalization factor $Z_A$ by calculating the average over two ranges
of $t$: $4 \leq t \leq 14$ and $18 \leq t \leq 28$, chosen to avoid the
largest time separation $t \sim 16$ where the errors are quite large.
A jackknife error is determined, to compensate for possible correlation
between the numerator and denominator in Eq.~\ref{eq:Za_def}.

The results for $Z_A$ at $\beta=6.0$, $16^3 \times 32$, $M_5=1.8$ and
with different values of $L_s$ are listed in Table \ref{tab:Z_A_b6_0}.
The data shows little $L_s$ dependence, as should be expected.  Figure
\ref{fig:za}, also shows our result of $Z_A=0.7732(14)$ found for
the $8^3 \times 32$ lattice with $\beta=5.7$, $L_s=48$, $M_5=1.65$,
$m_f=0.02$.

The results for the amplitudes for the axial vector current correlator
and the pseudoscalar density correlator at $\beta=5.7$ and $6.0$ are
given in
Tables~\ref{tab:fpi_832_b57k165ns32}-\ref{tab:fpi_1632_b60k18ns24}.
They are obtained from the point-source correlators using a conventional
2-parameter fit with the pion masses extracted concurrently.  We also
list in the same tables the results for $f_\pi$ as a function of $m_f$
determined from the corresponding correlators with the help of $Z_A$
and $m_{\rm res}$ (Tables~\ref{tab:m_res_b6_0}-\ref{tab:m_res_b5_7}).
These values of $f_\pi$ have been converted to physical units using the
measured $\rho$ mass discussed in Section~\ref{sec:sim_results},
extrapolated to the chiral limit $m_f+\mres = 0$.

Next, we use a linear fit in $m_f$ to evaluate $f_\pi$ for two values
of $m_f$.  To obtain a value of $f_\pi$ close to that for the physical
pion, we go to the chiral limit $m_f+\mres=0$.  For $f_K$ we choose
for $m_f$ that value which gives $m_\pi/m_\rho = 0.645$.  In determining
$f_\pi$ for the physical pion state, we did not attempt to use a value
of $m_f$ giving the physical value for the ratio $m_\pi/m_\rho=0.18$
since we do not adequately know the $m_f$ dependence of this ratio 
in the relevant region.  These linear fit parameters as well as
the resulting values for $f_\pi$ and $f_K$ are summarized in
Table~\ref{tab:fpi_extrap}.  The errors given in the tables are
obtained from the jackknife method.

\begin{figure}
\epsfxsize=\hsize
\epsfbox{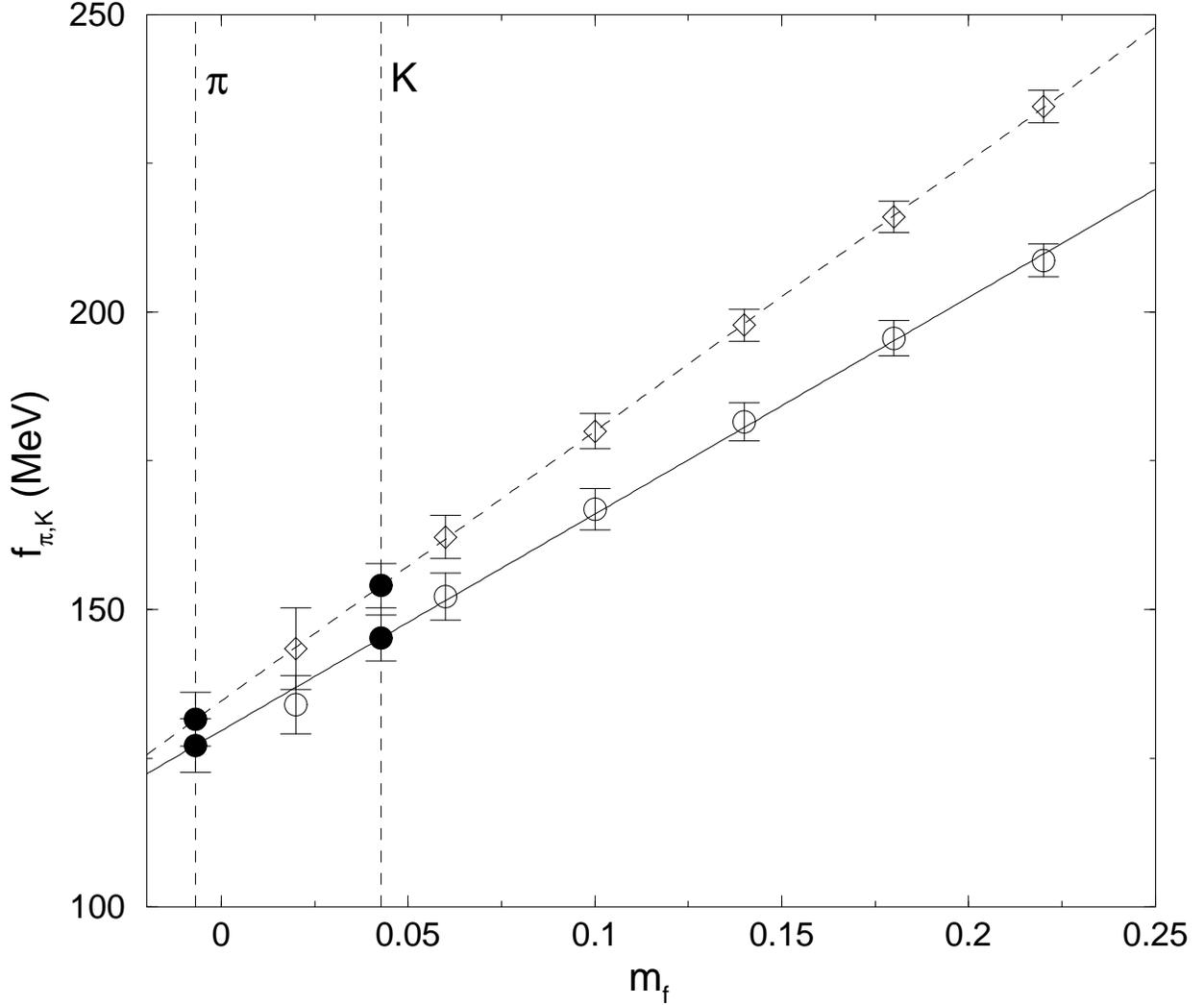}
\caption{Results for $f_\pi$ at $\beta=5.7$ with a $8^3 \times 32$ 
lattice and $L_s=48$ plotted as a function of $m_f$.  The open circles 
are obtained from the axial vector current correlator, while the open 
diamonds are obtained from the pseudoscalar density correlator. We also 
show the linear fits which are used to determine our estimate for 
$f_\pi$ and $f_K$.  The vertical dashed lines identify the values for 
$m_f$ which locate the chiral limit, $m_f=-\mres$ and give the physical 
ratio for $m_K/m_\rho$.  The solid symbols represent the extrapolations to 
the point $m_f=-\mres$ and interpolations to the kaon mass.}
\label{fig:fpi_b5_7_8nt32_ls48}
\end{figure}

\begin{figure}
\epsfxsize=\hsize
\epsfbox{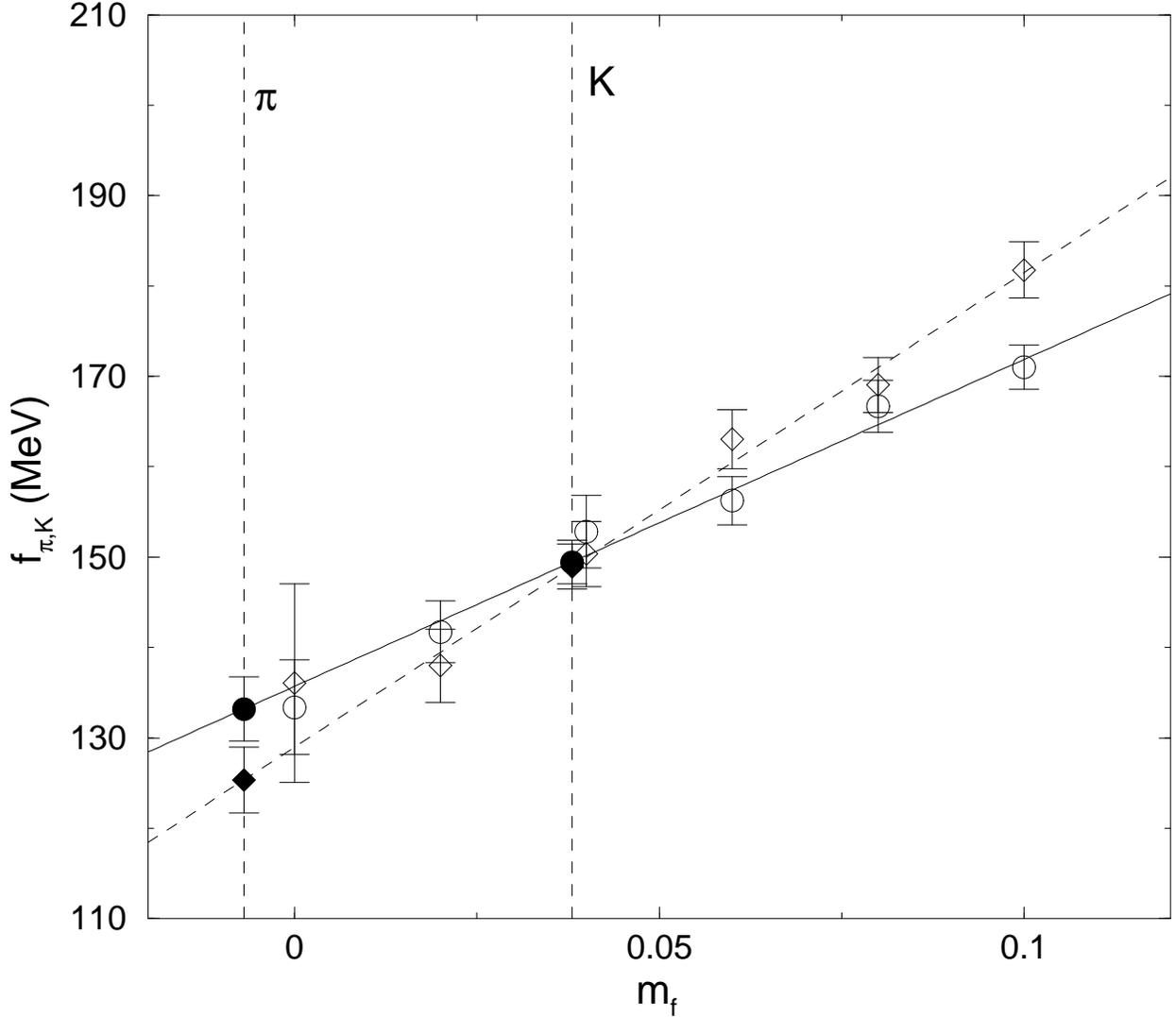}
\caption{Results for $f_\pi$ at $\beta=5.7$ with a $16^3 \times 32$
lattice and $L_s=48$ plotted as a function of $m_f$. The open circles
are obtained from the axial vector current correlator, while the open
diamonds are obtained from the pseudoscalar density correlator. We also
show the linear fits which are used to determine our estimate for
$f_\pi$ and $f_K$.  The fits are done to the points with
$m_f=0.02-0.10$.  The vertical dashed lines identify the values for
$m_f$ which locate the chiral limit, $m_f=-\mres$ and give the physical 
ratio for $m_K/m_\rho$.  The solid symbols represent the extrapolations to 
the point $m_f=-\mres$ and interpolations to the kaon mass.}
\label{fig:fpi_b5_7_16nt32_ls48}
\end{figure}

Figure \ref{fig:fpi_b5_7_8nt32_ls48} shows the values for $f_\pi$
at $\beta=5.7$, $8^3 \times 32$, $L_s=48$ as a function of $m_f$ and
the linear fits through all the $m_f$ points. The results obtained from
the pseudoscalar-pseudoscalar correlator are higher than those from the
axial-axial correlator.  The two linear fits give $f_\pi=127(4)\ {\rm
MeV}$, $f_K=145(4)\ {\rm MeV}$ and $f_\pi=132(4)\ {\rm MeV}$,
$f_K=154(4)\ {\rm MeV}$ respectively.  When the lattice volume is
increased to $16^3 \times 32$ (Figure
\ref{fig:fpi_b5_7_16nt32_ls48}), the difference between the linear
fits from the two methods becomes smaller.   We obtain
$f_\pi=133(4)\ {\rm MeV}$, $f_K=149(2)\ {\rm MeV}$ and
$f_\pi=125(4)\ {\rm MeV}$, $f_K=149(2)\ {\rm MeV}$ from the two
correlators.  The values for $f_\pi(m_f)$ obtained from the two methods
should agree for all values of $m_f$ since they are related by a
Ward-Takahashi identity that should become exact in the continuum
limit.  Presumably the visibly different slopes seen in
Figures~\ref{fig:fpi_b5_7_8nt32_ls48} and
\ref{fig:fpi_b5_7_16nt32_ls48} are the result of order $a^2$
errors.

We also calculate $f_\pi$ at a weaker coupling. Figure
\ref{fig:fpi_b6_0_2parm} shows our results for $\beta=6.0$, $16^3
\times 32$, $L_s=16$ on 85 configurations. The two independent
calculations give very consistent results. We have $f_\pi=137(11)\ {\rm
MeV}$, $f_K=156(8)\ {\rm MeV}$ from the axial vector current correlator
and almost the same values from the pseudoscalar correlator.  Our
results for $f_\pi$ at both $\beta=5.7$ and $6.0$ agree well with the
experimental value of $\approx 130\ {\rm MeV}$, while the values for
$f_K$ may be somewhat smaller than the experimental value of $\approx
160\ {\rm MeV}$ as is expected from quenched chiral perturbation theory
arguments~\cite{Bernard:1992mk} and naive scaling 
considerations~\cite{Butler:1994zx}.  Note, in Table~\ref{tab:fpi_extrap}
we also list $f_K/f_\pi$ with jackknifed errors for the ratio.  Here the 
statistical errors are now well below the systematic errors that might be 
expected in the $m_f+\mres \rightarrow 0$ extrapolation.  The values shown 
for $f_K/f_\pi$ agree on the 5\% level between methods of determination and 
different lattice spacings but are systematically below the experimental 
value of 1.21.

This same analysis was done using the amplitudes calculated from the
point-source correlators but making a 1-parameter fit using the
pre-determined pion masses computed from the more mass accurate
measurements based on the wall-source correlators. This method gives
consistent results with slightly smaller errors.  The results are not
listed here.

The reasonable agreement of our domain wall results with the
experimental values and their relative insensitivity to $a$ is 
encouraging.  Similar results were obtained at $\beta=6.0$ for smaller 
values of $L_s$ with somewhat larger errors in Ref.~\cite{Aoki:2000pc}.  
Of special interest here is the comparison that we make between the
two methods of determining $f_\pi$, which is done here for the
first time.  As can be seen from Eq.~\ref{eq:euclid_lat_PP}, the 
determination of $f_\pi$ from $\corrpp$ depends directly on 
$\mres$.  Thus, the comparison of these two methods is an important check
of our understanding of the chiral properties of the domain wall
formulation.  The ratio of these two quantities extrapolated to the
point $m_f+\mres=0$ provides an interesting figure of merit for the present 
calculation.  We find $(f_\pi)_{PP} / (f_\pi)_{AA} = 1.00(10)$ and
0.96(10) for $L_s=16$ and 24 respectively.  However, if instead
of the values of $\mres$ given in Table~\ref{tab:m_res_b6_0},
we use the $x$-intercepts -0.0031(7) and -0.0030(9) quoted earlier
and obtained from the $\corraa$ values of $m_\pi^2$, we find
$(f_\pi)_{PP} / (f_\pi)_{AA} =1.20(12)$ for both the $L_s=16$ and 24
cases.  While these ratios each differ from 1 by two standard deviations,
they are independent calculations and demonstrate the good chiral
properties of domain wall fermions.


\subsection{Continuum limit of $m_N/m_\rho$}

Here we combine the hadron mass results tabulated in
Section~\ref{sec:sim_results} to examine the behavior of the nucleon to
$\rho$ mass ratio as $\beta$ varies between 5.7 to 6.0.  First we
evaluate $m_N$ and $m_\rho$ in the limit $m_f+\mres=0$.  We did
not use the value of $m_f$ which gives the physical ratio, 
$m_\pi/m_\rho = 0.18$ for the reasons outlined in the previous section.  
In Table~\ref{tab:mN_over_mrho} we give the resulting mass ratios as
well as the lattice spacings in physical units as determined from 
$m_\rho$ evaluated at $m_f+\mres=0$.  Note, no contribution to the 
quoted error for these mass ratios arising from the uncertainty 
in this choice of $m_f$ has been included.

The relatively large variation of $m_N/m_\rho$ with $\beta$ suggests
that the errors shown in Table~\ref{tab:mN_over_mrho} may be
underestimated and makes a simple $a^2$ extrapolation to the continuum
limit somewhat uncertain.  Nevertheless the result of such an
extrapolation to $a \rightarrow 0$ is $m_N/m_\rho = 1.37(5)$.  Perhaps
more interesting is a comparison with similar quantities computed at
comparable lattice spacings and volumes using Wilson and staggered 
fermions.   For staggered fermions at $\beta=6.0$ on comparable volumes, 
one finds \cite{Chen:1997jj,Chen:1996thesis} $m_N/m_\rho = 1.47(3)$, a
somewhat larger and less physical value than the 1.42(4) and 1.38(4) 
results obtained here for $L_s=16$ and 24.  However, this comparison 
is made somewhat ambiguous by the 
significant finite size effects seen in staggered calculations 
when going from our $16^3 \times 32$ to larger volumes 
\cite{Gottlieb:1997hy}.  For Wilson fermions, as reported in 
Ref.~\cite{Butler:1994em}, one deduces $m_N/m_\rho = 1.37(2)$ 
by linear interpolation between the $\beta=5.93$ and 6.17 values 
presented, a number remarkably close to our domain wall value.  
When comparing these values, it is important to recall that our 
$8^3$ and $16^3$ spatial volumes are not yet infinite and, as 
discussed in Section~\ref{sec:sim_results}, corrections on the
order of a few percent are expected.


\subsection{Determining the chiral condensate $\qbq$}

Finally we use the results presented earlier to estimate the size of
the chiral condensate $\qbq$.  Naively, one might expect that a physical
value for $\qbq$ could be easily identified in Table~\ref{tab:qbq_phenom}
as the $m_f$-independent term $a_0$, defined in Eq.~\ref{eq:qbq_phenom}.  
This quantity represents a simple extrapolation of $\qbq(m_f)$ from 
large mass down to the point $m_f=0$.  Given the volume independence 
seen for the parameter $a_0$ when comparing the $\beta=5.7$, $8^3$ and 
$16^3$ volumes in Table~\ref{tab:qbq_phenom}, it is natural to expect 
that such a choice minimizes the sensitivity to the finite-volume zero 
mode effects that give rise to the more singular $a_{-1}$ term.

However, there are other issues that must be addressed.  Perhaps
most obvious is the fact that the point $m_f=0$ is not the physical chiral
limit because the effects of $\mres$ have been ignored.  This is easily
remedied by using the slope $a_1$, to extrapolate to the physical point
$m_f+\mres=0$.  The resulting estimate of $\qbq$, in lattice units, is 
given as the fourth column in Table~\ref{tab:pbp_GMOR}.  However, because
$\qbq$ is a quadratically divergent quantity, we cannot expect that all
the chiral symmetry breaking effects of domain wall mixing are removed
by this choice of $m_f$.  In contrast to many physical quantities, 
$\qbq$ receives contributions from energy scales much larger than 
those for which $\mres$ represents the complete effect of chiral
symmetry breaking.  Thus, we should expect additional contributions to
$\qbq$ of order $e^{-\alpha L_s}/a^3  \sim \mres/a^2$.  This is born
out in Table~\ref{tab:pbp_GMOR} where we see that the differences between
$\qbq$ for the two different values of $L_s$ at a given $\beta$ are of
the same order as the difference between the values with and without
the extrapolation to $m_f=-\mres$.

This unwanted $\sim \mres/a^2$ contribution to $\qbq$ can only be 
controlled by explicitly taking the limit $L_s \rightarrow \infty$.  We do 
not at present have the numerical results to permit such an extrapolation.  
Therefore, we will use the $\beta=6.0$, $L_s=24$ result as our best 
approximation to such a limit and interpret the difference between 
the $L_s=16$ and 24 values as an estimate of the systematic error, 
$\approx 10\%$.  Given the value of 
$Z_S({\rm \overline{MS}, 2GeV}) = 0.619(25)$ for $\beta=6.0$ 
quoted earlier and the results for the lattice spacing in physical 
units in Table~\ref{tab:mN_over_mrho}, we can determine $\qbq$ in 
physical units.  The results for $L_s=16$ and 24, $(245(7){\rm MeV})^3$
and $(256(8){\rm MeV})^3$, are included in Table~\ref{tab:pbp_GMOR},
where only the statistical error is displayed.  The agreement between
these numbers and phenomenological estimates of the chiral condensate 
is satisfactory, for example the value of 
${1 \over 2}(\bar u u + \bar d d)_{\rm \overline{MS}, 1 GeV}
=(229 \pm 9 {\rm MeV})^3$ 
obtained in Ref.~\cite{Narison:1995hz}.  Note the $e^{-\alpha L_s}/a^3 $
uncertainty present in our calculation does not have an analogue in
the properly regulated continuum theory.  While 
$\langle\overline\psi\psi\rangle$ does contain a quadratically 
divergent piece in the continuum theory, this is eliminated for 
the chirally symmetric choice $m_{\rm quark}=0$.  This choice is
not available in a domain wall fermion calculation without taking
the $L_s \rightarrow \infty$ limit.  Of course, the other lattice 
methods for directly computing $\langle\overline\psi\psi\rangle$ have
equal or more severe difficulties.

Finally it is interesting to compare the $\beta=5.7$ and $\beta=6.0$
results for $\qbq$.  Since we do not at present have a reliable determination
of the needed renormalization constant, $Z_S$, for the stronger $\beta=5.7$
coupling, we do not attempt to quote a physical value.  However, the
ratio of the unrenormalized lattice numbers given in Table~\ref{tab:pbp_GMOR} 
for $\qbq_{(L_s=32,\,\beta=5.7)}/\qbq_{(L_s=24,\,\beta=6.0)}= 4.8(2)$ is 
reasonably consistent with the ratio expected from naive scaling 
$a^3_{(L_s=32,\,\beta=5.7)}/a^3_{(L_s=24,\,\beta=6.0)} = 7.4(4)$.

Given the values now determined for $\qbq$, $f_\pi$ and quark mass, it
is natural to test the degree to which the Gell-Mann-Oakes-Renner
relation \cite{Gell-Mann:1968rz}
\begin{equation}
f_\pi^2 {m_\pi^2 \over 48( m_f + \mres)} = -\qbq
\label{eq:GMOR}
\end{equation}
is obeyed.  However, the form of this equation reveals an
important difficulty.  At what value of $m_f$ should the ratio 
$m_\pi^2/( m_f + \mres)$ be computed?  In full QCD, this ratio 
becomes a constant for small quark mass.  As we have seen earlier,
this is not the case in the quenched approximation where one expects 
non-linearities.  

We might try to determine the proper treatment of
these non-linearities by returning to the underlying equation, Eq.~\ref{eq:int_ward_tak_id}, from which the Gell-Mann-Oakes-Renner
relation is derived.  However, this is somewhat complex.  Both sides
of this original equation have a mass dependence which comes from 
the contribution of the pion pole term and other physical states, all
influenced by the quenched approximation, as well as the quadratically
divergent terms in $\qbq$ and the contact term in $\chi_{\pi\pi}$.
Thus, while the underlying Eq.~\ref{eq:int_ward_tak_id} will be 
obeyed exactly in our calculation, there is considerable ambiguity
in deciding how to extract a quenched generalization of the
Gell-Mann-Oakes-Renner relation, Eq.~\ref{eq:GMOR}.

Here we will simply compare the right- and left-hand-sides of 
Eq.~\ref{eq:GMOR} by replacing the ratio $m_\pi^2/( m_f + \mres)$
by the slope $b$ obtained at larger masses, $m_f \ge 0.01$ and given in 
Tables~\ref{tab:extrap_832_b5_7}, \ref{tab:extrap_1632_b5_7} and
\ref{tab:extrap_1632_b6_0}.  The results from the left hand side
of Eq.~\ref{eq:GMOR} are given in Table~\ref{tab:pbp_GMOR}.  Given
our uncertainty in determining $\qbq$ and the significant 
non-linearities we see in $m_\pi^2$, the agreement seen between
the fourth and fifth columns in Table~\ref{tab:pbp_GMOR} is within 
our errors.


\section{Conclusions}
\label{sec:conclusions}

We have presented the results of detailed studies of quenched lattice
QCD using domain wall fermions, with particular attention paid to the
lowest order chiral symmetry breaking effects of finite $L_s$ and the
behavior of the theory for small values of $m_f$.  A major difficulty
in studying the small $m_f$ behavior of the theory is the presence of
topological near-zero modes which are unsuppressed in the quenched
theory.  These are a result of the improved character of the domain
wall fermion operator, which has an Atiyah-Singer index at finite lattice
spacing and $L_s \rightarrow \infty$.  However, these zero-modes 
complicate the quenched theory and demonstrate that the quenched 
approximation is considerably more treacherous than might
have been originally expected.  We have seen how these modes produce 
the expected $1/m_f$
divergence in $\qbq$ for small $m_f$ and distort correlation functions
used to measure the properties of the pion.  By working on larger
volumes, we found that the effects of these modes were dramatically
reduced, as expected.  We were then able to see a common pion mass
determined from different correlators.

We have determined or constrained the value for the residual mass,
$\mres$, which enters the effective quark mass for low-energy physics
as $m_{\rm eff} = m_f + \mres$, a number of ways and found good
agreement.  The residual mass was measured from the extra, finite $L_s$
term in the divergence of the conserved axial current and from
the explicitly determined lowest eigenvalues of the
hermitian domain wall fermion operator.  These two determinations agree
within errors.  We have also determined the difference in $\mres$ for
two values of $L_s$ from the pion mass and find this agrees with the
results from our explicitly calculated $\mres$.  Lastly, agreement for
$f_\pi$ as calculated from axial vector and pseudoscalar correlators
requires knowledge of $\mres$ and the agreement serves as a further
check.

While our data for weaker couplings does not clearly demonstrate that
$\mres \rightarrow 0$, we have seen it fall to 1 MeV for $L_s = 48$
at $\beta = 6.0$.  For $L_s = 16$, a practical value for studies
of low energy hadronic physics and matrix elements, $\mres$ has a
value of 3.87(16) MeV, roughly 1/30 of the strange quark mass.  Even at
stronger couplings, where the lattice spacing is $a^{-1} \sim 1$ GeV,
we have measured $\mres$ to also be about 1/14 of the strange quark
mass, although here $L_s = 48$ was required.  Thus, we see domain
wall fermions producing the desired light surface states with small
mixing, even for relatively strong couplings.

We have measured hadron masses and $f_\pi$ for lattice scales 1 GeV $ <
a^{-1} < $ 2 GeV and have studied scaling in this region.  Our
determinations of $f_\pi$ involve not only $\mres$ as mentioned above
but also the measurement of the Z-factor for the local axial current.
We find $f_\pi/m_\rho$ evaluated at the $m_f + \mres =0$ point
to be scaling very well, while for $m_N/m_\rho$ the scaling violations
may be at the 6\% level.  However, scaling seems at least as good
as that seen for staggered fermions at similar lattice spacings and
similar to that found for Wilson fermions with a clover term 
\cite{Gottlieb:1997hy}.  This is in accord with general expectations 
that finite lattice spacing errors will enter domain wall fermion 
amplitudes at $O(a^2)$ \cite{Blum:1997mz,Kikukawa:1997md}.

Our results demonstrate that quenched domain wall fermions do exhibit
the desired good chiral properties, even at relatively strong
couplings.  The residual quark mass effects, which break the 
full global symmetries to leading order in $a$, can be eliminated 
by an appropriate choice of $m_f$, so that low energy physics 
should be well described by an effective theory with the 
continuum global symmetries.  Quenched chiral logarithm
effects may appear for quenched domain wall fermion simulations, as
they do for other fermion formulations, but present no new difficulties.  
For large enough volumes, the effects of topological near-zero 
modes are suppressed and the small $m_f$ region can be investigated.  
For larger values of $m_f$, where these zero mode effects
are suppressed by the quark mass, one has a formulation of lattice QCD
with the full global symmetries realized to order $a^2$ and 
an effective quark mass of $m_f + \mres$.  Thus, the domain wall 
formulation provides a powerful new tool which can be used, even 
within the quenched approximation, to study many of the outstanding 
problems in particle and nuclear physics for which chiral 
symmetry plays an important role.

Note added: After this paper was essentially complete, the recent
work of the CP-PACS collaboration became available \cite{Khan:2000iv}.  
The reader is referred to this paper for another discussion of some of the 
topics presented here.


\section*{Acknowledgments}

The authors would like to acknowledge useful discussion with Shoichi Sasaki,
Thomas Manke, T.~D.~Lee, Robert Edwards and Mike Creutz.   We thank RIKEN, 
Brookhaven National Laboratory and the U.S. Department of Energy for 
providing the facilities essential for the completion of this work.
Finally, we acknowledge with gratitude the Ritz diagonalization program
provided by Robert Edwards.

The numerical calculations at $\beta = 5.7$ and 6.0 were done on the
400 Gflops QCDSP computer \cite{Chen:1998cg} at Columbia University and
the 600 Gflops QCDSP computer at the RIKEN-BNL Research Center
\cite{Mawhinney:2000fx}.  The $\beta = 5.85$ results were calculated at
NERSC.  We acknowledge the MILC collaboration \cite{MILC:1991aa} whose 
software provided the basis for this $\beta=5.85$ calculation.
This research was supported in part by the DOE under grant \#
DE-FG02-92ER40699 (Columbia), in part by the NSF under grant \#
NSF-PHY96-05199 (Vranas), in part by the DOE under grant
DE-AC02-98CH10886 (Soni-Dawson), in part by the RIKEN-BNL Research
Center (Blum-Wingate-Ohta) and in part by the Max-Kade Foundation
(Siegert).



%
\begin{table}
\caption{Simulation parameters for the quenched results obtained using 
the hybrid Monte Carlo method.  The mass ranges referred to are specified
in Table~\protect{\ref{tab:mass_ranges}}.  The spectrum column contains 
the number of configurations on which hadron mass measurements were 
performed while the $\qbq$ column shows the number of configurations 
used to compute the chiral condensate.  Finally, within parenthesis 
in the last column we specify the number of random noise sources 
(hits) that were used in each of these $\qbq$ measurements.  All of 
the calculations described in this table used the domain wall height 
parameter $M_5=1.65$.}
\label{tab:run_parameters_hmc}

\begin{tabular}{cccccc}
$\beta$ & $L^3 \times N_t$ & $L_s$ & mass range & spectrum& $\qbq$(hits) \\
\hline
5.70    & $8^3 \times 32$ & 10 & 0.02--0.20 & 87      & 87(1)        \\
\hline
5.70    & $8^3 \times 32$ & 16 & 0.02--0.20 & 67      & 67(1)        \\
\hline
5.70    & $8^3 \times 32$ & 24 & 0.02--0.22 & 84      & 84(1)        \\
\hline
5.70    & $8^3 \times 32$ & 32 & 0.02--0.22 & 94      & 94(1)        \\
\hline
5.70    & $8^3 \times 32$ & 48 & 0.02--0.22 & 81      & 81(1)        \\
\end{tabular}
\end{table}


\begin{table}
\caption{Simulation parameters for the quenched results obtained using the
heatbath method.  The column labeled sweeps records the number of Monte Carlo
sweeps between successive measurements.   The remaining notation is the
same as that used in Table~\protect{\ref{tab:run_parameters_hmc}}.}
\label{tab:run_parameters_hb}
\begin{tabular}{cccccccc}
$\beta$
     & $L^3 \times N_t$ & $L_s$& $M_5$&sweeps& mass range & spectrum& $\qbq$(hits) \\
\hline
5.70 & $8^3 \times 32$  & 32 & 1.65 & 2000 & 0.00025--0.008,
                                            0.00--0.04         & 0   & 210(1)\\
\hline
5.70 & $8^3 \times 32$  & 32 & 1.65 & 2000 & 0.02, 0.04         & 184 & 0     \\
\hline
5.70 & $8^3 \times 32$  & 48 & 1.65 & 200  & 0.02--0.22         & 46  & 0     \\
\hline
5.70 & $8^3 \times 32$  & 48 & 1.65 & 5000 & 0.001--0.01        &  0  & 42(3)\\
     &                  &    &      &      & 0.02--0.22         & 42  & 42(3)\\
\hline
5.70 & $8^3 \times 32$  & 48 & 1.65 & 2000 & 0.0, 0.04          & 336 & 0     \\
\hline
5.70 & $8^3 \times 32$  & 48 & 1.65 & 2000 & 0.00025--0.008,
                                            0.00--0.04         & 0   & 141(1)\\
\hline
5.70 & $8^3 \times 32$  & 64 & 1.65 & 5000  & 0.02--0.22        & 76  & 0     \\
\hline
5.70 & $16^3 \times 32$ & 24 & 1.65 & 5000  & 0.02--0.22        & 73  & 70(1) \\
\hline
5.70 & $16^3 \times 32$ & 32 & 1.65 & 2000  & 0.00025--0.008,
                                          0.00--0.04        & 0   & 60(1) \\
\hline
5.70 & $16^3 \times 32$ & 48 & 1.65 & 2000  & 0.001--0.01       & 0   & 10(3) \\
     &                  &    &      &       & 0.02--0.22        & 61  & 10(3) \\
\hline
5.70 & $16^3 \times 32$ & 48 & 1.65 & 2000  & 0.0, 0.04         & 106 & 0     \\
\hline
5.70 & $16^3 \times 32$ & 48 & 1.65 & 2000  & 0.02, 0.06, 0.1   & 45 & 0     \\
\hline
5.70 & $16^3 \times 32$ & 48 & 1.65 & 2000  & 0.08              & 106 & 0     \\
\hline
5.85 & $12^3 \times 64^\dagger$  
                        & 20 & 1.9  & 5000   & 0.025--0.075     &100  & 0     \\
\hline
5.85 & $8^3 \times 32$  & 32 & 1.65 & ---    & 0.001--0.01,
                                           0.02--0.10           & 0   & 200 \\
\hline
5.85 & $16^3 \times 32$ & 32 & 1.65 & 1000   & 0.001--0.01,
                                              0.02--0.10        & 0   &  91(1) \\
\hline
6.0  & $16^4$           & 16 & 1.8  & 2000  & ---               & 32  & ---     \\
\hline
6.0  & $16^3 \times 32$ & 12 & 1.8  & 2000  & 0.02              & 56  & 0     \\
\hline
6.0  & $16^3 \times 32$ & 16 & 1.8  & 5000  & 0.01--0.04        & 85  & 85(1)\\
\hline
6.0  & $16^3 \times 32$ & 16 & 1.8  & 2000  & 0.000             & 216 & 0     \\
\hline
6.0  & $16^3 \times 32$ & 16 & 1.8  & 2000  & 0.001             & 229 & 0     \\
\hline
6.0  & $16^3 \times 32$ & 16 & 1.8  & 2000  & 0.02              & 56  & 0     \\
\hline
6.0  & $16^3 \times 32$ & 16 & 1.8  & 2000  & 0.00025--0.008,
                                          0.00--0.04        & 0   & 120(1)\\
\hline
6.0  & $16^3 \times 64^\dagger$  
                        & 16 & 1.8  & 2000  & 0.01--0.05,
                                          0.075, 0.1, 0.125 & 98  & 0     \\
\hline
6.0  & $16^3 \times 32$ & 24 & 1.8  & 2000  & 0.01--0.04        & 76  & 0     \\
\hline
6.0  & $16^3 \times 32$ & 24 & 1.8  & 2000  & 0.02              & 56  & 0     \\
\hline
6.0  & $16^3 \times 32$ & 24 & 1.8  & 2000  & 0.00025--0.008
                                          0.00--0.04        & 0   & 110(1)\\
\hline
6.0  & $16^3 \times 32$ & 32 & 1.8  & 2000  & 0.02              & 72   & 0    \\
\hline
6.0  & $16^3 \times 32$ & 48 & 1.8  & 2000  & 0.02              & 64   & 0    \\
\end{tabular}
$\dagger$  This extent of 64 in the time direction was achieved by ``doubling''
$N_t=32$ lattice configurations in the time direction so the resulting 
gauge field background has an unphysical $t \rightarrow t+32$ periodicity. 
\end{table}

\begin{table}
\caption{\label{tab:mass_ranges}
Here we list the explicit masses that are included in the various mass
ranges referred to in Tables~\protect{\ref{tab:run_parameters_hmc}} 
and~\protect{\ref{tab:run_parameters_hb}}.}
\begin{tabular}{ll}
mass range     & \multicolumn{1}{c}{mass values} \\
\hline
0.00--0.04     & 0.00, 0.005, 0.01, 0.015, 0.02, 0.025, 0.03, 0.035 and 0.04\\
0.00025--0.008 & 0.00025, 0.0005, 0.001, 0.002, 0.004 and 0.008             \\
0.001--0.01    & 0.001, 0.004, 0.007 and 0.01                               \\
0.01--0.04     & 0.01, 0.015, 0.02, 0.025, 0.03, 0.035 and 0.04             \\
0.01--0.05     & 0.01, 0.02, 0.03, 0.04, 0.05                               \\
0.02--0.20     & 0.02, 0.04, 0.06, 0.08, 0.10, 0.12, 0.14, 0.16, 0.18 and 0.20\\
0.02--0.22     & 0.02, 0.06, 0.10, 0.14, 0.18 and 0.22                      \\
0.02--0.10     & 0.02, 0.04, 0.06, 0.08 and 0.10                            \\
0.025--0.075   & 0.025, 0.0325, 0.05, 0.0625 and 0.075			    \\
\end{tabular}
\end{table}
\begin{table}
\caption{\label{tab:mhad_832_b57k165ns10} 
Results for hadron masses at $\beta=5.7$, $8^3\times 32$, $M_5=1.65$, 
$L_s=10$ from 87 configurations.}
\begin{tabular}{cccc}
$m_f$ & $m_\pi$ & $m_\rho$ & $m_N$ \\ \hline
0.02 & 0.528(12) & 0.82(3) & 1.24(6) \\ 
0.04 & 0.604(12) & 0.874(19) & 1.32(4) \\ 
0.06 & 0.676(14) & 0.922(18) & 1.40(3) \\ 
0.08 & 0.744(12) & 0.971(15) & 1.48(3) \\ 
0.10 & 0.808(10) & 1.017(12) & 1.56(2) \\ 
0.12 & 0.869(9) & 1.063(10) & 1.63(2) \\ 
0.14 & 0.929(8) & 1.108(9)  & 1.705(18) \\ 
0.16 & 0.987(5) & 1.153(8)  & 1.779(15) \\ 
0.18 & 1.043(5) & 1.199(7)  & 1.853(14) \\ 
0.20 & 1.097(5) & 1.244(6)  & 1.928(13) \\ 
\end{tabular}
\end{table}
\begin{table}
\caption{\label{tab:mhad_832_b57k165ns16} 
Results for hadron masses at
$\beta=5.7$, $8^3\times 32$, $M_5=1.65$, $L_s=16$ from 67
configurations.}
\begin{tabular}{cccc}
$m_f$ & $m_\pi$ & $m_\rho$ & $m_N$ \\ \hline
0.02 & 0.483(18) & 0.87(12) & 1.25(14) \\ 
0.04 & 0.562(12) & 0.89(6) & 1.36(10) \\ 
0.06 & 0.635(7) & 0.93(4) & 1.44(8) \\ 
0.08 & 0.702(5) & 0.98(2) & 1.52(7) \\ 
0.10 & 0.768(5) & 1.019(18) & 1.59(6) \\ 
0.12 & 0.831(5) & 1.064(15) & 1.65(4) \\ 
0.14 & 0.892(6) & 1.109(13) & 1.71(4) \\ 
0.16 & 0.954(5) & 1.153(12) & 1.78(3) \\ 
0.18 & 1.012(5) & 1.199(10) & 1.85(3) \\ 
0.20 & 1.072(6) & 1.243(10) & 1.93(2) \\ 
\end{tabular}
\end{table}
\begin{table}
\caption{\label{tab:mhad_832_b57k165ns24} 
Results for hadron masses at
$\beta=5.7$, $8^3\times 32$, $M_5=1.65$, $L_s=24$ from 84
configurations.}
\begin{tabular}{cccc}
$m_f$ & $m_\pi$ & $m_\rho$ & $m_N$ \\ \hline
0.02 & 0.44(2) & 0.84(8) & 1.4(4) \\ 
0.06 & 0.613(12) & 0.94(3) & 1.31(6) \\ 
0.10 & 0.756(8) & 1.016(15) & 1.50(3) \\ 
0.14 & 0.882(5) & 1.102(11) & 1.662(16) \\ 
0.18 & 0.999(4) & 1.189(9) & 1.817(12) \\ 
0.22 & 1.108(4) & 1.278(8) & 1.959(11) \\ 
\end{tabular}
\end{table}
\begin{table}
\caption{\label{tab:mhad_832_b57k165ns32} 
Results for hadron masses at
$\beta=5.7$, $8^3\times 32$, $M_5=1.65$, $L_s=32$. The results for
$m_f=0.02$ are obtained from 278 configurations, those for
$m_f=0.04$ are from 184 configurations, while the others are
from 94 configurations.}
\begin{tabular}{cccc}
$m_f$ & $m_\pi$ & $m_\rho$ & $m_N$ \\ \hline
0.02 & 0.405(6) & 0.83(5) & 1.17(11) \\
0.04 & 0.502(5) & 0.87(4) & 1.16(5) \\ 
0.06 & 0.595(9) & 0.92(2) & 1.36(8) \\ 
0.10 & 0.743(7) & 0.995(16) & 1.50(3) \\ 
0.14 & 0.872(6) & 1.082(11) & 1.66(2) \\ 
0.18 & 0.991(5) & 1.178(8) & 1.822(16) \\ 
0.22 & 1.104(4) & 1.274(7) & 1.970(15) \\ 
\end{tabular}
\end{table}
\begin{table}
\caption{\label{tab:mhad_832_b57k165ns48} 
Results for hadron masses at
$\beta=5.7$, $8^3\times 32$, $M_5=1.65$, $L_s=48$. The results for
$m_f=0.04$ are obtained from 335 configurations, while the others
are from 169 configurations.  One of the original 336 configurations had 
eigenvalues very close to zero requiring nearly 11,000 conjugate gradient 
iterations to converge. The resulting pion propagator was so large 
as to dominate the average of the $\corrpp$ correlator for the 
$m_f=0.0$ case.  We omitted this single configuration from this 
analysis.}
\begin{tabular}{cccc}
$m_f$ & $m_\pi$ & $m_\rho$ & $m_N$ \\ \hline
0.02 & 0.374(10) & 0.99(12) & 1.07(19) \\ 
0.04 & 0.490(4) & 0.87(2) & 1.22(6) \\ 
0.06 & 0.580(7) & 0.95(3) & 1.38(5)   \\ 
0.10 & 0.730(5) & 1.016(13) & 1.52(2) \\ 
0.14 & 0.860(4) & 1.098(8) & 1.658(17) \\ 
0.18 & 0.981(4) & 1.184(6) & 1.809(14) \\ 
0.22 & 1.093(4) & 1.272(5) & 1.962(12) \\ 
\end{tabular}
\end{table}
\begin{table}
\caption{\label{tab:mhad_832_b57k165ns64} 
Results for hadron masses at
$\beta=5.7$, $8^3\times 32$, $M_5=1.65$, $L_s=64$ from 76
configurations.}
\begin{tabular}{cccc}
$m_f$ & $m_\pi$ & $m_\rho$ & $m_N$ \\ \hline
0.02 & 0.364(14) & 0.98(17) & 1.2(3) \\ 
0.06 & 0.563(8) & 0.96(5) & 1.18(4) \\ 
0.10 & 0.719(8) & 1.01(2) & 1.42(3) \\ 
0.14 & 0.854(7) & 1.089(12) & 1.62(2) \\ 
0.18 & 0.978(6) & 1.176(9) & 1.784(17) \\ 
0.22 & 1.097(5) & 1.265(7) & 1.938(14) \\ 
\end{tabular}
\end{table}
\begin{table}
\caption{\label{tab:mhad_1632_b57k165ns24} 
Results for hadron masses at
$\beta=5.7$, $16^3\times 32$, $M_5=1.65$, $L_s=24$ from 73
configurations.}
\begin{tabular}{cccc}
$m_f$ & $m_\pi$ & $m_\rho$ & $m_N$ \\ \hline
0.02 & 0.412(5) & 0.85(4) & 1.17(9) \\ 
0.06 & 0.597(4) & 0.909(12) & 1.29(4) \\ 
0.10 & 0.743(4) & 0.990(9) & 1.47(2) \\ 
0.14 & 0.873(4) & 1.082(6) & 1.642(14) \\ 
0.18 & 0.994(3) & 1.175(5) & 1.800(16) \\ 
0.22 & 1.105(4) & 1.265(4) & 1.953(16) \\ 
\end{tabular}
\end{table}
\begin{table}
\caption{\label{tab:mhad_1632_b57k165ns48_b} 
Results for hadron masses at
$\beta=5.7$, $16^3\times 32$, $M_5=1.65$, $L_s=48$. The results for
$m_f= 0.02, 0.04, 0.06, 0.08$ and 0.10 are obtained from 106 
configurations, while the others are from 61 configurations.}
\begin{tabular}{cccc}
$m_f$ & $m_\pi$ & $m_\rho$ & $m_N$ \\ \hline
0.02 & 0.383(4) & 0.88(5) & 1.00(4) \\
0.04 & 0.482(4) & 0.86(2) & 1.23(3) \\ 
0.06 & 0.577(2) & 0.918(18) & 1.260(15) \\
0.08 & 0.650(3) & 0.951(9) & 1.375(16) \\ 
0.10 & 0.729(2) & 0.989(9) & 1.449(10) \\
0.14 & 0.865(3) & 1.083(7) & 1.623(14) \\ 
0.18 & 0.986(3) & 1.173(5) & 1.786(12) \\ 
0.22 & 1.097(4) & 1.264(5) & 1.942(12) \\ 
\end{tabular}
\end{table}

\begin{table}
\caption{\label{tab:mhad_1232_b585k19ns20}
Results for hadron masses at
$\beta=5.85$, $12^3\times32$, $M_5=1.9$, $L_s=20$, 100 config.}
\begin{tabular}{cccc}
$m_f$ & $m_\pi$ & $m_\rho$ & $m_N$ \\ \hline
0.0250  & 0.359(4) & 0.627(15) & 0.844(24) \\
0.0375  & 0.426(3) & 0.650(10) & 0.919(16) \\
0.0500  & 0.483(3) & 0.675(8)  & 0.985(13) \\
0.0625  & 0.536(3) & 0.706(6)  & 1.044(10) \\
0.0750  & 0.585(3) & 0.738(5)  & 1.106(9) \\
\end{tabular}
\end{table}

\begin{table}
\caption{\label{tab:mhad_1632_b60k18ns16} 
Results for hadron masses at
$\beta=6.0$, $16^3\times32$, $M_5=1.8$, $L_s=16$ from 85
configurations.}
\begin{tabular}{cccc}
$m_f$ & $m_\pi$ & $m_\rho$ & $m_N$ \\ \hline
0.010 & 0.203(3) & 0.442(10) & 0.621(30) \\ 
0.015 & 0.239(3) & 0.451(7)  & 0.648(21) \\
0.020 & 0.270(3) & 0.462(6)  & 0.668(15) \\
0.025 & 0.298(3) & 0.475(5)  & 0.686(12) \\
0.030 & 0.324(2) & 0.488(5)  & 0.706(10) \\
0.035 & 0.348(2) & 0.502(4)  & 0.729(9) \\
0.040 & 0.371(2) & 0.515(4)  & 0.752(9) \\
\end{tabular}
\end{table}

\begin{table}
\caption{\label{tab:mnd_1632_b60k18ns16} 
Results for hadron masses with nondegenerate valence quarks at
$\beta=6.0$, $16^3\times32$, $M_5=1.8$, $L_s=16$ from 98
configurations.}
\begin{tabular}{cccc}
$m_f(1)$ & $m_f(2)$ & $m_\pi$ & $m_\rho$ \\ \hline
0.010 & 0.020 & 0.238(2) & 0.441(12) \\
0.030 & 0.020 & 0.298(2) & 0.471(8) \\
0.040 & 0.020 & 0.325(2) & 0.487(7) \\
0.050 & 0.020 & 0.349(2) & 0.502(7) \\
0.075 & 0.020 & 0.406(2) & 0.538(6) \\
0.100 & 0.020 & 0.457(2) & 0.574(5) \\
0.125 & 0.020 & 0.505(2) & 0.608(4) \\
\end{tabular}
\end{table}

\begin{table}
\caption{\label{tab:mhad_1632_b60k18ns24} 
Results for hadron masses at
$\beta=6.0$, $16^3\times32$, $M_5=1.8$, $L_s=24$ from 76
configurations.}
\begin{tabular}{cccc}
$m_f$ & $m_\pi$ & $m_\rho$   & $m_N$ \\ \hline
0.010 & 0.201(5) & 0.423(11) & 0.664(43) \\
0.015 & 0.236(4) & 0.441(9)  & 0.644(25) \\
0.020 & 0.267(3) & 0.458(7)  & 0.653(18) \\
0.025 & 0.295(3) & 0.473(6)  & 0.674(15) \\
0.030 & 0.320(3) & 0.487(5)  & 0.698(12) \\
0.035 & 0.345(3) & 0.502(5)  & 0.723(11) \\
0.040 & 0.368(2) & 0.516(4)  & 0.748(10) \\
\end{tabular}
\end{table}

%

\begin{table}
\caption{\label{tab:extrap_832_b5_7}
Valence extrapolations ($a + b m_f$) for $m_\pi^2$,
$m_\rho$ and $m_N$ at $\beta = 5.7$, $8^3 \times 32$, $M_5=1.65$.
The fitting ranges used are described in 
Section~\protect{\ref{ssec:hdm_mass_rslts}}.}
\begin{tabular}{ccccc}
mass      & $L_s$ & $a$ & $b$ & $\chi^2/{\rm dof}$ \\ \hline
$m_\pi^2$ & 10 & 0.201(17) & 4.55(17) & 5.35 \\ 
$m_\rho$  & 10 & 0.792(19) & 2.26(6) & 0.55 \\ 
$m_N$     & 10 & 1.18(4) & 3.74(17) & 0.13 \\ 
$m_\pi^2$ & 16 & 0.147(32) & 4.32(36) & 1.59 \\ 
$m_\rho$  & 16 & 0.798(24) & 2.23(10) & 0.17 \\ 
$m_N$     & 16 & 1.21(8) & 3.60(29) & 0.28  \\ 
$m_\pi^2$ & 24 & 0.093(16) & 4.82(12) & 1.47 \\ 
$m_\rho$  & 24 & 0.786(21) & 2.24(7)  & 0.34 \\ 
$m_N$     & 24 & 1.17(3)   & 3.60(14) & 1.01 \\ 
$m_\pi^2$ & 32 & 0.076(10) & 4.83(12) & 4.81 \\ 
$m_\rho$  & 32 & 0.753(26) & 2.36(11) & 0.43 \\ 
$m_N$     & 32 & 1.14(5) & 3.78(22) & 1.14 \\ 
$m_\pi^2$ & 48 & 0.042(8)  & 4.90(9) & 0.012 \\ 
$m_\rho$  & 48 & 0.790(13) & 2.18(5) & 0.50 \\ 
$m_N$     & 48 & 1.13(4)   & 3.79(15) & 0.73 \\ 
$m_\pi^2$ & 64 & 0.039(11) & 4.75(21) & 6.28 \\ 
$m_\rho$  & 64 & 0.756(22) & 2.28(8) & 0.65 \\ 
$m_N$     & 64 & 1.03(6) & 4.06(22) & 3.23 \\ 
\end{tabular}
\end{table}


\begin{table}
\caption{\label{tab:extrap_1632_b5_7}
Valence extrapolations ($a + b m_f$) for $m_\pi^2$,
$m_\rho$ and $m_N$ at $\beta = 5.7$, $16^3 \times 32$, $M_5=1.65$.
The fitting ranges used are described in 
Section~\protect{\ref{ssec:hdm_mass_rslts}}.}
\begin{tabular}{ccccc}
mass      & $L_s$ & $a$ & $b$ & $\chi^2/{\rm dof}$ \\ \hline
$m_\pi^2$ & 24 & 0.072(6)  & 4.83(9) & 7.26 \\ 
$m_\rho$  & 24 & 0.764(12) & 2.27(5) & 1.10 \\ 
$m_N$     & 24 & 1.10(3)  & 3.87(15) & 0.69 \\ 
$m_\rho$  & 48 & 0.775(18) & 2.20(7) & 1.04 \\ 
$m_N$     & 48 & 1.03(4) & 4.13(17) & 3.58 \\ 
\end{tabular}
\end{table}


\begin{table}
\caption{\label{tab:extrap_1232_b5_85}
Valence extrapolations ($a + b m_f$) for  $m_\pi^2$,
$m_\rho$ and $m_N$ at $\beta = 5.85$, $12^3 \times 32$,
$M_5=1.9$. The masses from all five values of $m_f$ are included
in the fits.}
\begin{tabular}{ccccc}
mass      & $L_s$ & $a$ & $b$  & $\chi^2/{\rm dof}$ \\ \hline
$m_\pi^2$ & 20    & 0.024(3)   & 4.27(4)   & 2.8 \\ 
$m_\rho$  & 20    & 0.549(14)  & 2.50(17)  & 0.88 \\ 
$m_N$     & 20    & 0.74(2)    & 4.9(3)    & 0.81 \\ 
\end{tabular}
\end{table}


\begin{table}
\caption{\label{tab:extrap_1632_b6_0}
Valence extrapolations ($a + b m_f$) for $m_\pi^2$,
$m_\rho$ and $m_N$ at $\beta = 6.0$, $16^3 \times 32$, $M_5=1.8$.
The fitting ranges used are described in 
Section~\protect{\ref{ssec:hdm_mass_rslts}}.}
\begin{tabular}{ccccc}
mass      & $L_s$ & $a$ & $b$ & $\chi^2/{\rm dof}$ \\ \hline
$m_\pi^2$ & 16 & 0.0098(20) & 3.14(9) & 0.029 \\ 
$m_\rho$  & 16 & 0.404(8) & 2.78(11) & 0.48 \\ 
$m_N$     & 16 & 0.566(21) & 4.66(29) & 0.34 \\ 
$m_\pi^2$ & 24 & 0.0094(26) & 3.09(7) & 0.32 \\ 
$m_\rho$  & 24 & 0.400(10)  & 2.86(12) & 0.38 \\ 
$m_N$     & 24 & 0.546(19) & 5.05(33) & 0.65 \\ 
\end{tabular}
\end{table}



\begin{table}
\caption{\label{tab:qbq_phenom} 
Results for fits of $\qbq$ to the form given in Eq.\
\ref{eq:qbq_phenom}.}
\begin{tabular}{cccccc}
$L^3 \times N_t \times L_s$ & $\beta$ & $a_{-1}$ & $a_0$ & $a_1$ &
$\dmqbq$
  \\ \hline
$8^3 \times 32 \times 32$ & 5.7 & $ 6.0(6) \times 10^{-6} $ &
  $ 1.76(3) \times 10^{-3} $ & $ 6.53(4) \times 10^{-2}$ &
  $ 4.0(4) \times 10^{-3}$ \\
$8^3 \times 32 \times 48$ & 5.7 & $ 6.8(7) \times 10^{-6} $ &
  $ 1.92(5) \times 10^{-3} $ & $ 6.04(14) \times 10^{-2}$ &
  $ 1.7(2) \times 10^{-3}$ \\
$16^3 \times 32 \times 32$ & 5.7 & $ 2.5(4) \times 10^{-6} $ &
  $ 1.86(2) \times 10^{-3} $ & $ 6.53(2) \times 10^{-2}$ &
  $ 9.3(9) \times 10^{-3}$ \\
$16^3 \times 32 \times 16$ & 6.0 & $ 1.0(1) \times 10^{-6} $ &
  $ 3.87(8) \times 10^{-4} $ & $ 8.64(1) \times 10^{-2}$ &
  $ 5.6(3) \times 10^{-4}$ \\
$16^3 \times 32 \times 24$ & 6.0 & $ 9.1(10) \times 10^{-7} $ &
  $ 3.62(9) \times 10^{-4} $ & $ 8.64(2) \times 10^{-2}$ &
  $ 1.1(1) \times 10^{-4}$ \\
\end{tabular}
\end{table}



\begin{table}
\caption{\label{tab:mpi_mf_small} 
Results for $m_\pi$ from different correlators for $m_f < 0.01$.
There should be a single value for the pion mass, determined at
asymptotically large times, irrespective of the correlator used.
Since the correlators give different masses for the fitting
ranges used, the localized topological near-zero mode effects
are important.  Here PP, AA and PP+SS represent the correlators
$\corrpp$, $\corraa$ and $\corrpp+\corrss$ respectively.}
\begin{tabular}{ccccccc}
$\beta$ & V & $L_s$ & correlator & $m_f$ & $m_\pi$ & $\chi^2/{\rm dof}$
  \\ \hline
5.7  & $8^3 \times 32$ & 48 & PP & 0.0 & 0.273(39) & $1.3 \pm 0.6$ \\
5.7  & $8^3 \times 32$ & 48 & AA & 0.0 & 0.197(11) & $1.5 \pm 1.0$\\
5.7  & $8^3 \times 32$ & 48 & PP+SS & 0.0 & 0.128(21) & $1.8 \pm 0.8$\\
\hline
5.7  & $16^3 \times 32$ & 48 & PP & 0.0 & 0.200(5) & $1.9 \pm 0.9$\\
5.7  & $16^3 \times 32$ & 48 & AA & 0.0 & 0.193(7) & $1.6 \pm 0.9$\\
5.7  & $16^3 \times 32$ & 48 & PP+SS & 0.0 & 0.191(7) & $1.1 \pm 0.6$\\
\hline
6.0  & $16^3 \times 32$ & 16 & PP & 0.0 & 0.151(8) & $0.6 \pm 0.6$\\
6.0  & $16^3 \times 32$ & 16 & AA & 0.0 & 0.098(7) & $0.6 \pm 0.5$\\
6.0  & $16^3 \times 32$ & 16 & PP+SS & 0.0 & 0.017(43) & $1.0 \pm 0.6$\\
\hline
6.0  & $16^3 \times 32$ & 16 & PP & 0.001 & 0.141(6) & $0.9 \pm 0.6$\\
6.0  & $16^3 \times 32$ & 16 & AA & 0.001 & 0.118(6) & $1.1 \pm 0.9$\\
6.0  & $16^3 \times 32$ & 16 & PP+SS & 0.001 & 0.082(11) & $0.5 \pm 0.4$
  \\
\end{tabular}
\end{table}



\begin{table}
\caption{\label{tab:m_res_b6_0} 
Results for residual mass at $\beta=5.85$ and 6.0.  The $\beta=5.85$
calculation was performed on a $12^3\times32$ with anti-periodic
boundary conditions, $M_5=1.9$ and the ratio $R(t)$ from 
Eq.~\protect{\ref{eq:mres_ratio}} averaged over the time range
$6 \le t \le 26$.  The $\beta = 6.0$ calculation, described in the
text, was performed on a $16^3 \times 32$ lattice with $M_5=1.8$.}
\begin{tabular}{ccccc}
$\beta$& $L_s$ & $m_f$ & \# of config. & $m_{\rm res}$ \\ \hline
5.85   & 20 & 0.05 & 100& 0.00281(8) \\
6.0    & 12 & 0.02 & 56 & 0.00239(6) \\
6.0    & 16 & 0.02 & 56 & 0.00124(5) \\
6.0    & 24 & 0.02 & 56 & 0.00059(4) \\
6.0    & 32 & 0.02 & 72 & 0.00044(4) \\
6.0    & 48 & 0.02 & 64 & 0.00027(3) \\
\end{tabular}
\end{table}


\begin{table}
\caption{\label{tab:m_res_b5_7} 
Results for residual masses at $\beta = 5.7$, $M_5=1.65$.  }
\begin{tabular}{ccccc}
Lattice size     & $L_s$ & $m_f$ & \# of config. & $m_{\rm res}$ \\ \hline
$8^3 \times 32$  & 32    & 0.02  & 184 & 0.0106(2)   \\
$8^3 \times 32$  & 32    & 0.04  & 184 & 0.0105(2)   \\
$8^3 \times 32$  & 48    & 0.04  & 335 & 0.00688(13) \\
$16^3 \times 32$ & 48    & 0.02  &  50 & 0.0072(9) \\
$16^3 \times 32$ & 48    & 0.04  &  50 & 0.0071(4) \\
$16^3 \times 32$ & 48    & 0.06  &  50 & 0.0066(6) \\
$16^3 \times 32$ & 48    & 0.08  &  50 & 0.0065(5) \\
$16^3 \times 32$ & 48    & 0.10  &  50 & 0.0063(4) \\
\end{tabular}
\end{table}


\begin{table}
\caption{\label{tab:q_chiral_log} 
Results for fits to the form predicted for a quenched chiral logarithm,
Eq.\ \ref{eq:q_chiral_log} for the $16^3 \times 32$ simulations
at $\beta = 5.7$ with $L_s = 48$.  For comparison, we have also 
included the column $x_{\rm intcpt}$ which gives the $x$ intercepts
predicted by the simple linear fits of 
\protect{Eq.~\ref{eq:large_vol_0.02_0.08_fit}} in the heavier mass 
range $0.02 \le m_f \le 0.08$.  Here PP, AA and PP+SS represent the 
correlators $\corrpp$, $\corraa$ and $\corrpp+\corrss$ respectively.}
\begin{tabular}{cccccc}
Correlator & $a_0$ & $a_1$ & $a_2$ & $\chi^2/{\rm dof}$ & $x_{\rm intcpt}$ \\ 
\hline

 PP     & 4.7(3) & 0.0085(7)  & -0.008(28) & $1.9 \pm 4.7$ & -0.0092(12)  \\
 AA     & 4.1(3) & 0.0073(10) & -0.07(4)   & $3.6 \pm 2.4$ & -0.0108(7)   \\
 PP+SS  & 4.3(3) & 0.0075(11) & -0.05(4)   & $4.8 \pm 3.2$ & -0.0104(7)
\end{tabular}
\end{table}



\begin{table}
\caption{\label{tab:Z_A_b6_0} 
Results for $Z_A$ at $\beta = 6.0$, $16^3 \times 32$, $M_5=1.8$.  }
\begin{tabular}{cccc}
$L_s$ & $m_f$ & configurations & $Z_A$ \\ \hline
12 & 0.02 & 56 & 0.7560(3) \\
16 & 0.02 & 56 & 0.7555(3) \\
24 & 0.02 & 56 & 0.7542(3) \\
32 & 0.02 & 72 & 0.7535(3) \\
48 & 0.02 & 64 & 0.7533(3) \\
\end{tabular}
\end{table}


\begin{table}
\caption{\label{tab:fpi_832_b57k165ns32}
Results for the correlator amplitude $A$ and $f_\pi$
at $\beta=5.7$, $8^3\times 32$, $M_5=1.65$, $L_s=32$, 94 configurations.
Parameters with subscript $AA$ are obtained from the axial
vector current correlator. Parameters with subscript $PP$ are obtained
from the pseudoscalar density correlator. $Z_A = 0.7732$ and
$m_{\rm res} = 0.0105$ are used in the $f_\pi$ calculation as
described in the text.}
\begin{tabular}{ccccc}
$m_f$ & $A_{AA}$ & ${(f_\pi)}_{AA}$ & $A_{PP}$ & ${(f_\pi)}_{PP}$ \\
\hline
0.02 & 0.0077(11) & 158(9) & 0.176(18) & 145(8) \\
0.06 & 0.0132(13) & 172(8) & 0.145(10) & 173(6) \\
0.10 & 0.0197(15) & 188(7) & 0.143(7) & 194(4) \\
0.14 & 0.0267(15) & 202(5) & 0.151(6) & 214(4) \\
0.18 & 0.0346(15) & 216(4) & 0.164(5) & 233(3) \\
0.22 & 0.0432(16) & 229(4) & 0.180(5) & 252(3) \\
\end{tabular}
\end{table}


\begin{table}
\caption{\label{tab:fpi_832_b57k165ns48}
Results for the correlator amplitude $A$ and $f_\pi$ at $\beta=5.7$,
$8^3\times 32$, $M_5=1.65$, $L_s=48$, 169 configurations. 
Parameters with subscript $AA$ are obtained from the axial
vector current correlator. Parameters with subscript $PP$ are obtained
from the pseudoscalar density correlator. $Z_A = 0.7732$ and
$m_{\rm res} = 0.00688$ are used in the $f_\pi$ calculation as
described in the text.}
\begin{tabular}{ccccc}
$m_f$ & $A_{AA}$ & ${(f_\pi)}_{AA}$ & $A_{PP}$ & ${(f_\pi)}_{PP}$ \\
\hline
0.02 & 0.0058(5) & 134(5) & 0.202(13) & 143(7) \\
0.06 & 0.0114(7) & 152(4) & 0.144(7) & 162(4) \\
0.10 & 0.0172(8) & 167(3) & 0.138(5) & 180(3) \\
0.14 & 0.0240(9) & 182(3) & 0.146(4) & 198(3) \\
0.18 & 0.0317(10) & 196(3) & 0.159(4) & 216(2) \\
0.22 & 0.0403(12) & 209(3) & 0.176(5) & 234(3) \\
\end{tabular}
\end{table}


\begin{table}
\caption{\label{tab:fpi_1632_b57k165ns48}
Results for the correlator amplitude $A$ and $f_\pi$
at $\beta=5.7$, $16^3\times 32$, $M_5=1.65$, $L_s=48$. 
Parameters with subscript $AA$ are obtained from the axial
vector current correlator. Parameters with subscript $PP$ are obtained
from the pseudoscalar density correlator. $Z_A = 0.7732$ and
$m_{\rm res} = 0.00688$ are used in the $f_\pi$ calculation as
described in the text.
}
\begin{tabular}{ccccc}
$m_f$ & $A_{AA}$ & ${(f_\pi)}_{AA}$ & $A_{PP}$ & ${(f_\pi)}_{PP}$ \\ \hline
0.00 & 0.0028(2) & 133(5) & 0.34(4) & 136(11) \\
0.02 & 0.0063(3) & 142(3) & 0.180(10) & 138(4) \\
0.04 & 0.0092(5) & 153(4) & 0.140(6) & 150(4) \\
0.06 & 0.0115(4) & 156(3) & 0.139(6) & 163(3) \\
0.08 & 0.0147(5) & 167(3) & 0.127(4) & 169(3) \\
0.10 & 0.0174(5) & 171(2) & 0.136(5) & 182(3) \\
0.14 & 0.0248(10) & 188(4) & 0.147(8) & 201(5) \\
0.18 & 0.0326(12) & 202(4) & 0.161(7) & 220(4) \\
0.22 & 0.0411(15) & 214(4) & 0.178(7) & 239(4) \\
\end{tabular}
\end{table}


\begin{table}
\caption{\label{tab:fpi_1632_b60k18ns16}
Results for the correlator amplitude $A$ and $f_\pi$
at $\beta=6.0$, $16^3\times 32$, $M_5=1.8$, $L_s=16$, 85 configurations.
Parameters with subscript $AA$ are obtained from the axial
vector current correlator. Parameters with subscript $PP$ are obtained
from the pseudoscalar density correlator. $Z_A = 0.7555$ and
$m_{\rm res} = 0.00124$ are used in the $f_\pi$ calculation as
described in the text.}
\begin{tabular}{ccccc}
$m_f$ & $A_{AA}$ & ${(f_\pi)}_{AA}$ & $A_{PP}$ & ${(f_\pi)}_{PP}$ \\
\hline
0.010 & 0.00100(13) & 144(10) & 0.050(6) & 149(9) \\
0.015 & 0.00132(14) & 153(8) & 0.040(4) & 151(8) \\
0.020 & 0.00161(15) & 159(8) & 0.036(4) & 156(8) \\
0.025 & 0.00189(16) & 164(7) & 0.034(3) & 161(8) \\
0.030 & 0.00216(17) & 168(7) & 0.032(3) & 166(8) \\
0.035 & 0.00243(18) & 172(6) & 0.032(3) & 171(8) \\
0.040 & 0.00269(18) & 175(6) & 0.031(3) & 176(7) \\
 \end{tabular}
  \end{table}


\begin{table}
\caption{\label{tab:fpi_1632_b60k18ns24}
Results for the correlator amplitude $A$ and $f_\pi$
at $\beta=6.0$, $16^3\times 32$, $M_5=1.8$, $L_s=24$, 76 configurations.
Parameters with subscript $AA$ are obtained from the axial
vector current correlator. Parameters with subscript $PP$ are obtained
from the pseudoscalar density correlator. $Z_A = 0.7542$ and
$m_{\rm res} = 0.00059$ are used in the $f_\pi$ calculation as
described in the text.}
\begin{tabular}{ccccc}
$m_f$ & $A_{AA}$ & ${(f_\pi)}_{AA}$ & $A_{PP}$ & ${(f_\pi)}_{PP}$ \\
\hline
0.010 & 0.00089(14) & 138(11) & 0.043(6) & 134(10) \\
0.015 & 0.00110(11) & 141(7) & 0.034(4) & 138(8) \\
0.020 & 0.00130(11) & 144(6) & 0.031(3) & 144(6) \\
0.025 & 0.00152(11) & 148(5) & 0.029(2) & 149(6) \\
0.030 & 0.00176(12) & 153(5) & 0.0280(18) & 154(5) \\
0.035 & 0.00201(14) & 157(5) & 0.0275(16) & 159(5) \\
0.040 & 0.00227(15) & 162(5) & 0.0272(15) & 164(5) \\
\end{tabular}
\end{table}

\begin{table}
\caption{\label{tab:fpi_extrap}
Linear fit parameters, $f_\pi$, $f_K$ and $f_K/f_\pi$ determined
from the axial vector current correlator and the pseudoscalar density
correlator. }
\begin{tabular}{ccccccccc}
$\beta$ & $V$ & $L_s$ & correlator & intercept & slope  & $f_\pi$ &
$f_K$ & $f_K/f_\pi$\\
\hline
5.7 & $8^3\times 32$ & 32 & axial & 152(9) & 352(38) & 148(10) &
162(8) & 1.094(16) \\
5.7 & $8^3\times 32$ & 32 & pseudoscalar & 142(6) & 505(25) & 137(6) &
157(5) & 1.146(14) \\
5.7 & $8^3\times 32$ & 48 & axial & 130(4) & 364(18) & 127(4) & 145(4)
& 1.142(11) \\
5.7 & $8^3\times 32$ & 48 & pseudoscalar & 135(4) & 453(21) & 132(4) &
154(4) & 1.171(13)\\
5.7 & $16^3\times 32$ & 48 & axial & 136(3) & 362(33) & 133(4) &
149(2) & 1.122(14) \\
5.7 & $16^3\times 32$ & 48 & pseudoscalar & 129(3) & 525(36) & 125(4)
& 149(2) & 1.188(18) \\
6.0 & $16^3\times 32$ & 16 & axial & 138(10) & 958(195) & 137(11) &
156(8) & 1.134(37) \\
6.0 & $16^3\times 32$ & 16 & pseudoscalar & 138(10) & 938(235) &
137(10)  & 155(8) & 1.131(40) \\
6.0 & $16^3\times 32$ & 24 & axial & 128(10) & 847(222) & 127(10) &
143(7) & 1.124(41) \\
6.0 & $16^3\times 32$ & 24 & pseudoscalar & 123(10) & 1031(232) &
122(11)  & 142(7) & 1.156(49) \\
\end{tabular}
\end{table}
 

\begin{table}
\caption{\label{tab:mN_over_mrho}
Results for $a^{-1}$ using $m_\rho$ extrapolated to $m_f+\mres=0$ to
set the scale and for $m_N/m_\rho$ extrapolated to this same value of $m_f$.
}
\begin{tabular}{ccccc}
$\beta$ & $V$ & $L_s$ & $a^{-1}({\rm GeV})$ & $m_N/m_\rho$ \\
\hline
5.7 & $8^3\times 32$  & 32 & 1.058(40) & 1.51(5) \\
5.7 & $8^3\times 32$  & 48 & 0.994(18) & 1.45(3) \\
5.7 & $16^3\times 32$ & 48 & 1.013(24) & 1.40(5) \\
5.85& $12^3\times 32$ & 20 & 1.419(34) & 1.34(5) \\
6.0 & $16^3\times 32$ & 16 & 1.922(40) & 1.42(4) \\
6.0 & $16^3\times 32$ & 24 & 1.933(50) & 1.38(4) \\
\end{tabular}
\end{table}

\begin{table}
\caption{\label{tab:pbp_GMOR} 
A variety of expressions for $\qbq$ in lattice and physical units.  The
quantity $b$ in the final column comes from our earlier $(a + b m_f)$ fits 
to $m_\pi^2$.  The final column gives a conventionally normalized value of
the chiral condensate which is to be compared with the phenomenological
value of $(229\pm9)^3$.}
\begin{tabular}{cccccc}
$L^3 \times N_t \times L_s$ & $\beta$ & $a_0$ & $-\qbq_{m_f=-\mres}$ 
         & $b\,f_\pi^2 /48$ & $(-12\qbq_{\rm \overline{MS}, 2GeV})^{1/3}$ \\ \hline
$8^3 \times 32\times 32$& 5.7 &$1.76(3) \,10^{-3}$ &$1.07(3) \, 10^{-3}$ 
        & $1.7(2) \,10^{-3}$ & --- \\
$8^3 \times 32\times 48$& 5.7 &$1.92(5) \, 10^{-3}$ &$1.50(5) \, 10^{-3}$
        & $1.8(1) \,10^{-3}$ & --- \\
$16^3\times 32\times 16$& 6.0 &$3.87(8) \, 10^{-4}$ &$2.80(9) \, 10^{-4}$
        & $3.3(5) \,10^{-4}$  & $245(7){\rm MeV}$ \\
$16^3\times 32\times 24$& 6.0 &$3.62(9) \, 10^{-4}$ &$3.11(10)\, 10^{-4}$
        & $2.6(5) \,10^{-4}$ & $256(8){\rm MeV}$ \\
\end{tabular}
\end{table}




\clearpage

\clearpage

\clearpage

\clearpage

\clearpage



\begin{thebibliography}{10}
\expandafter\ifx\csname bibnamefont\endcsname\relax
  \def\bibnamefont#1{#1}\fi
\expandafter\ifx\csname bibfnamefont\endcsname\relax
  \def\bibfnamefont#1{#1}\fi
\expandafter\ifx\csname url\endcsname\relax
  \def\url#1{\texttt{#1}}\fi
\expandafter\ifx\csname urlprefix\endcsname\relax\def\urlprefix{URL }\fi
\expandafter\ifx\csname bibinfo\endcsname\relax \def\bibinfo#1#2{#2}\fi
\expandafter\ifx\csname eprint\endcsname\relax \def\eprint#1{#1}\fi

\bibitem{Kaplan:1992bt}
\bibinfo{author}{\bibfnamefont{D.~B.} \bibnamefont{Kaplan}},
  \bibinfo{journal}{Phys. Lett.} \textbf{\bibinfo{volume}{B288}},
  \bibinfo{pages}{342} (\bibinfo{year}{1992}), \eprint{hep-lat/9206013}.

\bibitem{Narayanan:1993wx}
\bibinfo{author}{\bibfnamefont{R.}~\bibnamefont{Narayanan}} \bibnamefont{and}
  \bibinfo{author}{\bibfnamefont{H.}~\bibnamefont{Neuberger}},
  \bibinfo{journal}{Phys. Lett.} \textbf{\bibinfo{volume}{B302}},
  \bibinfo{pages}{62} (\bibinfo{year}{1993}), \eprint{hep-lat/9212019}.

\bibitem{Narayanan:1993ss}
\bibinfo{author}{\bibfnamefont{R.}~\bibnamefont{Narayanan}} \bibnamefont{and}
  \bibinfo{author}{\bibfnamefont{H.}~\bibnamefont{Neuberger}},
  \bibinfo{journal}{Phys. Rev. Lett.} \textbf{\bibinfo{volume}{71}},
  \bibinfo{pages}{3251} (\bibinfo{year}{1993}), \eprint{hep-lat/9308011}.

\bibitem{Narayanan:1994sk}
\bibinfo{author}{\bibfnamefont{R.}~\bibnamefont{Narayanan}} \bibnamefont{and}
  \bibinfo{author}{\bibfnamefont{H.}~\bibnamefont{Neuberger}},
  \bibinfo{journal}{Nucl. Phys.} \textbf{\bibinfo{volume}{B412}},
  \bibinfo{pages}{574} (\bibinfo{year}{1994}), \eprint{hep-lat/9307006}.

\bibitem{Narayanan:1995gw}
\bibinfo{author}{\bibfnamefont{R.}~\bibnamefont{Narayanan}} \bibnamefont{and}
  \bibinfo{author}{\bibfnamefont{H.}~\bibnamefont{Neuberger}},
  \bibinfo{journal}{Nucl. Phys.} \textbf{\bibinfo{volume}{B443}},
  \bibinfo{pages}{305} (\bibinfo{year}{1995}), \eprint{hep-th/9411108}.

\bibitem{Shamir:1993zy}
\bibinfo{author}{\bibfnamefont{Y.}~\bibnamefont{Shamir}},
  \bibinfo{journal}{Nucl. Phys.} \textbf{\bibinfo{volume}{B406}},
  \bibinfo{pages}{90} (\bibinfo{year}{1993}), \eprint{hep-lat/9303005}.

\bibitem{Narayanan:1993gq}
\bibinfo{author}{\bibfnamefont{R.}~\bibnamefont{Narayanan}},
  \bibinfo{journal}{Nucl. Phys. Proc. Suppl.} \textbf{\bibinfo{volume}{34}},
  \bibinfo{pages}{95} (\bibinfo{year}{1994}), \eprint{hep-lat/9311014}.

\bibitem{Creutz:1994px}
\bibinfo{author}{\bibfnamefont{M.}~\bibnamefont{Creutz}},
  \bibinfo{journal}{Nucl. Phys. Proc. Suppl.} \textbf{\bibinfo{volume}{42}},
  \bibinfo{pages}{56} (\bibinfo{year}{1995}), \eprint{hep-lat/9411033}.

\bibitem{Shamir:1995zx}
\bibinfo{author}{\bibfnamefont{Y.}~\bibnamefont{Shamir}},
  \bibinfo{journal}{Nucl. Phys. Proc. Suppl.} \textbf{\bibinfo{volume}{47}},
  \bibinfo{pages}{212} (\bibinfo{year}{1996}), \eprint{hep-lat/9509023}.

\bibitem{Blum:1998ud}
\bibinfo{author}{\bibfnamefont{T.}~\bibnamefont{Blum}}, \bibinfo{journal}{Nucl.
  Phys. Proc. Suppl.} \textbf{\bibinfo{volume}{73}}, \bibinfo{pages}{167}
  (\bibinfo{year}{1999}), \eprint{hep-lat/9810017}.

\bibitem{Neuberger:1999ry}
\bibinfo{author}{\bibfnamefont{H.}~\bibnamefont{Neuberger}},
  \bibinfo{journal}{Nucl. Phys. Proc. Suppl.} \textbf{\bibinfo{volume}{83-84}},
  \bibinfo{pages}{67} (\bibinfo{year}{2000}), \eprint{hep-lat/9909042}.

\bibitem{Chen:2000zu}
\bibinfo{author}{\bibfnamefont{P.}~\bibnamefont{Chen}} \emph{et~al.}
  (\bibinfo{year}{2000}), \eprint{hep-lat/0006010}.

\bibitem{Narayanan:1997sa}
\bibinfo{author}{\bibfnamefont{R.}~\bibnamefont{Narayanan}} \bibnamefont{and}
  \bibinfo{author}{\bibfnamefont{P.}~\bibnamefont{Vranas}},
  \bibinfo{journal}{Nucl. Phys.} \textbf{\bibinfo{volume}{B506}},
  \bibinfo{pages}{373} (\bibinfo{year}{1997}), \eprint{hep-lat/9702005}.

\bibitem{Chen:1998ne}
\bibinfo{author}{\bibfnamefont{P.}~\bibnamefont{Chen}} \emph{et~al.},
  \bibinfo{journal}{Phys. Rev.} \textbf{\bibinfo{volume}{D59}},
  \bibinfo{pages}{054508} (\bibinfo{year}{1999}), \eprint{hep-lat/9807029}.

\bibitem{Edwards:1998dj}
\bibinfo{author}{\bibfnamefont{R.~G.} \bibnamefont{Edwards}},
  \bibinfo{author}{\bibfnamefont{U.~M.} \bibnamefont{Heller}},
  \bibnamefont{and}
  \bibinfo{author}{\bibfnamefont{R.}~\bibnamefont{Narayanan}},
  \bibinfo{journal}{Phys. Lett.} \textbf{\bibinfo{volume}{B438}},
  \bibinfo{pages}{96} (\bibinfo{year}{1998}), \eprint{hep-lat/9806011}.

\bibitem{Edwards:1998gk}
\bibinfo{author}{\bibfnamefont{R.~G.} \bibnamefont{Edwards}},
  \bibinfo{author}{\bibfnamefont{U.~M.} \bibnamefont{Heller}},
  \bibnamefont{and}
  \bibinfo{author}{\bibfnamefont{R.}~\bibnamefont{Narayanan}},
  \bibinfo{journal}{Nucl. Phys.} \textbf{\bibinfo{volume}{B522}},
  \bibinfo{pages}{285} (\bibinfo{year}{1998}), \eprint{hep-lat/9801015}.

\bibitem{Edwards:1998sh}
\bibinfo{author}{\bibfnamefont{R.~G.} \bibnamefont{Edwards}},
  \bibinfo{author}{\bibfnamefont{U.~M.} \bibnamefont{Heller}},
  \bibnamefont{and}
  \bibinfo{author}{\bibfnamefont{R.}~\bibnamefont{Narayanan}},
  \bibinfo{journal}{Nucl. Phys.} \textbf{\bibinfo{volume}{B535}},
  \bibinfo{pages}{403} (\bibinfo{year}{1998}), \eprint{hep-lat/9802016}.

\bibitem{Gadiyak:2000kz}
\bibinfo{author}{\bibfnamefont{V.}~\bibnamefont{Gadiyak}},
  \bibinfo{author}{\bibfnamefont{X.-D.} \bibnamefont{Ji}}, \bibnamefont{and}
  \bibinfo{author}{\bibfnamefont{C.-W.} \bibnamefont{Jung}}
  (\bibinfo{year}{2000}), \eprint{hep-lat/0002023}.

\bibitem{Blum:1997mz}
\bibinfo{author}{\bibfnamefont{T.}~\bibnamefont{Blum}} \bibnamefont{and}
  \bibinfo{author}{\bibfnamefont{A.}~\bibnamefont{Soni}},
  \bibinfo{journal}{Phys. Rev. Lett.} \textbf{\bibinfo{volume}{79}},
  \bibinfo{pages}{3595} (\bibinfo{year}{1997}), \eprint{hep-lat/9706023}.

\bibitem{Blum:1997jf}
\bibinfo{author}{\bibfnamefont{T.}~\bibnamefont{Blum}} \bibnamefont{and}
  \bibinfo{author}{\bibfnamefont{A.}~\bibnamefont{Soni}},
  \bibinfo{journal}{Phys. Rev.} \textbf{\bibinfo{volume}{D56}},
  \bibinfo{pages}{174} (\bibinfo{year}{1997}), \eprint{hep-lat/9611030}.

\bibitem{Mawhinney:1998ut}
\bibinfo{author}{\bibfnamefont{R.}~\bibnamefont{Mawhinney}} \emph{et~al.},
  \bibinfo{journal}{Nucl. Phys. Proc. Suppl.} \textbf{\bibinfo{volume}{73}},
  \bibinfo{pages}{204} (\bibinfo{year}{1999}), \eprint{hep-lat/9811026}.

\bibitem{Fleming:1998cc}
\bibinfo{author}{\bibfnamefont{G.~T.} \bibnamefont{Fleming}} \emph{et~al.},
  \bibinfo{journal}{Nucl. Phys. Proc. Suppl.} \textbf{\bibinfo{volume}{73}},
  \bibinfo{pages}{207} (\bibinfo{year}{1999}), \eprint{hep-lat/9811013}.

\bibitem{Kaehler:1998xx}
\bibinfo{author}{\bibfnamefont{A.~L.} \bibnamefont{Kaehler}} \emph{et~al.},
  \bibinfo{journal}{Nucl. Phys. Proc. Suppl.} \textbf{\bibinfo{volume}{73}},
  \bibinfo{pages}{405} (\bibinfo{year}{1999}).

\bibitem{Blum:1999xi}
\bibinfo{author}{\bibfnamefont{T.}~\bibnamefont{Blum}},
  \bibinfo{author}{\bibfnamefont{A.}~\bibnamefont{Soni}}, \bibnamefont{and}
  \bibinfo{author}{\bibfnamefont{M.}~\bibnamefont{Wingate}},
  \bibinfo{journal}{Phys. Rev.} \textbf{\bibinfo{volume}{D60}},
  \bibinfo{pages}{114507} (\bibinfo{year}{1999}), \eprint{hep-lat/9902016}.

\bibitem{Fleming:1999eq}
\bibinfo{author}{\bibfnamefont{G.~T.} \bibnamefont{Fleming}},
  \bibinfo{journal}{Nucl. Phys. Proc. Suppl.} \textbf{\bibinfo{volume}{83-84}},
  \bibinfo{pages}{363} (\bibinfo{year}{2000}), \eprint{hep-lat/9909140}.

\bibitem{Wu:1999cd}
\bibinfo{author}{\bibfnamefont{L.}~\bibnamefont{Wu}}
  (\bibinfo{collaboration}{RIKEN-BNL-CU}), \bibinfo{journal}{Nucl. Phys. Proc.
  Suppl.} \textbf{\bibinfo{volume}{83-84}}, \bibinfo{pages}{224}
  (\bibinfo{year}{2000}), \eprint{hep-lat/9909117}.

\bibitem{AliKhan:1999zn}
\bibinfo{author}{\bibfnamefont{A.~A.} \bibnamefont{Khan}} \emph{et~al.}
  (\bibinfo{collaboration}{CP-PACS}), \bibinfo{journal}{Nucl. Phys. Proc.
  Suppl.} \textbf{\bibinfo{volume}{83-84}}, \bibinfo{pages}{591}
  (\bibinfo{year}{2000}), \eprint{hep-lat/9909049}.

\bibitem{Edwards:1999bm}
\bibinfo{author}{\bibfnamefont{R.~G.} \bibnamefont{Edwards}},
  \bibinfo{author}{\bibfnamefont{U.~M.} \bibnamefont{Heller}},
  \bibnamefont{and}
  \bibinfo{author}{\bibfnamefont{R.}~\bibnamefont{Narayanan}},
  \bibinfo{journal}{Phys. Rev.} \textbf{\bibinfo{volume}{D60}},
  \bibinfo{pages}{034502} (\bibinfo{year}{1999}), \eprint{hep-lat/9901015}.

\bibitem{Aoki:1999uv}
\bibinfo{author}{\bibfnamefont{S.}~\bibnamefont{Aoki}},
  \bibinfo{author}{\bibfnamefont{T.}~\bibnamefont{Izubuchi}},
  \bibinfo{author}{\bibfnamefont{Y.}~\bibnamefont{Kuramashi}},
  \bibnamefont{and}
  \bibinfo{author}{\bibfnamefont{Y.}~\bibnamefont{Taniguchi}},
  \bibinfo{journal}{Nucl. Phys. Proc. Suppl.} \textbf{\bibinfo{volume}{83-84}},
  \bibinfo{pages}{624} (\bibinfo{year}{2000}), \eprint{hep-lat/9909154}.

\bibitem{Aoki:2000pc}
\bibinfo{author}{\bibfnamefont{S.}~\bibnamefont{Aoki}},
  \bibinfo{author}{\bibfnamefont{T.}~\bibnamefont{Izubuchi}},
  \bibinfo{author}{\bibfnamefont{Y.}~\bibnamefont{Kuramashi}},
  \bibnamefont{and} \bibinfo{author}{\bibfnamefont{Y.}~\bibnamefont{Taniguchi}}
   (\bibinfo{year}{2000}), \eprint{hep-lat/0004003}.

\bibitem{Chen:1998xw}
\bibinfo{author}{\bibfnamefont{P.}~\bibnamefont{Chen}} \emph{et~al.}
  (\bibinfo{year}{1998}), \eprint{hep-lat/9812011}.

\bibitem{Vranas:1997tj}
\bibinfo{author}{\bibfnamefont{P.}~\bibnamefont{Vranas}},
  \bibinfo{journal}{Nucl. Phys. Proc. Suppl.} \textbf{\bibinfo{volume}{53}},
  \bibinfo{pages}{278} (\bibinfo{year}{1997}), \eprint{hep-lat/9608078}.

\bibitem{Vranas:1998vm}
\bibinfo{author}{\bibfnamefont{P.}~\bibnamefont{Vranas}} \emph{et~al.},
  \bibinfo{journal}{Nucl. Phys. Proc. Suppl.} \textbf{\bibinfo{volume}{73}},
  \bibinfo{pages}{456} (\bibinfo{year}{1999}), \eprint{hep-lat/9809159}.

\bibitem{Vranas:1999nx}
\bibinfo{author}{\bibfnamefont{P.}~\bibnamefont{Vranas}},
  \bibinfo{author}{\bibfnamefont{I.}~\bibnamefont{Tziligakis}},
  \bibnamefont{and} \bibinfo{author}{\bibfnamefont{J.}~\bibnamefont{Kogut}}
  (\bibinfo{year}{1999}), \eprint{hep-lat/9905018}.

\bibitem{Vranas:1999pm}
\bibinfo{author}{\bibfnamefont{P.}~\bibnamefont{Vranas}}
  (\bibinfo{year}{1999}), \eprint{hep-lat/9903024}.

\bibitem{Neuberger:1997fp}
\bibinfo{author}{\bibfnamefont{H.}~\bibnamefont{Neuberger}},
  \bibinfo{journal}{Phys. Lett.} \textbf{\bibinfo{volume}{B417}},
  \bibinfo{pages}{141} (\bibinfo{year}{1998}), \eprint{hep-lat/9707022}.

\bibitem{Edwards:1998wx}
\bibinfo{author}{\bibfnamefont{R.~G.} \bibnamefont{Edwards}},
  \bibinfo{author}{\bibfnamefont{U.~M.} \bibnamefont{Heller}},
  \bibnamefont{and}
  \bibinfo{author}{\bibfnamefont{R.}~\bibnamefont{Narayanan}},
  \bibinfo{journal}{Phys. Rev.} \textbf{\bibinfo{volume}{D59}},
  \bibinfo{pages}{094510} (\bibinfo{year}{1999}), \eprint{hep-lat/9811030}.

\bibitem{Hernandez:1998et}
\bibinfo{author}{\bibfnamefont{P.}~\bibnamefont{Hernandez}},
  \bibinfo{author}{\bibfnamefont{K.}~\bibnamefont{Jansen}}, \bibnamefont{and}
  \bibinfo{author}{\bibfnamefont{M.}~\bibnamefont{Luscher}},
  \bibinfo{journal}{Nucl. Phys.} \textbf{\bibinfo{volume}{B552}},
  \bibinfo{pages}{363} (\bibinfo{year}{1999}), \eprint{hep-lat/9808010}.

\bibitem{Borici:1999zw}
\bibinfo{author}{\bibfnamefont{A.}~\bibnamefont{Borici}},
  \bibinfo{journal}{Nucl. Phys. Proc. Suppl.} \textbf{\bibinfo{volume}{83-84}},
  \bibinfo{pages}{771} (\bibinfo{year}{2000}), \eprint{hep-lat/9909057}.

\bibitem{Edwards:1999yi}
\bibinfo{author}{\bibfnamefont{R.~G.} \bibnamefont{Edwards}},
  \bibinfo{author}{\bibfnamefont{U.~M.} \bibnamefont{Heller}},
  \bibnamefont{and} \bibinfo{author}{\bibfnamefont{R.}~\bibnamefont{Narayanan}}
   (\bibinfo{year}{1999}), \eprint{hep-lat/0001013}.

\bibitem{Giusti:1999be}
\bibinfo{author}{\bibfnamefont{L.}~\bibnamefont{Giusti}},
  \bibinfo{author}{\bibfnamefont{C.}~\bibnamefont{Hoelbling}},
  \bibnamefont{and} \bibinfo{author}{\bibfnamefont{C.}~\bibnamefont{Rebbi}},
  \bibinfo{journal}{Nucl. Phys. Proc. Suppl.} \textbf{\bibinfo{volume}{83-84}},
  \bibinfo{pages}{896} (\bibinfo{year}{2000}), \eprint{hep-lat/9906004}.

\bibitem{Neuberger:1999rz}
\bibinfo{author}{\bibfnamefont{H.}~\bibnamefont{Neuberger}},
  \bibinfo{journal}{Nucl. Phys. Proc. Suppl.} \textbf{\bibinfo{volume}{83-84}},
  \bibinfo{pages}{813} (\bibinfo{year}{2000}), \eprint{hep-lat/9909043}.

\bibitem{Dong:2000mr}
\bibinfo{author}{\bibfnamefont{S.~J.} \bibnamefont{Dong}},
  \bibinfo{author}{\bibfnamefont{F.~X.} \bibnamefont{Lee}},
  \bibinfo{author}{\bibfnamefont{K.~F.} \bibnamefont{Liu}}, \bibnamefont{and}
  \bibinfo{author}{\bibfnamefont{J.~B.} \bibnamefont{Zhang}}
  (\bibinfo{year}{2000}), \eprint{hep-lat/0006004}.

\bibitem{Edwards:2000qv}
\bibinfo{author}{\bibfnamefont{R.~G.} \bibnamefont{Edwards}} \bibnamefont{and}
  \bibinfo{author}{\bibfnamefont{U.~M.} \bibnamefont{Heller}}
  (\bibinfo{year}{2000}), \eprint{hep-lat/0005002}.

\bibitem{Hernandez:2000sb}
\bibinfo{author}{\bibfnamefont{P.}~\bibnamefont{Hernandez}},
  \bibinfo{author}{\bibfnamefont{K.}~\bibnamefont{Jansen}}, \bibnamefont{and}
  \bibinfo{author}{\bibfnamefont{L.}~\bibnamefont{Lellouch}}
  (\bibinfo{year}{2000}), \eprint{hep-lat/0001008}.

\bibitem{Narayanan:2000qx}
\bibinfo{author}{\bibfnamefont{R.}~\bibnamefont{Narayanan}} \bibnamefont{and}
  \bibinfo{author}{\bibfnamefont{H.}~\bibnamefont{Neuberger}}
  (\bibinfo{year}{2000}), \eprint{hep-lat/0005004}.

\bibitem{Furman:1995ky}
\bibinfo{author}{\bibfnamefont{V.}~\bibnamefont{Furman}} \bibnamefont{and}
  \bibinfo{author}{\bibfnamefont{Y.}~\bibnamefont{Shamir}},
  \bibinfo{journal}{Nucl. Phys.} \textbf{\bibinfo{volume}{B439}},
  \bibinfo{pages}{54} (\bibinfo{year}{1995}), \eprint{hep-lat/9405004}.

\bibitem{Kalkreuter:1996mm}
\bibinfo{author}{\bibfnamefont{T.}~\bibnamefont{Kalkreuter}} \bibnamefont{and}
  \bibinfo{author}{\bibfnamefont{H.}~\bibnamefont{Simma}},
  \bibinfo{journal}{Comput. Phys. Commun.} \textbf{\bibinfo{volume}{93}},
  \bibinfo{pages}{33} (\bibinfo{year}{1996}), \eprint{hep-lat/9507023}.

\bibitem{Symanzik:1983dc}
\bibinfo{author}{\bibfnamefont{K.}~\bibnamefont{Symanzik}},
  \bibinfo{journal}{Nucl. Phys.} \textbf{\bibinfo{volume}{B226}},
  \bibinfo{pages}{187} (\bibinfo{year}{1983}).

\bibitem{Symanzik:1983gh}
\bibinfo{author}{\bibfnamefont{K.}~\bibnamefont{Symanzik}},
  \bibinfo{journal}{Nucl. Phys.} \textbf{\bibinfo{volume}{B226}},
  \bibinfo{pages}{205} (\bibinfo{year}{1983}).

\bibitem{Sheikholeslami:1985ij}
\bibinfo{author}{\bibfnamefont{B.}~\bibnamefont{Sheikholeslami}}
  \bibnamefont{and} \bibinfo{author}{\bibfnamefont{R.}~\bibnamefont{Wohlert}},
  \bibinfo{journal}{Nucl. Phys.} \textbf{\bibinfo{volume}{B259}},
  \bibinfo{pages}{572} (\bibinfo{year}{1985}).

\bibitem{Wilson:1974sk}
\bibinfo{author}{\bibfnamefont{K.~G.} \bibnamefont{Wilson}},
  \bibinfo{journal}{Phys. Rev.} \textbf{\bibinfo{volume}{D10}},
  \bibinfo{pages}{2445} (\bibinfo{year}{1974}).

\bibitem{Gottlieb:1987mq}
\bibinfo{author}{\bibfnamefont{S.}~\bibnamefont{Gottlieb}},
  \bibinfo{author}{\bibfnamefont{W.}~\bibnamefont{Liu}},
  \bibinfo{author}{\bibfnamefont{D.}~\bibnamefont{Toussaint}},
  \bibinfo{author}{\bibfnamefont{R.~L.} \bibnamefont{Renken}},
  \bibnamefont{and} \bibinfo{author}{\bibfnamefont{R.~L.} \bibnamefont{Sugar}},
  \bibinfo{journal}{Phys. Rev.} \textbf{\bibinfo{volume}{D35}},
  \bibinfo{pages}{2531} (\bibinfo{year}{1987}).

\bibitem{Creutz:1980zw}
\bibinfo{author}{\bibfnamefont{M.}~\bibnamefont{Creutz}},
  \bibinfo{journal}{Phys. Rev.} \textbf{\bibinfo{volume}{D21}},
  \bibinfo{pages}{2308} (\bibinfo{year}{1980}).

\bibitem{Cabibbo:1982zn}
\bibinfo{author}{\bibfnamefont{N.}~\bibnamefont{Cabibbo}} \bibnamefont{and}
  \bibinfo{author}{\bibfnamefont{E.}~\bibnamefont{Marinari}},
  \bibinfo{journal}{Phys. Lett.} \textbf{\bibinfo{volume}{B119}},
  \bibinfo{pages}{387} (\bibinfo{year}{1982}).

\bibitem{Kennedy:1985nu}
\bibinfo{author}{\bibfnamefont{A.~D.} \bibnamefont{Kennedy}} \bibnamefont{and}
  \bibinfo{author}{\bibfnamefont{B.~J.} \bibnamefont{Pendleton}},
  \bibinfo{journal}{Phys. Lett.} \textbf{\bibinfo{volume}{B156}},
  \bibinfo{pages}{393} (\bibinfo{year}{1985}).

\bibitem{MILC:1991aa}
\bibinfo{author}{\bibfnamefont{C.}~\bibnamefont{Bernard}} \emph{et~al.}
  \bibinfo{note}{In {\it Workshop on Fermion Algorithms}, edited by
  H.~J.~Hermann and F.~Karsch, (World Scientific, Singapore, 1991)}.

\bibitem{Brown:1987rr}
\bibinfo{author}{\bibfnamefont{F.~R.} \bibnamefont{Brown}} \bibnamefont{and}
  \bibinfo{author}{\bibfnamefont{T.~J.} \bibnamefont{Woch}},
  \bibinfo{journal}{Phys. Rev. Lett.} \textbf{\bibinfo{volume}{58}},
  \bibinfo{pages}{2394} (\bibinfo{year}{1987}).

\bibitem{Creutz:1987xi}
\bibinfo{author}{\bibfnamefont{M.}~\bibnamefont{Creutz}},
  \bibinfo{journal}{Phys. Rev.} \textbf{\bibinfo{volume}{D36}},
  \bibinfo{pages}{515} (\bibinfo{year}{1987}).

\bibitem{Wingate:1999yr}
\bibinfo{author}{\bibfnamefont{M.}~\bibnamefont{Wingate}},
  \bibinfo{journal}{Nucl. Phys. Proc. Suppl.} \textbf{\bibinfo{volume}{83-84}},
  \bibinfo{pages}{221} (\bibinfo{year}{2000}), \eprint{hep-lat/9909101}.

\bibitem{Malureanu:1998thesis}
\bibinfo{author}{\bibfnamefont{C.}~\bibnamefont{Malureanu}}
  (\bibinfo{year}{1998}), \bibinfo{note}{\uppercase{P}h.D. thesis,
  unpublished}.

\bibitem{Schwarz:1977az}
\bibinfo{author}{\bibfnamefont{A.~S.} \bibnamefont{Schwarz}},
  \bibinfo{journal}{Phys. Lett.} \textbf{\bibinfo{volume}{B67}},
  \bibinfo{pages}{172} (\bibinfo{year}{1977}).

\bibitem{Brown:1977bj}
\bibinfo{author}{\bibfnamefont{L.~S.} \bibnamefont{Brown}},
  \bibinfo{author}{\bibfnamefont{R.~D.} \bibnamefont{Carlitz}},
  \bibnamefont{and} \bibinfo{author}{\bibfnamefont{C.}~\bibnamefont{Lee}},
  \bibinfo{journal}{Phys. Rev.} \textbf{\bibinfo{volume}{D16}},
  \bibinfo{pages}{417} (\bibinfo{year}{1977}).

\bibitem{Banks:1980yr}
\bibinfo{author}{\bibfnamefont{T.}~\bibnamefont{Banks}} \bibnamefont{and}
  \bibinfo{author}{\bibfnamefont{A.}~\bibnamefont{Casher}},
  \bibinfo{journal}{Nucl. Phys.} \textbf{\bibinfo{volume}{B169}},
  \bibinfo{pages}{103} (\bibinfo{year}{1980}).

\bibitem{Chandrasekharan:1998yx}
\bibinfo{author}{\bibfnamefont{S.}~\bibnamefont{Chandrasekharan}}
  \emph{et~al.}, \bibinfo{journal}{Phys. Rev. Lett.}
  \textbf{\bibinfo{volume}{82}}, \bibinfo{pages}{2463} (\bibinfo{year}{1999}),
  \eprint{hep-lat/9807018}.

\bibitem{Aoki:1994gi}
\bibinfo{author}{\bibfnamefont{S.}~\bibnamefont{Aoki}} \emph{et~al.},
  \bibinfo{journal}{Phys. Rev.} \textbf{\bibinfo{volume}{D50}},
  \bibinfo{pages}{486} (\bibinfo{year}{1994}).

\bibitem{Mawhinney:1996qy}
\bibinfo{author}{\bibfnamefont{R.~D.} \bibnamefont{Mawhinney}},
  \bibinfo{journal}{Nucl. Phys. Proc. Suppl.} \textbf{\bibinfo{volume}{47}},
  \bibinfo{pages}{557} (\bibinfo{year}{1996}), \eprint{hep-lat/9603019}.

\bibitem{Sharpe:1990me}
\bibinfo{author}{\bibfnamefont{S.~R.} \bibnamefont{Sharpe}},
  \bibinfo{journal}{Phys. Rev.} \textbf{\bibinfo{volume}{D41}},
  \bibinfo{pages}{3233} (\bibinfo{year}{1990}).

\bibitem{Bernard:1992mk}
\bibinfo{author}{\bibfnamefont{C.~W.} \bibnamefont{Bernard}} \bibnamefont{and}
  \bibinfo{author}{\bibfnamefont{M.~F.~L.} \bibnamefont{Golterman}},
  \bibinfo{journal}{Phys. Rev.} \textbf{\bibinfo{volume}{D46}},
  \bibinfo{pages}{853} (\bibinfo{year}{1992}), \eprint{hep-lat/9204007}.

\bibitem{Sharpe:1992ft}
\bibinfo{author}{\bibfnamefont{S.~R.} \bibnamefont{Sharpe}},
  \bibinfo{journal}{Phys. Rev.} \textbf{\bibinfo{volume}{D46}},
  \bibinfo{pages}{3146} (\bibinfo{year}{1992}), \eprint{hep-lat/9205020}.

\bibitem{Leutwyler:1992yt}
\bibinfo{author}{\bibfnamefont{H.}~\bibnamefont{Leutwyler}} \bibnamefont{and}
  \bibinfo{author}{\bibfnamefont{A.}~\bibnamefont{Smilga}},
  \bibinfo{journal}{Phys. Rev.} \textbf{\bibinfo{volume}{D46}},
  \bibinfo{pages}{5607} (\bibinfo{year}{1992}).

\bibitem{Dawson:1999yx}
\bibinfo{author}{\bibfnamefont{C.}~\bibnamefont{Dawson}},
  \bibinfo{journal}{Nucl. Phys. Proc. Suppl.} \textbf{\bibinfo{volume}{83-84}},
  \bibinfo{pages}{854} (\bibinfo{year}{2000}), \eprint{hep-lat/9909107}.

\bibitem{Npr:2000zz}
 \bibinfo{note}{RIKEN-Brookhaven-Columbia collaboration, in preparation}.

\bibitem{Kim:1995tg}
\bibinfo{author}{\bibfnamefont{S.}~\bibnamefont{Kim}} \bibnamefont{and}
  \bibinfo{author}{\bibfnamefont{D.~K.} \bibnamefont{Sinclair}},
  \bibinfo{journal}{Phys. Rev.} \textbf{\bibinfo{volume}{D52}},
  \bibinfo{pages}{2614} (\bibinfo{year}{1995}), \eprint{hep-lat/9502004}.

\bibitem{Duncan:1996ma}
\bibinfo{author}{\bibfnamefont{A.}~\bibnamefont{Duncan}},
  \bibinfo{author}{\bibfnamefont{E.}~\bibnamefont{Eichten}},
  \bibinfo{author}{\bibfnamefont{S.}~\bibnamefont{Perrucci}}, \bibnamefont{and}
  \bibinfo{author}{\bibfnamefont{H.}~\bibnamefont{Thacker}},
  \bibinfo{journal}{Nucl. Phys. Proc. Suppl.} \textbf{\bibinfo{volume}{53}},
  \bibinfo{pages}{256} (\bibinfo{year}{1997}), \eprint{hep-lat/9608110}.

\bibitem{Bernard:1998db}
\bibinfo{author}{\bibfnamefont{C.}~\bibnamefont{Bernard}} \emph{et~al.}
  (\bibinfo{collaboration}{MILC}), \bibinfo{journal}{Phys. Rev. Lett.}
  \textbf{\bibinfo{volume}{81}}, \bibinfo{pages}{3087} (\bibinfo{year}{1998}),
  \eprint{hep-lat/9805004}.

\bibitem{Aoki:1999yr}
\bibinfo{author}{\bibfnamefont{S.}~\bibnamefont{Aoki}} \emph{et~al.}
  (\bibinfo{collaboration}{CP-PACS}), \bibinfo{journal}{Phys. Rev. Lett.}
  \textbf{\bibinfo{volume}{84}}, \bibinfo{pages}{238} (\bibinfo{year}{2000}),
  \eprint{hep-lat/9904012}.

\bibitem{Bardeen:2000cz}
\bibinfo{author}{\bibfnamefont{W.}~\bibnamefont{Bardeen}},
  \bibinfo{author}{\bibfnamefont{A.}~\bibnamefont{Duncan}},
  \bibinfo{author}{\bibfnamefont{E.}~\bibnamefont{Eichten}}, \bibnamefont{and}
  \bibinfo{author}{\bibfnamefont{H.}~\bibnamefont{Thacker}}
  (\bibinfo{year}{2000}), \eprint{hep-lat/0007010}.

\bibitem{Gockeler:2000pg}
\bibinfo{author}{\bibfnamefont{M.}~\bibnamefont{Gockeler}} \emph{et~al.}
  (\bibinfo{year}{2000}), \eprint{hep-lat/0002013}.

\bibitem{eigen:2000aa}
 \bibinfo{note}{RIKEN-Brookhaven-Columbia collaboration, in preparation}.

\bibitem{Butler:1994zx}
\bibinfo{author}{\bibfnamefont{F.}~\bibnamefont{Butler}},
  \bibinfo{author}{\bibfnamefont{H.}~\bibnamefont{Chen}},
  \bibinfo{author}{\bibfnamefont{J.}~\bibnamefont{Sexton}},
  \bibinfo{author}{\bibfnamefont{A.}~\bibnamefont{Vaccarino}},
  \bibnamefont{and}
  \bibinfo{author}{\bibfnamefont{D.}~\bibnamefont{Weingarten}},
  \bibinfo{journal}{Nucl. Phys.} \textbf{\bibinfo{volume}{B421}},
  \bibinfo{pages}{217} (\bibinfo{year}{1994}), \eprint{hep-lat/9310009}.

\bibitem{Chen:1997jj}
\bibinfo{author}{\bibfnamefont{D.}~\bibnamefont{Chen}} \bibnamefont{and}
  \bibinfo{author}{\bibfnamefont{R.~D.} \bibnamefont{Mawhinney}},
  \bibinfo{journal}{Nucl. Phys. Proc. Suppl.} \textbf{\bibinfo{volume}{53}},
  \bibinfo{pages}{216} (\bibinfo{year}{1997}), \eprint{hep-lat/9705029}.

\bibitem{Chen:1996thesis}
\bibinfo{author}{\bibfnamefont{D.}~\bibnamefont{Chen}}  (\bibinfo{year}{1996}),
  \bibinfo{note}{\uppercase{P}h.D. thesis, unpublished}.

\bibitem{Gottlieb:1997hy}
\bibinfo{author}{\bibfnamefont{S.}~\bibnamefont{Gottlieb}},
  \bibinfo{journal}{Nucl. Phys. Proc. Suppl.} \textbf{\bibinfo{volume}{53}},
  \bibinfo{pages}{155} (\bibinfo{year}{1997}), \eprint{hep-lat/9608107}.

\bibitem{Butler:1994em}
\bibinfo{author}{\bibfnamefont{F.}~\bibnamefont{Butler}},
  \bibinfo{author}{\bibfnamefont{H.}~\bibnamefont{Chen}},
  \bibinfo{author}{\bibfnamefont{J.}~\bibnamefont{Sexton}},
  \bibinfo{author}{\bibfnamefont{A.}~\bibnamefont{Vaccarino}},
  \bibnamefont{and}
  \bibinfo{author}{\bibfnamefont{D.}~\bibnamefont{Weingarten}},
  \bibinfo{journal}{Nucl. Phys.} \textbf{\bibinfo{volume}{B430}},
  \bibinfo{pages}{179} (\bibinfo{year}{1994}), \eprint{hep-lat/9405003}.

\bibitem{Narison:1995hz}
\bibinfo{author}{\bibfnamefont{S.}~\bibnamefont{Narison}},
  \bibinfo{journal}{Phys. Lett.} \textbf{\bibinfo{volume}{B358}},
  \bibinfo{pages}{113} (\bibinfo{year}{1995}), \eprint{hep-ph/9504333}.

\bibitem{Gell-Mann:1968rz}
\bibinfo{author}{\bibfnamefont{M.}~\bibnamefont{Gell-Mann}},
  \bibinfo{author}{\bibfnamefont{R.~J.} \bibnamefont{Oakes}}, \bibnamefont{and}
  \bibinfo{author}{\bibfnamefont{B.}~\bibnamefont{Renner}},
  \bibinfo{journal}{Phys. Rev.} \textbf{\bibinfo{volume}{175}},
  \bibinfo{pages}{2195} (\bibinfo{year}{1968}).

\bibitem{Kikukawa:1997md}
\bibinfo{author}{\bibfnamefont{Y.}~\bibnamefont{Kikukawa}},
  \bibinfo{author}{\bibfnamefont{R.}~\bibnamefont{Narayanan}},
  \bibnamefont{and}
  \bibinfo{author}{\bibfnamefont{H.}~\bibnamefont{Neuberger}},
  \bibinfo{journal}{Phys. Lett.} \textbf{\bibinfo{volume}{B399}},
  \bibinfo{pages}{105} (\bibinfo{year}{1997}), \eprint{hep-th/9701007}.

\bibitem{Khan:2000iv}
\bibinfo{author}{\bibfnamefont{A.~A.} \bibnamefont{Khan}} \emph{et~al.}
  (\bibinfo{collaboration}{CP-PACS})  (\bibinfo{year}{2000}),
  \eprint{hep-lat/0007014}.

\bibitem{Chen:1998cg}
\bibinfo{author}{\bibfnamefont{D.}~\bibnamefont{Chen}} \emph{et~al.},
  \bibinfo{journal}{Nucl. Phys. Proc. Suppl.} \textbf{\bibinfo{volume}{73}},
  \bibinfo{pages}{898} (\bibinfo{year}{1999}), \eprint{hep-lat/9810004}.

\bibitem{Mawhinney:2000fx}
\bibinfo{author}{\bibfnamefont{R.~D.} \bibnamefont{Mawhinney}},
  \bibinfo{journal}{Parallel Comput.} \textbf{\bibinfo{volume}{25}},
  \bibinfo{pages}{1281} (\bibinfo{year}{1999}), \eprint{hep-lat/0001033}.

\end{thebibliography}
\end{document}